\documentclass{LMCS}

\usepackage{graphics}
\usepackage[all]{xypic}
\usepackage{color}

\usepackage{afterpage}

\usepackage{seplog,tech}
\usepackage{environments}

\usepackage{bussproofs,enumerate,hyperref}

\theoremstyle{definition}

\newcommand{\srn}[1]{\Small\rn{#1}}

\def\doi{4 (1:6) 2008}
\lmcsheading%
{\doi}
{1--68}
{}
{}
{Apr.~10, 2007}
{Mar.~19, 2008}
{}

\begin{document}
\title{Independence and Concurrent Separation Logic\rsuper *}
\author[J.~Hayman]{Jonathan Hayman}      
\address{Computer Laboratory\\
  University of Cambridge\\
  William Gates Building\\
  15 JJ Thomson Avenue\\
  Cambridge CB£ 0FD\\
  United Kingdom}
\vskip-6 pt

\author[G.~Winskel]{Glynn Winskel}        
\email{\{jonathan.hayman,glynn.winskel\}@cl.cam.ac.uk}

\titlecomment{{\lsuper *}Extended version of \cite{lics}}

\begin{abstract}
  A compositional Petri net-based semantics is given to a simple
  language allowing pointer manipulation and parallelism.  The model
  is then applied to give a notion of validity to the judgements made
  by concurrent separation logic that emphasizes the
  process-environment duality inherent in such rely-guarantee
  reasoning.  Soundness of the rules of concurrent separation logic
  with respect to this definition of validity is shown.  The
  independence information retained by the Petri net model is then
  exploited to characterize the independence of parallel processes
  enforced by the logic.  This is shown to permit a refinement
  operation capable of changing the granularity of atomic actions.
\end{abstract}

\maketitle

\section{Introduction}
The foundational work of Hoare on parallel programming
\cite{hoare:tatpp} identified the fact that attributing an interleaved
semantics to parallel languages is problematic.  Three areas of
difficulty were isolated, quoted directly: {\it
\begin{vtitemize}
\item That of defining a `unit of action'.
\item That of implementing the interleaving on genuinely parallel hardware.
\item That of designing programs to control the fantastic number of
  combinations involved in arbitrary interleaving.
\end{vtitemize}}
The significance of these problems increases with developments in
hardware, such as multiple-core processors, that allow primitive
machine actions to occur at the same time. 

As Hoare went on to explain, a feature of concurrent systems in the
physical world is that they are often \emph{spatially separated},
operating on completely different resources and not interacting.  When
this is so, the systems are \emph{independent} of each other, and
therefore it is unnecessary to consider how they interact.  This
perspective can be extended by regarding computer processes as
spatially separated if they operate on different memory locations.
The problems above are resolved if the occurrence of non-independent
parallel actions is prohibited except in rare cases where atomicity
may be assumed, as might be enforced using the constructs proposed in
\cite{dijkstra, pbhansen:structmulti}.

Independence models for concurrency allow semantics to be given to
parallel languages in a way that can tackle the problems associated
with an interleaved semantics.  The common core of independence models
is that they record when actions are independent, and that independent
actions can be run in either order or even concurrently with no
consequence on their effect.  This mitigates the increase in the state
space since unnecessary interleavings of independent actions need not
be considered (see \eg~\cite{clarkeetal:pmcheck} for applications to
model checking).  Independence models also permit easier notions of
refinement which allow the assumed atomicity of actions to be changed.

It is surprising that, to our knowledge, there has been no
\emph{comprehensive} study of the semantics of programming languages
inside an independence model.  The first component of our work gives
such a semantics in terms of a well-known independence model, namely
Petri nets.  Our model isolates the specification of the control flow
of programs from their effect on the shared state.  It indicates what
appears to be a general method (an alternative to Plotkin's structural
operational semantics) for giving a \emph{structural Petri net
semantics} to a variety of languages --- see the Conclusion, Section
\ref{sec:conclusion}.

The language that we consider is motivated by the emergence of
\emph{concurrent separation logic} \cite{ohearn:rclr}, the rules of
which form a partial correctness judgement about the execution of
pointer-manipulating concurrent programs.  Reasoning about such
programs has traditionally proved difficult due to the problem of
\emph{variable aliasing}.  For instance, Owicki and Gries' system for
proving properties of parallel programs that do not manipulate
pointers \cite{owicki:acm} essentially requires that the programs
operate on disjoint collections of variables, thereby allowing
judgements to be composed.  In the presence of pointers, the same
syntactic condition cannot be imposed to yield a sound logic since
distinct variables may point to the same memory location, thereby
allowing arbitrary interaction between the processes.  To give a
specific example, Owicki and Gries' system would allow a judgement of
the form
\[ \ho{x\mapsto 0 \land y\mapsto 0}{x:= 1\pll y:= 2}{x\mapsto 1\land y\mapsto 2},\]
indicating that the result of assigning $1$ to the program variable
$x$ concurrently with assigning $2$ to $y$ from a state where $x$ and
$y$ both initially hold value $0$ is a state where $x$ holds value $1$
and $y$ holds value $2$.  The judgement is sound because the variables
$x$ and $y$ are distinct.  If pointers are introduced to the language,
however, it is not sound to conclude that
\[ \ho{[x]\mapsto 0 \land [y]\mapsto 0}{[x]:= 1\pll 
  [y]:= 2}{[x]\mapsto 1\land [y]\mapsto 2},\] which would indicate
that assigning $1$ to the location pointed to by $x$ and $2$ to the
location pointed to by $y$ yields a state in which $x$ points to a
location holding $1$ and $y$ points to a location holding $2$, since
$x$ and $y$ may both point to the same location.

At the core of separation logic \cite{REYNOLDS00,iohearn},
initially presented for non-concurrent programs, is the
\emph{separating conjunction}, $\phi\mand\psi$, which asserts that the
state in which processes execute may be split into two parts,
one part satisfying $\phi$ and the other $\psi$.  The separating
conjunction was used by O'Hearn to adapt Owicki and Gries' system to
provide a rule for parallel composition suitable for
pointer-manipulating programs \cite{ohearn:rclr}.

As we shall see, the rule for parallel composition is informally
understood by
splitting the initial state into two parts, one \emph{owned} by the
first process and the other by the second.  Ownership can be seen as a
dynamic constraint on the interference to be assumed: parallel
processes always own disjoint sets of locations and only ever act on
locations that they own.  As processes evolve, ownership of locations
may be transferred using a system of \emph{invariants} (an example is
presented in Section \ref{sec:seplogic}).  A consequence of this
notion of ownership is that the rules discriminate between the
parallel composition of processes and their interleaved expansion.
For example, the logic does \emph{not} allow the judgement
\[\ho{\lcon\mapsto 0}{[\lcon]:=1\pll [\lcon]:=1}{\lcon\mapsto 1},\]
which informally means that the effect of two processes acting in
parallel which both assign the value $1$ to the memory location
$\lcon$ from a state in which $\lcon$ holds $0$ is to yield a state in
which $\lcon$ holds $1$.  However, if we adopt the usual rule for the
nondeterministic sum of processes, the corresponding judgement
\emph{is} derivable for their interleaved expansion,
\[ ([\lcon]:= 1;[\lcon]:=1) + ([\lcon]:=1;[\lcon]:=1). \] One would
hope that the distinction that the logic makes between concurrent
processes and their interleaved expansion is captured by the
semantics; the Petri net model that we give does so directly.

The rules of concurrent separation logic contain a good deal of
subtlety, and so lacked a completely formal account until the
pioneering proof of their soundness due to Brookes
\cite{brookes:soundness}.  The proof that Brookes gives
is based on a form of interleaved trace semantics.  The presence of
pointers within the model alongside the possibility that ownership of
locations is transferred means, however, that the way in which
processes are separated is absolutely non-trivial, which motivates
strongly the study of the language within an independence model.  We
therefore give a proof of soundness using our net model and then
characterize entirely semantically the independence of concurrent
processes in Theorem \ref{theorem:separation}.

It should be emphasized that the model that we present is different
from Brookes' since it provides an \emph{explicit} account of the
intuitions behind ownership presented by O'Hearn.  It involves taking
the original semantics of the process and embellishing it to capture
the semantics of the logic.  The proof technique that we employ
defines validity of assertions in a way that captures the
rely-guarantee reasoning \cite{cliffjones} emanating from ownership in
separation logic directly, and in a way that might be applied in other
situations.

In \cite{REYNOLDS04C}, Reynolds argues that the separation of parallel
processes arising from the logic allows store actions that were
assumed to be atomic, in fact, to be implemented as composite actions
(seen as a change in their
\emph{granularity}) with no effect on the validity of the judgement.
Independence models are suited to modeling situations where actions
are not atomic, a perspective advocated by Lamport and Pratt
\cite{pratt:pomset, lamport:time}.  We introduce a novel form of
refinement, inspired by that of \cite{vanglabbeekgoltz}, and show how
this may be applied to address the issue of granularity using our
characterization of the independence of processes arising from the
logic.

\newcommand{\hvl}[2]{\lcon #1\mapsto #2}

\section{Terms and states}
Concurrent separation logic is a logic for programs that operate on a
\emph{heap}.  A heap is a structure recording the values held by
memory locations that allows the existence of pointers as well as
providing primitives for the allocation and deallocation of memory
locations.  A heap can be seen as a finite partial function from a set
of locations $\sn{Loc}$ to a set of values $\sn{Val}$:
\[ \sn{Heap} \quad\eqd\quad\sn{Loc}\parto_{\mathsf{fin}} \sn{Val} \] We will use
$\lcon$ to range over elements of $\sn{Loc}$ and $v$ to range over
elements of $\sn{Val}$.  As stated, a heap location can point to
another location, so we require that $\sn{Loc}\subseteq \sn{Val}$.  We
shall say that a location is \emph{current} (or \emph{allocated}) in a
heap if the heap is defined at that location.  The procedure of making
a non-current location current is \emph{allocation}, and the reverse
procedure is called \emph{deallocation}.  If $h$ is a heap and
$h(\lcon)=\lcon'$, there is no implicit assumption that $h(\lcon')$ is
defined.  Consequently, heaps may contain \emph{dangling} pointers.

In addition to operating on a heap, the programs that we shall
consider shall make use of \emph{critical regions} \cite{dijkstra}
protected by \emph{resources}.  The mutual exclusion property that
they provide is that no two parallel processes may be inside critical
regions protected by the same resource.  We will write $\sn{Res}$ for
the set of resources and use $\res$ to range over its elements.
Critical regions are straightforwardly implemented by recording, for
each resource, whether the resource is available or unavailable.  A
process may enter a critical region protected by $\res$ only if $\res$
is available; otherwise it is blocked and may not resume execution
until the resource becomes available.  The process makes $\res$
unavailable upon entering the critical region and makes $\res$
available again when it leaves the critical region.  The language also
has a primitive, $\resource\rvar\ldo t\lod$, which says that the
variable $\rvar$ represents a resource local to $t$.

\begin{figure}
\hrule
\begin{flushleft}
\textit{Terms:}
\end{flushleft}
\[
\begin{array}{rcl@{\qquad}l}
  t & \coloneqq& \alpha & \text{heap action}\\
  &\gor& \alloc\lcon & \text{heap allocation}\\
  &\gor& \dealloc\lcon & \text{heap disposal}\\
  &\gor& t_1;t_2 & \text{sequential composition}\\
  &\gor& t_1\pll t_2 & \text{parallel composition}\\
  &\gor& \alpha_1.t_1 + \alpha_2.t_2 & \text{guarded sum}\\
  &\gor& \while b\ldo t\lod &\text{iteration}\\
  &\gor& \resource\rvar\ldo t\lod & \text{resource declaration}\\
  &\gor& \with\res \ldo t\lod & \text{critical region}\\
  &\gor& \with\rvar\ldo t\lod &\text{critical region (local)}.\\
\end{array}
\]
\\~
\begin{flushleft}
\textit{Free variables and resources:}
\end{flushleft}
{\small
\[
\begin{array}{rcl@{\qquad}rcl}
\fv \alpha &=& \emptyset &
\fr \alpha &=& \emptyset \\
\fv{\alloc\lcon} &=& \emptyset &
\fr{\alloc\lcon} &=& \emptyset \\
\fv{\dealloc\lcon} &=& \emptyset &
\fr{\dealloc\lcon} &=& \emptyset \\
\fv{t_1;t_2} &=& \fv{t_1}\cup \fv{t_2} &
\fr{t_1;t_2} &=& \fr{t_1}\cup \fr{t_2} \\
\fv{t_1\pll t_2} &=& \fv{t_1}\cup \fv{t_2} &
\fr{t_1\pll t_2} &=& \fr{t_1}\cup \fr{t_2} \\
\fv{\alpha_1.t_1 + \alpha_2.t_2} &=& \fv{t_1}\cup \fv{t_2} &
\fr{\alpha_1.t_1;\alpha_2.t_2} &=& \fr{t_1}\cup \fr{t_2} \\
\fv{\while b\ldo t\lod } &=& \fv{t}&
\fr{\while b\ldo t\lod} &=& \fr{t} \\
\fv{\resource\rvar\ldo t\lod} &=& \fv{t}\setminus \{\rvar\} &
\fr{\resource\rvar\ldo t\lod} &=& \fr{t} \\
\fv{\with\res\ldo t\lod} &=& \fv{t} &
\fr{\with\res\ldo t\lod} &=& \fr{t}\cup \{\res\} \\
\fv{\with\rvar\ldo t\lod} &=& \fv{t}\cup \{\rvar\} &
\fr{\with\rvar\ldo t\lod} &=& \fr{t}\\
\end{array}
\]
}
\\~
\begin{flushleft}
\textit{Substitution:}
\end{flushleft}
{\newcommand{\rr}{\left[ \res/\rvar\right]}
\newcommand{\tif}{\text{if }}
{\small
\[
\begin{array}{llcl}
\rr & \alpha &=& \alpha\\
\rr & \alloc\lcon&=& \alloc\lcon\\
\rr & \dealloc\lcon &=& \dealloc\lcon\\
\rr &  t_1;t_2 &=& (\rr t_1);(\rr t_2)\\
\rr &  t_1\pll t_2 &=& (\rr t_1) \pll (\rr t_2)\\
\rr & \alpha_1.t_1 + \alpha_2.t_2 &=& \alpha_1.(\rr t_1) + \alpha_2.(\rr t_2)\\
\rr & \while b\ldo t\lod &=& \while b \ldo \rr t \lod\\
\rr & \resource{\rvar'}\ldo t\lod &=& \resource{\rvar'} \ldo \rr t \lod
\qquad\tif \rvar\neq \rvar'\\
\rr & \resource\rvar \ldo t\lod &=& \resource\rvar\ldo t \lod\\
\rr & \with{\res'} \ldo t\lod &=& \with{\res'} \ldo \rr t \lod\\
\rr & \with{\rvar'}\ldo t\lod &=& \left\{
  \begin{array}{ll}
    \with{\res}\ldo \rr t \lod &\qquad\tif \rvar=\rvar'\\
    \with{\rvar'}\ldo\rr t\lod &\qquad\text{otherwise}
  \end{array}\right.
\end{array}\]
}
}
\hrule
\caption{Syntax of terms}
\label{fig:defterms}
\end{figure}

The syntax of the language that we will consider is presented in
Figure \ref{fig:defterms}. The symbol $\alpha$ is used to range over
\emph{heap actions}, which are actions on the heap that might change
the values held at locations but do not affect the domain of
definition of the heap.  That is, they neither allocate nor deallocate
locations.  We reserve the symbol $b$ for \emph{boolean guards}, which
are heap actions that may proceed without changing the heap if the
boolean $b$ holds.

Provision for allocation within our language is made via the
$\alloc\lcon$ primitive for $\lcon\in \sn{Loc}$, which makes a
location current and sets $\lcon$ to point at this location.  For
symmetry, $\dealloc\lcon$ makes the location pointed to by $\lcon$
non-current if $\lcon$ points to a current location.  Writing a heap
as the set of values that it holds for each allocated location, the
effect of the command $\alloc \lcon$ on the heap $\{\hvl{}{0}\}$ might
be to form a heap $\{\hvl{}{\lcon'},\hvl{'}{1}\}$ if the location
$\lcon'$ is chosen to be allocated and is assigned initial value $1$.
The effect of the command $\dealloc\lcon$ on the heap
$\{\hvl{}{\lcon'},\hvl{'}{1}\}$ would be to form the heap
$\{\hvl{}{\lcon'}\}$.

The \emph{guarded sum} $\alpha.t+\alpha'.t'$ is a process that
executes as $t$ if $\alpha$ takes place or as $t'$ if $\alpha'$ takes
place.  We refer the reader to Section \ref{sec:gsum} for a brief
justification for disallowing non-guarded sums.

As mentioned earlier, critical regions are provided to control
concurrency: the sub-process $t$ inside $\with \res \ldo t\lod$ can
only run when no other process is inside a critical region protected
by $\res$.  The term $\resource\rvar\ldo t\lod$ has the resource
variable $\rvar$ bound within $t$, asserting that a resource is to be
chosen that is \emph{local} to $t$ and used for $\rvar$.
Consequently, in the process
\[ (\resource\rvar\ldo \with \rvar\ldo t_1 \lod\lod)
\pll (\resource \rvar \ldo \with \rvar\ldo t_2 \lod\lod) \] the
sub-processes $t_1$ and $t_2$ may run concurrently since they must be
protected by different resources, one local to the process on the left
and the other local to the process on the right. To model this, we
shall say that the construct $\resource\rvar \ldo t\lod$ \emph{binds}
the variable $\rvar$ within $t$, and the variable $\rvar$ is \emph{free}
in $\with \rvar\ldo t\lod$.  We write $\fv t$ for the free variables
in $t$ and say that a term \emph{closed} if it contains no free
resource variables; we shall restrict attention to such terms.  We
write $[\res/\rvar]t$ for the term obtained by substituting $\res$ for
free occurrences of the variable $\rvar$ within $t$.  As standard, we
will identify terms `up to' the standard alpha-equivalence $\equiv$
induced by renaming bound occurrences of variables.  The notation $\fr
t$ is adopted to represent the resources occurring in $t$.

The semantics of the term $\resource\rvar\ldo t\lod$ will involve
first picking a `fresh' resource $\res$ and then running
$[\res/\rvar]t$.  It will therefore be necessary to record during the
execution of processes which resources are \emph{current} (\ie{} not
fresh) as well as which current resources are \emph{available} (\ie{}
not held by any process).

The way in which we shall formally model the state in which processes
execute is motivated by the way in which we shall give the net
semantics to closed terms.  We begin by defining the following sets:
\[\begin{array}{rcl}
\cset D &\eqd& \sn{Loc}\times\sn{Val}\\
\cset L &\eqd& \{\curr(\lcon)\st \lcon\in\sn{Loc}\}\\
\cset R &\eqd& \sn{Res}\\
\cset N &\eqd& \{\curr(\res)\st \res\in\sn{Res}\}.
\end{array}\]
A state $\sigma$ is defined to be a tuple
\[(D,L,R,N)\]
where $D\subseteq \cset D$ represents the values held by locations in
the heap; $L\subseteq \cset{L}$ represents the set of current, or
allocated, locations of the heap; $R\subseteq \cset{R}$ represents the
set of available resources; and $N\subseteq \cset N$ represents the
set of current resources.
The sets $\cset{D}$,
$\cset{L}$, $\cset{R}$ and $\cset{N}$ are disjoint, so no ambiguity
arises from writing, for example, $(\lcon,v)\in \sigma$.

The interpretation of a state for the heap is that $(\lcon,v)\in D$ if
$\lcon$ holds value $v$ and that $\curr(\lcon)\in L$ if $\lcon$ is
current.  For resources, $\res\in R$ if the resource $\res$ is
available and $\curr(\res)\in N$ if $\res$ is current.  It is clear
that only certain such tuples of subsets are sensible.  In particular,
the heap must be defined precisely on the set of current locations,
and only current resources may be available.  
\begin{defi}[Consistent state]
  The state $(D, L, R, N)$ is \emph{consistent} if we have:
  \begin{enumerate}[$\bullet$]
  \item the sets $D$, $L$, $R$ and $N$ are all finite,
  \item $D$ is a partial function: for all $\lcon,v$ and $v'$, if
    $(\lcon,v)\in D$ and $(\lcon,v')\in D$ then $v=v'$,
  \item $L$ represents the domain of $D$: \quad$L=\{\curr(\lcon) \st
    \exists v:(\lcon,v)\in D\}$, and
  \item all available resources are current: $R\subseteq
    \{\res\mid\curr(\res)\in N\}$.
  \end{enumerate}
\end{defi}
It is clear to see that the $L$ component of any given consistent
state may be inferred from the $D$ component.  It will, however, be
useful to retain this information separately for when the net
semantics is given.  We shall call $D\subseteq \cset D$ a \emph{heap}
when it is a finite partial function from locations to values, and
shall write $\hvl{}v$ for its elements rather than $(\lcon,v)$.  We
shall frequently make use of the following definition of the domain of
a heap $D$:
\[ \dom(D) \eqd \{\lcon\st \exists v.(\hvl{}v)\in D\}. \]

\section{Process models}
The definition of state that we have adopted permits a net semantics
to be defined.  Before doing so, we shall define how heap actions are
to be interpreted and then give a transition semantics to closed terms.
\subsection{Actions}
The earlier definition of state allows a very general form of heap
action to be defined that forms a basis for both the transition and
net semantics.  We assume that we are given the semantics of primitive
actions $\alpha$ as $\asem{\alpha}$ comprising a set of heap pairs:
\[ \asem{\alpha} \subseteq \sn{Heap}\times \sn{Heap}.\] We
require that whenever $(D_1,D_2)\in\asem\alpha$, it is the case that
$D_1$ and $D_2$ are (the graphs of) partial functions with the same
domain.

The interpretation is that $\alpha$ can proceed in heap $D$ if there
are $(D_1,D_2)\in\asem{\alpha}$ such that $D$ has the same value as
$D_1$ wherever $D_1$ is defined.  The resulting heap is formed by
updating $D$ to have the same value as $D_2$ wherever it is defined.
It is significant that this definition allows us to infer precisely
the set of locations upon which an action depends.  The requirement on
the domains of $D_1$ and $D_2$ ensures that actions preserve
consistent markings (Lemma \ref{lemma:conspres}).

\begin{exa}[Assignment]
  For any two locations $\lcon$ and $\lcon'$, let $[\lcon]:=[\lcon']$
  represent the action that copies the value held at location $\lcon'$
  to location $\lcon$.  Its semantics is as follows:
  \[
  \asem{[\lcon]:= [\lcon']} \eqd
  \begin{array}{l@{}l}
    \{( & \{\hvl{}{v},\hvl{'}{v'}\},\\
    & \{\hvl{}{v'},\hvl{'}{v'}\}) \st v,v'\in\sn{Val}\}
  \end{array}
  \]
  Following the informal account above of the semantics of actions,
  because in the semantics we have
  \[(\{\hvl{_0} 0,\hvl{_1}1\},\{\hvl{_0} 1,\hvl{_1}1\})\in\asem{[\lcon_0]:=[\lcon_1]},\] the
  state $\{\hvl{_0}0,\hvl{_1}1,\hvl{_2}2\}$ is updated by
  $[\lcon_0]:=[\lcon_1]$ to $\{\hvl{_0}1,\hvl{_1}1,\hvl{_2}2\}$.\qed
\end{exa}
\begin{exa}[Booleans]
\label{example:booleans}
Boolean guards $b$ are actions that 
wait until the boolean expression holds and may then take place; they
do not update the state.  A selection of literals may be defined.  For
example:
\begin{eqnarray*}
\asem{[\lcon] = v} &\eqd& \{(\{\hvl{} v\},\{\hvl{} v\})\}\\
\asem{[\lcon]=[\lcon']} &\eqd& 
\{(\{\hvl{}v,\hvl{'}{v}\},\{\hvl{}v,\hvl{'}v\})\st v\in\sn{Val}\}
\end{eqnarray*}
The first gives the semantics of an action that proceeds only if
$\lcon$ holds value $v$ and the second gives the semantics of an
action that proceeds only if the locations $\lcon$ and $\lcon'$ hold
the same value.

\newcommand{\nuparrow}{\mathbin{\not{\!\uparrow}}}
Since boolean actions shall not modify the heap, they shall possess
the property that:
\[ \text{if }(D_1,D_2)\in\asem{b} \text{ then }D_1=D_2.\] This is
preserved by the operations defined below.  For heaps $D$ and $D'$, we
use $D\uparrow D'$ to mean that $D$ and $D'$ are compatible as partial
functions and $D\nuparrow D'$ otherwise, \ie{} if they disagree on
the values assigned to a common location.

\[
\begin{array}{rcl}
 \asem{\co{true}} &\eqd& \{(\emptyset,\emptyset) \}\\
 \asem{\co{false}} &\eqd& \emptyset\\
 \asem{b\land b'} &\eqd& 
 \{(\{D\cup D'\},\{D\cup D'\}) \st
   D\uparrow D' ~\tand~ (D,D)\in \asem{b}~\tand~ (D',D')\in\asem{b'} \}
 \\
 \asem{b\lor b'} &\eqd& \asem{b}\cup \asem{b'}\\
 \asem{\lnot b} &\eqd& \{(D,D)\st D\text{ is a $\subseteq$-minimal heap s.t.}~
 \forall D'.(D',D')\in\asem{b}: D\nuparrow D'\}
\end{array}
\]
By insisting on minimality in the clause for $\lnot b$, we form an
action that is defined at as few locations as possible to refute all
grounds for $b$.\qed
\end{exa}

\subsection{Transition semantics}
As an aid to understanding the net model, and in particular to give a
model with respect to which we can prove its correspondence, a
transition semantics for \emph{closed} terms (terms such that
$\fv{t}=\emptyset$) is given in Figure \ref{fig:opsem}.  A formal
relationship between the two semantics is presented in Theorem
\ref{theorem:semcorr}.  The transition semantics is given by means of
labelled transition relations of the forms $\ang{t,\sigma}\otr\lambda
\ang{t',\sigma'}$ and $\ang{t,\sigma}\otr\lambda\sigma'$.  As usual,
the first form of transition indicates that $t$ performs an action
labelled $\lambda$ in state $\sigma$ to yield a resumption $t'$ and a
state $\sigma'$.  The second indicates that $t$ in state $\sigma$
performs an action labelled $\lambda$ to terminate and yields a state
$\sigma'$.  Labels follow the grammar
\[
\begin{array}{rcl@{\qquad}l}
  \lambda& \coloneqq &
  \lact(D_1,D_2) & \text{heap action}\\
  &\gor& \lall(\lcon,v,\lcon',v') & \text{heap allocation}\\
  &\gor& \ldea(\lcon,\lcon',v) &\text{heap disposal}\\
  &\gor& \lnew(\res) & \text{resource declaration}\\
  &\gor& \lend(\res) & \text{end of resource scope}\\
  &\gor& \lget(\res)  & \text{resource acquisition (critical region entry)}\\
  &\gor& \lput(\res)  & \text{resource release (critical region exit)}.
\end{array}
\]
In the transition semantics, we write $\sigma\oplus\sigma'$ for the
union of the components of two states where they are disjoint and
impose the implicit side-condition that this is defined wherever it is
used.  For example, this implicit side-condition means, in the rule
$\rn{Alloc}$, that for $\lall(\lcon,v,\lcon',v')$ to occur we must
have $\curr(\lcon')\not\in \sigma$, and hence
$\lcon'$ was initially non-current.  Similarly, the rule $\rn{Res}$
can only be applied to derive a transition labelled $\lnew(\res)$ if
the resource $\res$ was not initially current.

The syntax of terms is extended temporarily to include $\rel\res$ and
$\finres\res$ which are special terms used in the rules $\rn{Rel}$ and
$\rn{End}$.  These, respectively, are attached to the ends of terms
protected by critical regions and the ends of terms in which a
resource was declared.

For conciseness, we do not give an error semantics to situations in
which non-current locations or resources are used; instead, the
process will become stuck.  We show in Section \ref{sec:error} that
such situations are excluded by the logic.

\begin{figure}[tbh]\small
  \hrule
  \[
  \begin{array}{cl@{\qquad}cl}
    \rn{Act}:&
    \multicolumn{3}{c}{
      \deriv{ (D_1,D_2)\in\asem{\alpha}\\
        D_1\subseteq D\qquad
        D' = (D\setminus D_1)\cup D_2}
      {\ang{\alpha,(D,L,R,N)}
        \otr{\lact(D_1,D_2)}
        (D',L,R,N)}
    }\\\\
    \rn{Alloc}:& 
    \multicolumn{3}{c}{
      \ang{\alloc{\lcon},\sigma\oplus\{\hvl{}v\}}
      \otr{\lall(\lcon,v,\lcon',v')}
      \sigma\oplus\{\hvl{}{\lcon'},\hvl{'}{v'},\curr(\lcon')\}}
    \\\\
    \rn{Dealloc}:& 
    \multicolumn{3}{c}{
      \ang{\dealloc{\lcon},\sigma\oplus\{\hvl{}{\lcon'},\hvl{'}{v'},\curr(\lcon')\}}
      \otr{\ldea(\lcon,\lcon',v')}
      \sigma\oplus\{\hvl{}{\lcon'}\}}
    \\\\
    \rn{Seq}:& 
    \deriv{\ang{t_1,\sigma}\otr \lambda \ang{t_1',\sigma'}}
    {\ang{t_1;t_2,\sigma}\otr \lambda \ang{t_1';t_2,\sigma'}}
    &
    \rn{Seq$'$}:&
    \deriv{\ang{t_1,\sigma}\otr \lambda \sigma'}
    {\ang{t_1;t_2,\sigma}\otr \lambda \ang{t_2,\sigma'}}
    \\\\
    \rn{Par-1}:&
    \deriv{\ang{t_1,\sigma}\otr\lambda \ang{t_1',\sigma'}}
    {\ang{t_1\pll t_2,\sigma}\otr\lambda\ang{t_1'\pll t_2,\sigma'}}
    &
    \rn{Par-2}:&
    \deriv{\ang{t_2,\sigma}\otr\lambda \ang{t_2',\sigma'}}
    {\ang{t_1\pll t_2,\sigma}\otr\lambda\ang{t_1\pll t_2',\sigma'}}
    \\\\
    \rn{Par$'$-1}:&
    \deriv{\ang{t_1,\sigma}\otr\lambda \sigma'}
    {\ang{t_1\pll t_2,\sigma}\otr{\lambda} \ang{t_2,\sigma'}}
    &
    \rn{Par$'$-2}:&
    \deriv{\ang{t_2,\sigma}\otr\lambda \sigma'}
    {\ang{t_1\pll t_2,\sigma}\otr{\lambda} \ang{t_1,\sigma'}}
    \\\\
    \rn{Sum-1}:&
    \deriv{\ang{\alpha_1,\sigma}\otr{\lambda}\sigma'}
    {\ang{\alpha_1.t_1 +\alpha_2.t_2,\sigma}\otr\lambda
      \ang{t_1,\sigma'}}
    &
    \rn{Sum-2}:&
    \deriv{\ang{\alpha_2,\sigma}\otr{\lambda}\sigma'}
    {\ang{\alpha_1.t_1 +\alpha_2.t_2,\sigma}\otr\lambda
      \ang{t_2,\sigma'}}
    \\\\
    \rn{While}:&
      \deriv{\ang{b,\sigma}\otr \lambda \sigma}
      {\ang{\while b\ldo t\lod,\sigma}\\\otr\lambda \ang{p;\while b\ldo t\lod,\sigma}}
      &
      \rn{While$'$}:&
      \deriv{\ang{\lnot b,\sigma}\otr{\lambda} \sigma}
      {\ang{\while b\ldo t\lod,\sigma}\otr{\lambda} \sigma}
    \\\\
    \rn{With}: &
    \multicolumn{3}{c}{
      \ang{\with\res\ldo t\lod,\sigma\oplus\{\res\})}
      \otr{\lget(\res)}
      \ang{t;\rel\res,\sigma}
    }
    \\\\
    \rn{Rel}: &
    \multicolumn{3}{c}{
      \ang{\rel\res,\sigma}
      \otr{\lput(\res)}
      \sigma\oplus\{\res\}
    }
    \\\\
    \rn{Res}: &
    \multicolumn{3}{c}{
      \ang{\resource\rvar\ldo t\lod,\sigma}
      \otr{\lnew(\res)}
      \ang{[\res/\rvar]t;\finres\res, \sigma\oplus\{\res,\curr(\res)\}}
    }
    \\\\
    \rn{End}: &
    \multicolumn{3}{c}{
      \ang{\finres\res,\sigma\oplus\{\res,\curr(\res)\}}
      \otr{\lend(\res)} \sigma
    }
  \end{array}
  \]
\hrule
  \caption{Transition semantics}
  \label{fig:opsem}
\end{figure}

\subsection{Petri nets}
Petri nets, introduced by Petri in his 1962 thesis \cite{petri}, are a
well-known model for concurrent computation.  It is beyond the scope
of the current article to provide a full account of the many variants
of Petri net and their associated theories; we instead refer the
reader to \cite{springer:acpn} for a good account.  Roughly, a Petri net can
be thought of as a transition system where, instead of a transition
occurring from a single global state, an occurrence of an event is
imagined to affect only the conditions in its neighbourhood. Petri
nets allow a derived notion of \emph{independence} of events; two
events are independent if their neighbourhoods of conditions do not
intersect.

We base our semantics on the following well-known variant of Petri net
(\textit{cf.}~the `basic' nets of \cite{spl} and \cite{wn:mfc}):

\begin{defi}[Petri net]
A Petri net is a five-tuple,
\[ (B,E,\pre(-),\post{(-)},M_0).\]
The set $B$ comprises the \emph{conditions} of the net, the set $E$
consists of the \emph{events} of the net, and $M_0$ is the subset of
$B$ of \emph{marked conditions} (the \emph{initial marking}).  The maps
\[ \pre(-),\post{(-)} \ty E \to \pow(B) \]
are the \emph{precondition} and \emph{postcondition} maps, respectively.
\end{defi}
Petri nets have an appealing graphical representation, with:
\begin{enumerate}[$\bullet$]
\item circles to represent conditions,
\item bold lines to represent events, 
\item arrows from conditions to events to represent the precondition map,  
\item arrows from events to conditions to represent the postcondition map, and
\item tokens (dots) inside conditions to represent the marking.
\end{enumerate}

Action within nets is defined according to a \emph{token game} which
defines how the marking of the net changes according to \emph{firing}
of the events.  An event $e$ can fire if all its preconditions are
marked and, following their un-marking, all the postconditions are not
marked.  That is, in marking $M$,
\begin{eqnarray*}
(1)&\quad& \pre e \subseteq M\\
(2)&\quad& (M\setminus \pre e)\cap \post e =\emptyset.
\end{eqnarray*}
Such an event is said to have \emph{concession} or to be
\emph{enabled}. 
The marking following the occurrence of $e$ is obtained by removing
the tokens from the preconditions of $e$ and placing a token in every
postcondition of $e$.  We write
$ M\ltr e M' $
where 
\[ M' = (M \setminus \pre e) \cup \post e. \]
If constraint $(2)$ does not hold but constraint (1) does, so the preconditions are all marked (have a token inside) but following removal of the tokens from the preconditions there is a token in some postcondition,
there is said to be \emph{contact} in the
marking and the event cannot fire.

Consider the following example Petri net, with its transition system
between markings derived according to the token game.
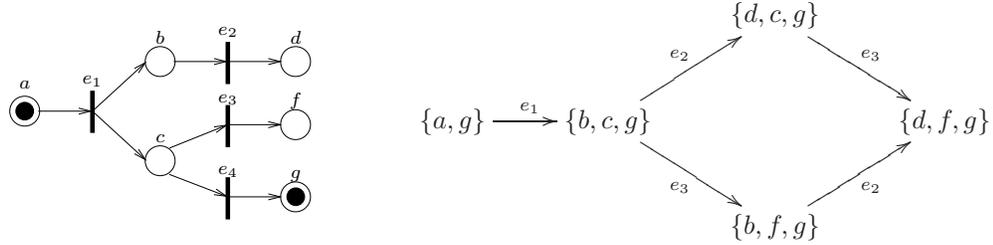
\begin{center}
\begin{tabular}{ll}
\begin{minipage}{4cm}\hskip-1 cm
\input{petri.pstex_t}
\end{minipage}\hskip-2 cm
&
\begin{minipage}{5cm}
\small $
\xymatrix{
 & & \{d,c,g\} \ar[dr]^{e_3}&\\
\{a,g\} \ar[r]^{e_1} & \{b,c,g\} \ar[ur]^{e_2} \ar[dr]_{e_3}& & \{d,f,g\}\\
& & \{b,f,g\}\ar[ur]_{e_2} &\\ 
}
$
\end{minipage}
\end{tabular}\medskip
\end{center}
The event $e_1$ is the only event with concession in the initial
marking $\{a,g\}$.  Its occurrence yields the marking obtained by
un-marking its preconditions and marking its postconditions, namely
$\{b,c,g\}$.  In the marking $\{b,c,g\}$, contact prevents the
occurrence of $e_4$ since its postcondition $g$ is marked following
removal of the token from its precondition $c$.  However, in the
marking $\{b,c,g\}$ both event $e_2$ and event $e_3$ can occur.  Note
that the occurrence of $e_2$ in marking $\{b,c,g\}$ does not affect
the occurrence of $e_3$ and {\it vice versa} since the two events operate
on completely disjoint sets of conditions.

For any event $e\in E$, define the notation
\[ \pre{\post e} \eqd \pre e \cup \post e. \]
The standard notion of independence within this form of Petri net
is to say that two events $e_1$ and $e_2$ are independent, written
$e_1 I e_2$, if their neighbourhoods are disjoint.  That is,
\[e_1 I e_2 \iff \pre{\post{e_1}} \cap \pre{\post{e_2}} = \emptyset.\]
It is easy to see in general that the occurrences of independent events in a
marking do not affect each other.
\begin{prop}
\label{lemma:indepswap}
Let $e_1$ and $e_2$ be events of the net $N$ and suppose that $e_1 I e_2$.
\begin{enumerate}[$\bullet$]
\item If there exist markings $M$, $M'$ and $M_1$ of $N$ such that
$M\ltr{e_1} M_1$ and $M_1\ltr{e_2} M'$ then there exists a marking
$M_2$ such that $M\ltr{e_2}M_2$ and $M_2\ltr{e_1} M'$.
\item If there exist markings $M$, $M_1$ and $M_2$ of $N$ such that
  $M\ltr{e_1}M_1$ and $M\ltr{e_2}M_2$ then there exists a marking $M'$
  such that $M_1\ltr{e_2} M'$ and $M_2\ltr{e_1} M'$.\qed
\end{enumerate}
\end{prop}

\subsection{Overview of net semantics}
{\newcommand{\swl}{\co{toggle}(\lcon,0,1)} Before giving the formal
  definition of the net semantics of closed terms, by means of an example we
  shall illustrate how our semantics shall be defined.  First, we
  shall draw the semantics of an action $\swl$ that toggles the value
  held at a location $\lcon$ between $0$ and $1$.

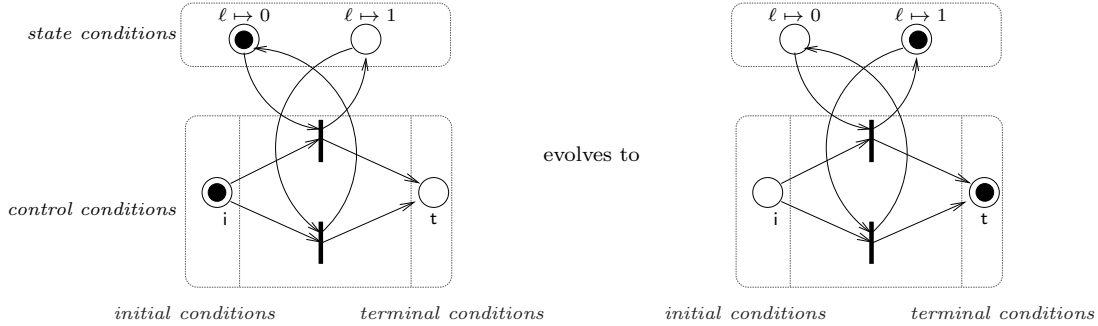
\begin{center}
\vspace{1em}
\input{firstex.pstex_t}
\end{center}
Notice that in the above net there are conditions to represent the
\emph{shared state} in which processes execute, including for example
the values held at locations (we have only drawn conditions that are
actually used by the net).  There are also conditions to represent the
\emph{control} point of the process.  The net pictured on the left is
in its \emph{initial marking} of control conditions and the net on the
right is in its \emph{terminal marking} of control conditions,
indicating successful completion of the process following the toggle
of the value; the marking of the net initially had the state condition
$\hvl{} 0$ marked and finished with the condition $\hvl{} 1$ marked.  There is an event present in the net for each way that the
action could take place: one event for toggling the value from $0$ to
$1$ and another event for toggling the value from $1$ to $0$.  Only
the first event could occur in the initial marking of the net on the
left, and no event can occur in the marking on the right
since the control conditions are not appropriately marked.

The parallel composition $\swl \pll \swl$ can be formed by taking two
copies of the net $\swl$ and forcing them to operate on disjoint sets
of control conditions.
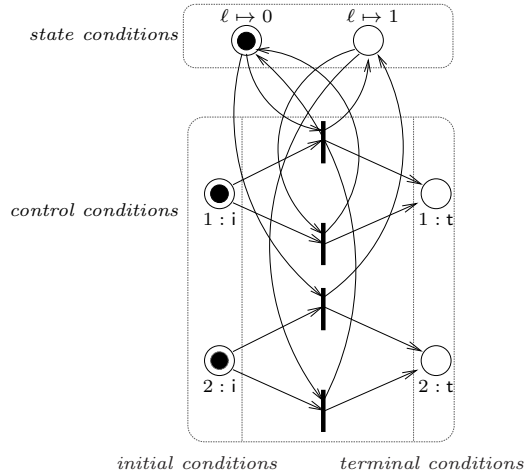
\begin{center}\input{parex.pstex_t}\end{center}
An example run of this net would involve first the top event changing
the value of $\lcon$ from $0$ to $1$ and then the bottom event
changing $\lcon$ back from $1$ to $0$.  The resulting marking of
control conditions would be equal to the terminal conditions of the net, so
no event would have concession in this marking.

The net representing the sequential composition \[(\swl \pll
\swl);(\swl \pll \swl)\] is formed by a `gluing' operation that joins
the terminal conditions of one copy of the net for $\swl$ to the
initial conditions of another copy of the net for $\swl$. (In this
example net, for clarity we shall not show the state conditions.)
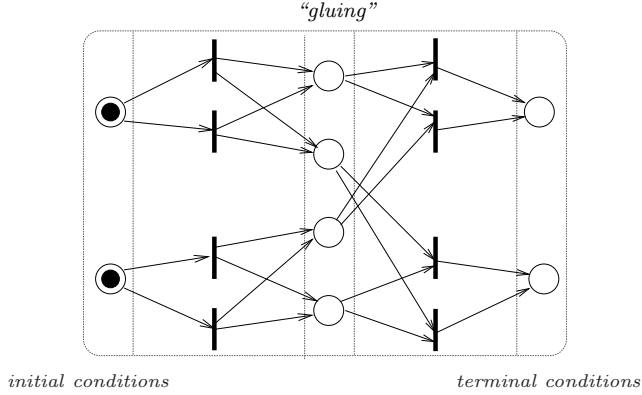
\begin{center}\input{seqex.pstex_t}\end{center}

}

\subsection{Net structure}
As outlined above, within the nets that we give for processes we
distinguish two forms of condition, namely \emph{control conditions}
and \emph{state conditions}.  The markings of these sets of conditions
determine the control point of the process and the state in which it
is executing, respectively.  When we give the net semantics, we will
make use of the closure of the set of control conditions under various
operations.
\begin{defi}[Conditions]
  {\newcommand{\ccon}{\cset{C}} Define the set of control conditions
    $\ccon$, ranged over by $c$, to be the least set such that:
    \begin{enumerate}[$\bullet$]
    \item $\ccon$ contains distinguished elements $\del$ and $\cel$, 
      standing for `initial' and `terminal', respectively.
    \item If $c\in\ccon$ then $\pref{\res}{c} \in \ccon$ for all
      $\res\in\sn{Res}$ and $\pref i c \in \ccon$ for all
      $i\in\{1,2\}$, to distinguish processes working on different
      resources or arising from different subterms.
    \item If $c,c'\in\ccon$ then $\tup{c,c'}\in\ccon$ to allow the
      `gluing' operation above.
    \end{enumerate}
}
\noindent Define the set of \emph{state conditions} $\cset S$ to be
$\cset{D}\cup \cset{L} \cup \cset{R} \cup \cset{N}$.
\end{defi}
A state $\sigma=(D,L,R,N)$ corresponds to the marking $D\cup L\cup
R\cup N$ of state conditions in the obvious way.  Similarly, if $C$ is
a marking of control conditions and $\sigma$ is a state, the pair
$(C,\sigma)$ corresponds to the marking $C\cup\sigma$.  We therefore
use the notations interchangeably.

The nets that we form shall be \emph{extensional} in the sense that two
events are equal if they have the same preconditions and the same
postconditions.  An event can therefore be regarded as a tuple
\[ e = (C,\sigma,C',\sigma')\] with preconditions $\pre e \eqd C\cup
\sigma$ and postconditions $\post e \eqd C'\cup \sigma'$.  To obtain a
concise notation for working with events, we write $\pres C e$ for the
pre-control conditions of $e$:
\[ \pres C e \eqd \pre e\cap \cset{C}.\] We likewise define notations
$\poss C e$, $\pres D e$, $\pres L e$ \etc{}, and call these the
\emph{components} of $e$ by virtue of the fact that it is sufficient
to define an event through the definition of its components.  The
pre-state conditions of $e$ are $\pres S e=\pres D e \cup \pres L e
\cup \pres R e \cup \pres N e$, and we define $\poss S e$ similarly.

Two markings of control conditions are of particular importance: those
marked when the process starts executing and those marked when the
process has terminated.  We call these the \emph{initial} control
conditions $I$ and \emph{terminal} control conditions $T$,
respectively.  We shall call a net with a partition of its conditions
into control and state with the subsets of control conditions $I$ and
$T$ an \emph{embedded net}.  For an embedded net $N$, we write
$\ic{N}$ for $I$ and $\tc{N}$ for $T$, and we write $\ev{N}$ for its
set of events.  Observe that no initial marking of state conditions is
specified.

The semantics of a closed term $t$ shall be an embedded net, written
$\nsem{t}$.  No confusion arises, so we shall write $\ic{t}$ for
$\ic{\nsem{t}}$, and $\tc{t}$ and $\ev{t}$ for $\tc{\nsem{t}}$ and
$\ev{\nsem{t}}$, respectively.  The nets formed shall always have the
same sets of control and state conditions; the difference shall arise
in the events present in the nets. It would be a trivial matter to
restrict to the conditions that are actually used.

As we give the semantics of closed terms, we will make use of several
constructions on nets. For example, we wish the events of parallel
processes to operate on disjoint sets of control conditions.  This is
conducted using a \emph{tagging} operation on events. We define $\pref
1 e$ to be the event $e$ changed so that
\[
  \pres C (\pref 1 e) \eqd \{\pref 1 c\st c\in\pres C e\}\qquad
  \poss C{(\pref 1 e)} \eqd \{\pref 1 c\st c\in\poss C e\}
\]
but otherwise unchanged in its action on state conditions.  We define
the notations $\pref 2 e$ and $\pref \res e$ where $\res\in\sn{Res}$
similarly.  The notations are extended pointwise to sets of events:
\[ \pref 1 E \eqd \{\pref 1 e \st e\in E \}.\]

Another useful operation is what we call \emph{gluing} two embedded
nets together.  For example, when forming the sequential composition
of processes $t_1;t_2$, we want to enable the events of $t_2$ when
$t_1$ has terminated.  This is done by `gluing' the two nets together
at the terminal conditions of $t_1$ and the initial conditions of
$t_2$, having made them disjoint on control conditions using tagging.
Wherever a terminal condition $c$ of $\tc{t_1}$ occurs as a
pre- or a postcondition of an event of $t_1$, every element of the set
$\{\pref 1 c\}\times(\pref 2{\ic{t_2}})$ would occur in its place.
Similarly, the events of $t_2$ use the set of conditions $(\pref 1
{\tc{t_1}})\times\{\pref 2 c'\}$ instead of an initial condition $c'$
of $\ic{t_2}$.  A variety of control properties that the nets we form
possess (Lemma \ref{lemma:embstruc}), such as that all events have at
least one pre-control condition, allows us to infer that it is
impossible for an event of $t_2$ to occur before $t_1$ has terminated,
and thereon it is impossible for $t_1$ to
resume.  
An example follows shortly.

Assume a set $P\subseteq \cset{C}\times\cset{C}$.  Useful definitions
to represent gluing are:
\[
\begin{array}{rc@{\;}l@{\;}l}
  P\gluel C &\eqd&& \{(c_1,c_2)\st c_1\in C ~\tand~ (c_1,c_2)\in P\}\\
  &&\cup& \{c_1 \st c_1\in C~\tand~\nexists c_2 . (c_1,c_2)\in P\}
\end{array}\]
\[\begin{array}{rc@{\;}l@{\;}l}
  P\gluer C &\eqd&& \{(c_1,c_2)\st c_2\in C ~\tand~ (c_1,c_2)\in P\}\\
  &&\cup& \{c_2 \st c_2\in C~\tand~\nexists c_1 . (c_1,c_2)\in P\}
\end{array}
\]
The first definition, $P\gluel C$, indicates that an occurrence of
$c_1$ in $C$ is to be replaced by occurrences of $(c_1,c_2)$ for every
$c_2$ such that $(c_1,c_2)$ occurs in $P$.  The second definition,
$P\gluer C$, indicates that an occurrence of $c_2$ in $C$ is to be
replaced by occurrences of $(c_1,c_2)$ for every $c_1$ such that
$(c_1,c_2)$ occurs in $P$.

The notation is extended to events to give an event $P\gluel e$ in the
following way, recalling that gluing will only affect the control
conditions used by an event and in particular not its state conditions:
\[
\begin{array}{rcl@{\qquad}rcl}
  \pres C {(P \gluel e)} &\eqd& P\gluel (\pres C e)&
  \poss C {(P \gluel e)} &\eqd& P\gluel (\poss C e)\\
  \pres S {(P\gluel e)} &\eqd& \pres S e&
  \poss S {(P\gluel e)} &\eqd& \poss S e
\end{array}
\]
The notation $P\gluer e$ is defined similarly, and it is also extended
to sets of events in the obvious pointwise manner.  For any marking
$M=(C,\sigma)$, we will write $P\gluel M$ for $(P\gluel C,\sigma)$ and
similarly write $P\gluer M$ for $(P\gluer C,\sigma)$.

To give an example, consider the gluings $P\gluel C_1$ and $P\gluer
C_2$ where $C_1=\{a,b\}$ and $C_2=\{c,d\}$ are joined at $P=C_1\times
C_2$.  Applying $P\gluel C_1$ to the left net and $P\gluer C_2$ to the
right net below, this indicates how gluing is used to sequentially
compose embedded nets:\medskip

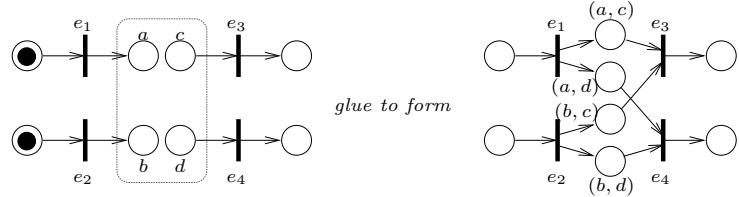
\begin{center}\input{seq.pstex_t}\end{center}

The operations of gluing and tagging affect only the control flow of
events, not their effect on the marking of state conditions.  
\begin{lem}
  \label{lemma:gluesim}
  Let $N$ be an embedded net with control conditions $\cset C$.
  Suppose that $P\subseteq \cset C\times \cset C$.  For any marking
  $M$ of $N$ and tag $x\in \sn{Res}\cup \{1,2\}$:
  \begin{enumerate}[$\bullet$]
  \item $M\ltr e M'$ iff $\pref x M \ltr{\pref x e} \pref x {M'}$.
  \item $M\ltr e M'$ iff $P\gluel M \ltr{P\gluel e} P\gluel M'$, and
  \item $M\ltr e M'$ iff $P\gluer M \ltr{P\gluer e} P\gluer M'$.
  \end{enumerate}
  Furthermore:
  \begin{enumerate}[$\bullet$]
  \item if $\pref 1 M \ltr{\pref 1 e} M_1'$ then $M_1'=\pref 1 M'$ for
    some $M'$, 
  \item if $P\gluel M \ltr{P\gluel e} M_1'$ then $M_1'=P\gluel M'$ for
    some $M'$, and
  \item if $P\gluer M \ltr{P\gluer e} M_2'$ then $M_2'=P\gluer M'$ for
    some $M'$.
  \end{enumerate}   
\begin{proof}
  The first and fourth items are straightforward to prove.  The
  remaining properties may be shown using the following
  easily-demonstrated equations, along with their counterparts for
  $\gluer$, for any subset of control conditions $C$:
  \begin{enumerate}[(1)]
  \item $C=\emptyset$ iff $P\gluel C=\emptyset$,
  \item $P\gluel (C\setminus C') = (P\gluel C)\setminus (P\gluel C')$,
  \item $P\gluel (C\cup C') = (P\gluel C)\cup (P\gluel C')$, and
  \item $P\gluel (C\cap C') = (P\gluel C)\cap (P\gluel C')$.
\qedhere
  \end{enumerate}
\end{proof}
\end{lem}

\subsection{Net semantics}
The net semantics that we now give for closed terms is defined by
induction on the \emph{size} of terms, given in the obvious way.  The reason
why it is not given by induction on terms is that the semantics of
$\resource\rvar\ldo t\lod$ is given according to the semantics of
$[\res/\rvar]t$ for all resources $\res$.
\begin{enumerate}[$\triangleright$]
\item{\bf Heap action:}\ Let $\eact{C,C'}(D_1,D_2)$ denote an event $e$ with
  \[ \pres C e = C \quad \poss C e = C' \quad \pres D e = D_1 \quad
  \poss D e = {D_2}
  \]
  and all other components empty,
\ie{} $\pres L e = \poss L e = \pres R e = \poss R e = \pres N e = \poss N e = \emptyset$.  For an action $\alpha$, we define: \begin{eqnarray*}
\ic\alpha&\eqd&\{\del\}\\ \tc\alpha&\eqd&\{\cel\}\\ \ev{\alpha}&\eqd&
\{\eact{\{\del\},\{\cel\}}(D_1,D_2)\st (D_1,D_2)\in\asem{\alpha}\}.
\end{eqnarray*} 
\end{enumerate}

\begin{exa}[$\nsem{[\lcon]:= 5}$] Recall that
\[\asem{[\lcon]:=5} = \{(\{\hvl{}{v}\},\{\hvl{}{5}\})\st v\in
\sn{Val}\},\] so \[\ev{[\lcon]:=5} =
\{\eact{\{\del\},\{\cel\}}(\{\hvl{}{v}\},\{\hvl{}{5}\})\st v\in
\sn{Val}\}.\] The definitions give the net $\nsem{[\lcon]:=5}$:\medskip
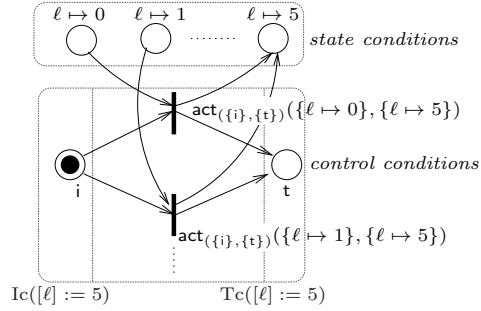
\begin{center} \input{act.pstex_t} \end{center} \end{exa}

\begin{enumerate}[$\triangleright$] 
\item{\bf Allocation and deallocation:}\ The command $\alloc\lcon$
  activates, by making current and assigning an arbitrary value to, a
  non-current location and sets $\lcon$ to point at it.  For symmetry,
  $\dealloc\lcon$ deactivates the current location pointed to by
    $\lcon$.

    We begin by defining two further event notations.
    First, $\eall{C,C'}(\lcon,v,\lcon',v')$ is the event $e$ such that 
    $\pres C e = C$ and $\poss C e = C'$ and
\[
\pres D e =\{\hvl{}v\} \qquad \poss D e =\{\hvl{}{\lcon'},\hvl'{v'}\}
\qquad \pres L e = \emptyset \qquad \poss L e =\{\curr(\lcon')\}, \]
and otherwise empty components, which changes $\lcon'$ from being
non-current to current, gives it value $v'$ and changes the value held
at $\lcon$ from $v$ to $\lcon'$.  If the condition $\curr(\lcon')$ is
marked before the event takes place, contact occurs, so the event has
concession only if the location $\lcon'$ is not initially current.
Second, $\edea{C,C'}(\lcon,\lcon',v')$ is the event $e$ such that
$\pres C e = C$ and $\poss C e = C'$ and
\[ \pres D e
    =\{\hvl{}{\lcon'},\hvl'{v'}\}\quad \poss D e
    =\{\hvl{}{\lcon'}\}\quad \pres L e =\{\curr(\lcon')\}, \] which
    does the converse of allocation.  The location $\lcon$ is left
    with a dangling pointer to $\lcon'$.  The two events may be drawn
    as: \begin{center} \begin{tabular}{l@{\qquad\qquad}l}
        $\eall{C,C'}(\lcon,v,\lcon',v')$: &
        $\edea{C,C'}(\lcon,\lcon',v')$:\\\\
        \qquad\input{alloc.pstex_t} &
        \qquad\input{dealloc.pstex_t}\\ \end{tabular}
    \end{center} The semantics of allocation is given by: \[
    \begin{array}{rcl} \ic{\alloc \lcon}&\eqd&\{\del\}\\ \tc{\alloc
    \lcon}&\eqd&\{\cel\}\\ \ev{\alloc \lcon}&\eqd&
    \{\eall{\{\del\},\{\cel\}}(\lcon,v,\lcon',v') \st
    \lcon'\in\sn{Loc} ~\tand~ v,v'\in\sn{Val}\}.  \end{array} \] Note
    that there is an event present for every value that $\lcon$ might
    initially hold and every value that $\lcon'$ might be assumed to
    take initially.
    
    The semantics of disposal is given by:
    \[
    \begin{array}{rcl}
      \ic{\dealloc \lcon}&\eqd&\{\del\}\\
      \tc{\dealloc \lcon}&\eqd&\{\cel\}\\
      \ev{\dealloc \lcon}&\eqd& 
      \{\edea{\{\del\},\{\cel\}}(\lcon,\lcon',v')
      \st \lcon'\in\sn{Loc} ~\tand~ v'\in\sn{Val}\}.
    \end{array}
    \]

    \item{\bf Sequential composition:}\  
    The sequential composition of terms involves gluing
    the terminal marking of the net for $t_1$ to the initial marking
    of the net for $t_2$.  The operation is therefore performed on the
    set
    \[P=\pref 1 {\tc{t_1}} \times \pref 2{\ic{t_2}}.\]
    Following the intuition above, we take
    \begin{eqnarray*}
      \ic{t_1;t_2} &\eqd&  \pref 1{\ic{t_1}} \\
      \tc{t_1;t_2} &\eqd&  \pref 2 {\tc{t_2}}\\
      \ev{t_1;t_2} &\eqd& (P\gluel \pref 1 {\ev{t_1}})
      \cup (P\gluer \pref 2 {\ev{t_2}}).
    \end{eqnarray*}
    The formation of the sequential composition on control conditions
    may be drawn schematically as:
    \begin{center}
      \input{seqcomp.pstex_t}
    \end{center}

  \item{\bf Parallel composition:}\ The control flow of the parallel
    composition of processes is autonomous; interaction occurs only
    through the state.  We therefore force the events of the two
    processes to work on disjoint sets of control conditions by giving
    them different tags:
    \begin{eqnarray*}
      \ic{t_1\pll t_2} &\eqd& \pref 1 {\ic{t_1}} \cup \pref 2 {\ic{t_2}}\\
      \tc{t_1\pll t_2} &\eqd& \pref 1 {\tc{t_1}} \cup \pref 2 {\tc{t_2}}\\ 
      \ev{t_1\pll t_2} &\eqd& \pref 1 {\ev{t_1}} \cup \pref 2 {\ev{t_2}}.
    \end{eqnarray*}
    Note that the definition of the semantics parallel composition is
    associative and commutative only if we regard nets up to
    isomorphism on the control conditions.
  \item{\bf Guarded sum:}\ 
    \label{sec:gsum}
    Let $t$ be the term $\alpha_1.t_1 + \alpha_2.t_2$.  The sum is
    formed by prefixing the actions onto the tagged nets representing
    the terms and then gluing the sets of terminal conditions.  Let
    $P=(\pref 1 {\tc{t_1}})\times(\pref 2{\tc{t_2}})$.  Define:
    \begin{eqnarray*}
      \ic{t} &\eqd& \{\del\}\\
      \tc{t} &\eqd& P\\      
      \ev{t} &\eqd&
      \quad\{\eact{\{\del\},\pref 1{\ic{t_1}}}(D_1,D_2) \st
      (D_1,D_2)\in\asem{\alpha_1}\}\\ 
      &&\cup\  \{\eact{\{\del\},\pref 2{\ic{t_2}}}(D_1,D_2) \st
      (D_1,D_2)\in\asem{\alpha_2}\}\\ 
      &&\cup\    P\gluel (\pref 1 {\ev{t_1}})
      \;\cup\;   P\gluer (\pref 2 {\ev{t_2}}).
    \end{eqnarray*}
    The net may be pictured schematically as follows, in which we have
    drawn only one representative event for each of $\alpha_1$ and $\alpha_2$,
    and have elided the effect of these events on state conditions.
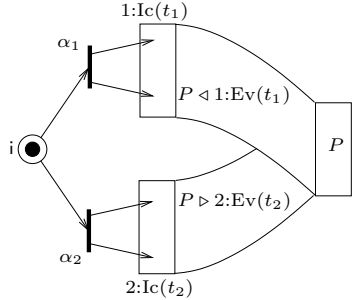
\begin{center}
\input{sum.pstex_t}
\end{center}

    On a technical point, one may wonder why the syntax of the
    language requires that sums possess guards.  This is seemingly
    curious since the category of safe Petri nets, which intuitively
    underlies a category of embedded nets, has a coproduct
    construction.  However, as remarked in Section 5 of
    \cite{winskel:pnamc}, there are cases where the coproduct of nets
    does not coincide with the usual interpretation of
    nondeterministic sum.  In Section 3.3 of \cite{winskel:evstr},
    this is explained as the occurrence net unfolding (the
    `behaviour') of the coproduct of two nets not being equal to the
    coproduct of their respective unfoldings.  To repeat an example
    given there, letting $+$ represent coproduct in the category of
    safe nets, we have: \begin{center} \input{coprod.pstex_t}
    \end{center} Consequently, using this coproduct as a definition of
    general sum, the runs of the net representing
    $\alpha+(\while{\co{true}}\ldo \alpha')$ would consist of some
    finite number of executions of $\alpha'$ followed, possibly, by
    one of $\alpha$.  Quite clearly, this does not correspond to the
    normal understanding of nondeterminism presented in the
    transition semantics.
    
    The restriction of processes to only use guarded sums allows us to
    recover the standard interpretation of sums (hence allowing the
    standard structural operational rule for sums).  As stated in
    \cite{winskel:pnamc,winskel:evstr}, another alternative would be
    to ensure that no event has a postcondition inside the initial
    conditions of the net.  This would necessitate a different
    semantics for $\co{while}$ loops, possibly along the lines of
    \cite{vanglabbeek:unfold} which would unfold one iteration of any
    loop.

  \item{\bf Iteration:}\ To form the net for $\while b\ldo t\lod$ we glue
  the initial and the terminal conditions of $b.t$ together and then
  add events to exit the loop when $\lnot b$ holds.  Let $P =
  \{\del\}\times \pref 1 {\tc{t}}$.  Define: \begin{eqnarray*}
  \ic{\while b\ldo t\lod} &\eqd& P\\ \tc{\while b\ldo t\lod} &\eqd&
  \{\cel\}\\ \ev{\while b\ldo t\lod} &\eqd& \quad \{\eact{P, \pref
  1{\ic{t}}}(D_{\sf t},D_{\sf t})\st (D_{\sf t},D_{\sf
  t})\in\asem{b}\}\\ &&\cup\ \{\eact{P,\{\cel\}}(D_{\sf f},D_{\sf
  f})\st (D_{\sf f},D_{\sf f})\in\asem{\lnot b}\}\\ &&\cup\ P\gluer
  (\pref 1 {\ev{t}}).  \end{eqnarray*} The loop can be visualized in
  the following way (in which we only present one event, $e_b$, for
  the boolean $b$ and one event, $e_{\lnot b}$, for the boolean $\lnot
  b$): \begin{center} \input{whilecomp.pstex_t} \end{center}

    \item{\bf Critical regions and local resources:}\ 
      We introduce the following notations for resource events. 
      \begin{center}\begin{tabular}{rl}
          $\enew{C,C'}(\res)$: & $\pres{\phantom{R}}{{\poss R e}}=\{\res\}$ 
          and $\pres{\phantom N}{\poss N e} = \{\curr(\res)\}$\\
          $\eend{C,C'}(\res)$: & $\pres{R}{\poss{\phantom{R}}e}=\{\res\}$ 
          and $\pres{N}{\poss {\phantom N} e} = \{\curr(\res)\}$\\
          $\eget{C,C'}(\res)$: & $\pres R {\poss {\phantom R} e}=\{\res\}$\\
          $\eput{C,C'}(\res)$: & $\pres {\phantom R} {\poss {R} e}=\{\res\}$
        \end{tabular}\end{center}
      These all have $\pres C e=C$ and $\poss C e=C'$, and the
      components other than those listed are empty.  Observe that the
      event $\enew{C,C'}(\res)$ will avoid contact, and thus be able
      to occur, only if the resource $\res$ is initially non-current.

      First consider $\resource\rvar\ldo t\lod$.
Its initial and
      terminal conditions are defined as:
      \begin{eqnarray*}
        \ic{\resource\rvar\ldo t\lod} &\eqd& \{\del\}\\
        \tc{\resource\rvar\ldo t\lod} &\eqd& \{\cel\}.
      \end{eqnarray*}
Its events are defined as:
\begin{eqnarray*}
&&{\ev{\resource\rvar\ldo t\lod}}\\
&\eqd&
\{\enew{\{\del\},\pref\res \ic{t'}}(\res),
\eend{\pref\res \tc{t'},\{\cel\}}(\res)
\st \res\in \sn{Res} ~\tand~ t'=[\res/\rvar]t\}\\
&&\cup
\bigcup\{\pref \res \ev{t'} \st \res\in \sn{Res} ~\tand~t'=[\res/\rvar]t\} 
\end{eqnarray*}
The net formed can be depicted:
        \begin{center}
          \input{resbind.pstex_t}
        \end{center}
As such, the semantics of resource variable binding is a
representation of the nondeterministic choice of resource to be
selected to be used for the variable.  Only one resource shall be
chosen for the variable, and it will initially have been non-current
thanks to contact described above.  Note that the semantics is
invariant under $\alpha$-equivalence $\equiv$.

      Now consider the term $\with\res\ldo t\lod$.  Its
      semantics is, informally, to acquire the resource $\res$, then
      to execute $t$, and finally to release the resource $\res$:
      \begin{eqnarray*}
        \ic{\with\res\ldo t\lod} &\eqd& \{\del\}\\
        \tc{\with\res\ldo t\lod} &\eqd& \{\cel\}\\
        \ev{\with\res\ldo t\lod} &\eqd&  
        \quad\{\eget{\{\del\},\pref \res{\ic{t}}}(\res)\} \cup \pref \res {\ev
          t}\\&& \cup\ \{\eput{\pref \res{\tc t},\{\cel\}}(\res)\}.
      \end{eqnarray*}

  \end{enumerate}
  
  \newcommand{\simbr}[1]{(#1)}  

  \subsection{Runs of nets}  
  A well-known property of independence models is that they support a
  form of run of the net in which independent actions are not
  interleaved: Given any sequence of events of the net between two
  markings, we can swap the consecutive occurrences of any two
  independent events to yield a run between the same two markings.  As
  seen in for example \cite{wn:mfc}, this allows us to form an
  equivalence class of runs between the same markings, generating a
  Mazurkiewicz trace.  This yields a partially ordered multiset, or
  \emph{pomset}, run \cite{pratt:pomset}, in which the independence of
  event occurrences is captured through them being incomparable.
  
  \begin{defi}
    A pomset path of a net $N=(B,E,\pre(-),\post{(-)},M_0)$ is a tuple
    $\pi=(X,\leq,\lambda)$ such that
    \begin{enumerate}[$\bullet$]
    \item $X$ is a finite set;
    \item $\leq$ is a partial order on $X$;
    \item $\lambda\ty X \to E$; and
    \item for all $x,x'\in X$, if $x\not\leq x'$ and $x'\not\leq x$ then
      $\lambda(x) \;I\; \lambda(x')$.
    \end{enumerate}
  \end{defi}
  The elements of $X$ can be thought of via $\lambda$ as occurrences
  of events.  Where two occurrences are unrelated through the order
  $\leq$, they can be thought of as occurring concurrently.  Their
  independence ensures that the effect of this is defined simply as
  any sequential occurrence of the events.
  \begin{defi}
    A \emph{sequence} is a path $\pi=(X,\leq,\lambda)$ in which $\leq$
    is a total order on $X$.  Let $x_1$ be the event occurrence least
    in $X$ according to $\leq$; let $x_2$ be the least event
    occurrence strictly greater than $x_1$; and so on, all the way up
    to $x_n$ which is the greatest event occurrence according to
    $\leq$ for $n$ equal to the size of $X$ (assumed to be finite).
    The sequence $\pi$ can be written as $e_1,\ldots, e_n$, where
    $\lambda(x_i)=e_i$ for all $0 < i\leq n$.  Say that a sequence
    $\pi=e_1,\ldots,e_n$ is \emph{from marking $M$ to marking $M'$ in
      $N$} if there exist $M_0,\ldots, M_n$ such that in $N$
    \[ M=M_0\ltr{e_1}M_1\ldots \ltr{e_n} M_n=M'. \]
  \end{defi}
  
  Note that the empty path is from marking $M$ to marking $M$ for any
  marking $M$.  We shall say that a pomset path $(X,\leq,\lambda)$ is
  from marking $M$ to $M'$ if there exists any extension of $\leq$ to
  a total order $\leq'$ such that $(X,\leq',\lambda)$ is a sequence
  from $M$ to $M'$.  As discussed, it is a standard result that any
  other extension of $\leq$ to a total order also yields a path from
  $M$ to $M'$.

  In fact, when we consider concurrent separation logic, we will only
  need to consider paths that are sequences, so in the rest of this
  paper we shall restrict attention to them;
  all our results generalize straightforwardly to pomsets.  From now
  on, we shall therefore use the terms `sequence', `path' and `run'
  interchangeably.  We have chosen to highlight pomset runs (for
  conciseness, we have not presented other forms of `run' of a net,
  such as causal nets) simply to show that Petri nets possess a notion
  of run that is non-interleaved.
  
  Write $()$ for the path comprising no events and write $e$ for the
  path with just a single event $e$.  We introduce the notation
  $\pi\ty M\ltr{}^* M'$ to mean that $\pi$ is a path from marking $M$
  to marking $M'$, and write $M\ltr{}^* M'$ if there exists a path
  from marking $M$ to marking $M'$.  We shall also write
  $\pi_1\cdot\pi_2$ for the composition of sequential paths; clearly,
  $M\ltr{\pi_1\cdot \pi_2} M'$ iff there exists $M''$ such that
  $M\ltr{\pi_1}M''$ and $M''\ltr{\pi_2} M'$.
  
  Finally, the tagging and gluing operations are extended to paths
  pointwise:
  \begin{eqnarray*}
    \pref x (e_1,\ldots, e_n) &\eqd& (\pref x e_1),\ldots,(\pref x e_n)\\
    P\gluel (e_1,\ldots,e_n) &\eqd& (P\gluel e_1),\ldots, (P\gluel e_n)\\
    P\gluer (e_1,\ldots,e_n) &\eqd& (P\gluer e_1),\ldots, (P\gluer e_n)
 \end{eqnarray*}

\subsection{Structural properties}
Here we establish characterizations of the runs of the net $\nsem{t}$
according to the structure of $t$.  The reader may wish to pass over
these technical, but important, details and go directly to Section
\ref{sec:coincidence}.

\newcommand{\ctr}[1]{\ltr{#1}_{\scriptscriptstyle \cset{C}}} 
\newcommand{\str}[1]{\ltr{#1}_{\scriptscriptstyle \cset{S}}} 

A complicating factor in characterizing the runs is that that we
cannot describe {\it a priori} the markings reachable in the net for
$t$ \emph{from an initial state} simply from the markings reachable
from the nets representing the subterms of $t$ (allowing for the
substitution of resources for resource names) running from suitable
initial states; this property, as one would expect, fails for parallel
composition.  However, we can establish properties about the
\emph{control flow} of programs.  Since such properties are
insensitive to the interaction through shared state of parallel
processes, they may be established inductively on (the size of) terms.
For an event $e$ and markings of control conditions $C$ and $C'$, we
write
$C\ctr e C'$ if the event $e$ has concession in the marking $C$ when
considering only its control conditions, and its occurrence would
result in the marking of control conditions $C'$:
\[ C\ctr{e} C' \iff  \pres C e \subseteq C ~\tand~
(C\setminus \pres C e) \cap \poss C e =\emptyset ~\tand~
C'=(C\setminus \pres C e)\cup \poss C e.\] We write $\sigma \str{e}
\sigma'$ if the event $e$ has concession on state conditions in the
marking $\sigma$ and its occurrence yields the marking of state
conditions $\sigma'$
\begin{lem}
\label{lemma:contsim}
    For any event $e$ and markings $C,C'$ of control conditions and
  $\sigma,\sigma'$ of state conditions, $(C,\sigma)\ltr{e}
  (C',\sigma')$ iff $C\ctr{e}C'$ and $\sigma\str{e}\sigma'$. \qed
\end{lem}
Following the above notation, we shall write $\pi\ty C\ctr{}^* C'$ if
the path $\pi$ is from the control marking $C$ to $C'$, defined in the
obvious way.  We shall say that a marking $C'$ is
\emph{control-reachable} from $C'$, written $C\ctr{}^* C'$, if there
exists a path $\pi$ such that $\pi\ty C\ctr{}^* C'$.  A particular
consequence of the above lemma is that the marking $(C',\sigma')$ is
reachable from $(C,\sigma)$ only if $C'$ is control-reachable from
$C$.

We begin with some fairly straightforward properties about the initial
and terminal markings and the sets of pre- and postconditions of each
event being nonempty.  The first and second items of the lemma below
could even be seen as part of the definition of embedded net since
nonemptiness is necessary for the constructions above to result in
nets with the expected behaviours.  With the final property, they can
be used to show that no event has concession in the terminal marking
of the net.  The third property eases the definitions constructing
$\nsem{t}$.
\begin{lem}
  \label{lemma:embstruc}
  For any closed term $t$ and event $e\in\ev{t}$:
  \begin{enumerate}[\em(1)]
  \item $\ic{t}\neq \emptyset$ and $\tc{t}\neq\emptyset$,
  \item $\pres C e\neq \emptyset$ and $\poss C  e\neq \emptyset$,
  \item $\ic{t}\cap\tc{t}=\emptyset$, and
  \item $\pres C e\cap \tc{t} = \emptyset$
  \end{enumerate}
  \begin{proof}
    The proof follows a simple induction on the size of terms. 
  \end{proof}
\end{lem}

The following property, that any event occurring from the initial
marking of a net has a precondition in the set of initial conditions
(and the corresponding property that any event into the terminal
marking of the net has a postcondition inside the terminal
conditions), follows immediately from the previous lemma.  It will be
used frequently; for instance, to show that in the net
$\nsem{t_1;t_2}$ if $e_1$ is an event from $\nsem{t_1}$ and $e_2$ is
an event from $\nsem{t_2}$ and $e_2$ immediately follows $e_1$ in some
sequential run, then there is a control condition that occurs in both
the postconditions of $e_1$ and the preconditions of $e_2$.  This
property is used in Theorem \ref{theorem:separation}.

\begin{lem}
    \label{lemma:ictc}
    For any closed term $t$, event $e$ and  marking $C$ of control conditions of
    $\nsem{t}$:
    \begin{enumerate}[$\bullet$]
    \item If $\ic{t} \ctr e C$
      then $\pre e\cap \ic{t}\neq \emptyset$.
    \item If $C\ctr{e}\tc t$ then $\post e \cap \tc t \neq \emptyset$.
      \qed
    \end{enumerate}
\end{lem}

Another important technical property that the embedded nets formed
possess is that the marking of control conditions is equal to the set
of initial conditions if either only initial conditions are marked or
if all initial conditions are marked, for any reachable marking, and
the similar statement for the terminal conditions of the net. 
\begin{defi}
  \label{def:dagger}
  Say that an embedded net $N$ is \emph{clear} if,
  for any marking of control conditions $C$ that is control-reachable
  from $\ic{N}$:
  \begin{enumerate}[(1)]
  \item if either $C\subseteq \ic{t}$ or $\ic{t}\subseteq C$ then $C=\ic{t}$, and
  \item if either $C\subseteq \tc{t}$ or $\tc{t}\subseteq C$ then $C=\tc{t}$.
  \end{enumerate}
\end{defi}
This is used in the proofs characterizing the markings reachable in
the net $\nsem{t}$ in terms of the markings reachable in the nets
representing $t$'s subterms (for instance, to show that any run to
completion of the net $\nsem{t_1;t_2}$ can be obtained as a run of the
net $\nsem{t_1}$ followed by a run of the net $\nsem{t_2}$ since when
$t_1$ in $\nsem{t_1;t_2}$ terminates, precisely the terminal control
conditions of $\nsem{t_1}$ will be marked).

Some care is necessary since the proof that, for any closed term $t$,
the net $\nsem{t}$ is clear itself requires understanding of the
markings reachable in the net $\nsem{t}$.  To resolve this apparent
`circularity', when proving the properties required of the net
$\nsem{t}$ required to show that the net is clear we shall assume that
the nets representing the subterms of $t$ are clear.  We shall
\emph{then} prove that any net $\nsem{t}$ is clear, allowing us to use
elsewhere the properties relating runs of the net $\nsem{t}$ to the
runs of the nets of subterms of $t$.  In effect, we will be proving
clearness and the structural properties simultaneously, by induction
on the size of terms.

\subsubsection{Sequential composition}
The technique that we use to relate the  runs of the net for
a term $t$ to the  runs of the nets of its subterms is to
establish a suitably strong invariant relating the markings arising
before and after the occurrence of any event present in $\nsem{t}$,
and then perform an induction on the length of sequence.  For
instance, for sequential composition, we prove:
\begin{lem}
  \label{lemma:seq}
  Let $P=\pref{1}{\tc{t_1}}\times\pref{2}{\ic{t_2}}$.  Assume that
  $\nsem{t_1}$ and $\nsem{t_2}$ are clear (Definition
  \ref{def:dagger}), and consider the net $\nsem{t_1;t_2}$.  For any
  event $e\in \ev{t_1;t_2}$ and any markings of control conditions
  $C_1$ and $C_2$:
  \begin{enumerate}[$\bullet$]
  \item $\ic{t_1;t_2}=P\gluel \pref{1}{\ic{t_1}}$ and
    $\tc{t_1;t_2}=P\gluer \pref{2}{\tc{t_2}}$.
  \item $P=P\gluel \pref 1 {C_1}$ iff $C_1={\tc{t_1}}$,
    and $P=P\gluer \pref 2 {C_2}$ iff $C_2={\ic{t_2}}$.
  \item Suppose that $C_1$ is control-reachable from
    $\ic{t_1}$ in $\nsem{t_1}$.  If $P\gluel \pref 1
    {C_1} \ctr{e} C'$ in $\nsem{t_1;t_2}$ then
    either $C_1=\tc{t_1}$ or there exist $C_1'$ and $e_1$
    such that $C_1\ctr{e_1}C_1'$ in $\nsem{t_1}$
    and $C'=P\gluel \pref{1}{C_1'}$ and $e=P\gluel \pref 1{e_1}$.
  \item Suppose that $C_2$ is control-reachable from
    $\ic{t_2}$ in $\nsem{t_2}$.  If $P\gluer \pref 2
    {C_2} \ctr{e} C'$ in $\nsem{t_1;t_2}$ then there
    exist $C_2'$ and $e_2$ such that
    $C_2\ctr{e_2}C_2'$ in $\nsem{t_2}$ and
    $C'=P\gluer \pref{2}{C_2'}$ and $e=P\gluer \pref{2}{e_2}$.
  \end{enumerate}
  \proof The first item is simply a re-statement of part of the
  definition of $\nsem{t_1;t_2}$ and the second item is easy to show.
  The remaining parts follow an analysis of the events of the net.\qed
\end{lem}

Using this result, it can be shown that any state reached in
$\nsem{t_1;t_2}$ is reached either as a run of $\nsem{t_1}$ or as a
run of $\nsem{t_1}$ to a terminal marking followed by a run of
$\nsem{t_2}$.
\begin{lem}
\label{lemma:seqpath}
Suppose that the nets $\nsem{t_1}$ and $\nsem{t_2}$ are clear.  If
$\pi\ty \ic{t_1;t_2} \ctr{}^* C$ in $\nsem{t_1;t_2}$ then either:
  \begin{enumerate}[$\bullet$]
  \item there exist $C_1$ and $\pi_1$ such that $C=P\gluel \pref 1
    {C_1}$ and $\pi=P\gluel \pref 1 \pi_1$ and $\pi_1\ty \ic{t_1}\ctr{}^* C_1$ in
    $\nsem{t_1}$, or
  \item there exist $C_2$, $\pi_1$ and $\pi_2$ such that $C=P\gluer
    \pref 2 {C_2}$ and $\pi=(P\gluel \pref 1 \pi_1)\cdot(P\gluer \pref
    2 \pi_2)$ and $\pi_1\ty \ic{t_1}\ctr{}^* \tc{t_1}$ in $\nsem{t_1}$
    and $\pi_2\ty \ic{t_2}\ctr{}^* C_2$ in $\nsem{t_2}$,
  \end{enumerate}
  where $P=\pref 1 {\tc{t_1}} \times \pref 2 {\ic{t_2}}$.
\begin{proof}
  A straightforward induction on the length of $\pi$ using Lemma
  \ref{lemma:seq}.
\end{proof}
\end{lem}
The above lemma can be extended straightforwardly using Lemma
\ref{lemma:contsim} to obtain the following result involving states,
using the fact that the operations of prefixing and tagging do not
affect the action of events on state conditions:
\begin{lem}
  \label{lemma:seqstate}
  Suppose that the nets
  $\nsem{t_1}$ and $\nsem{t_2}$ are clear.  If $\pi\ty (\ic{t_1;t_2},\sigma_0)
  \ltr{}^* (C,\sigma)$ in $\nsem{t_1;t_2}$ then either:
  \begin{enumerate}[$\bullet$]
  \item there exist $C_1$ and $\pi_1$ such that $C=P\gluel \pref 1
    {C_1}$ and $\pi=P\gluel {\pref 1 {\pi_1}}$ and $\pi_1\ty
    (\ic{t_1},\sigma_0)\ltr{}^* (C_1,\sigma)$ in $\nsem{t_1}$, or
  \item there exist $C_2$, $\sigma'$, $\pi_1$ and $\pi_2$ such that
    $C=P\gluer \pref 2 {C_2}$ and $\pi=(P\gluel \pref 1
    \pi_1)\cdot(P\gluer \pref 2 \pi_2)$ and $\pi_1\ty
    (\ic{t_1},\sigma_0)\ltr{}^* (\tc{t_1},\sigma')$ in $\nsem{t_1}$
    and $\pi_2\ty (\ic{t_2},\sigma')\ltr{}^* (C_2,\sigma)$ in
    $\nsem{t_2}$,
  \end{enumerate}
  where $P=\pref 1 {\tc{t_1}} \times \pref 2 {\ic{t_2}}$.
\qed
\end{lem}

The converse result, that runs of the nets $\nsem{t_1}$ and
$\nsem{t_2}$, with appropriate intermediate states, give rise to runs
of the net $\nsem{t_1;t_2}$ can also be shown.  

\subsubsection{Parallel composition}
Runs of control within
the net $\nsem{t_1\pll t_2}$ are amenable to a similar (though in fact
less complicated) analysis to that presented in Lemmas \ref{lemma:seq}
and \ref{lemma:seqpath}:

\begin{lem}
\label{lemma:paract}
  Consider the net $\nsem{t_1\pll t_2}$.
  \begin{enumerate}[$\bullet$]
  \item $\ic{t_1\pll t_2} = \pref 1 \ic{t_1} \cup \pref 2 \ic{t_2}$ and
    $\tc{t_1\pll t_2} = \pref 1 \tc{t_1} \cup \pref 2 \tc{t_2}$.
  \item For any markings $C_1, C_2$ and $C'$ of control conditions and
    any event $e\in \ev{t_1\pll t_2}$, if $\pref 1 C_1 \cup \pref 2
    C_2 \ctr{e} C'$ in $\nsem{t_1\pll t_2}$ then either:
    \begin{enumerate}[$-$]
    \item there exists $e_1 \in \ev{t_1}$ such that $e=\pref 1 e_1$
      and there exists $C_1'$ such that $C'=\pref 1 C_1' \cup \pref 2 C_2$ and
      $C_1\ctr{e_1} C_1'$ in $\nsem{t_1}$, or
    \item there exists $e_2 \in \ev{t_2}$ such that $e=\pref 2 e_2$
      and there exists $C_2'$ such that $C'=\pref 1 C_1 \cup \pref 2 C_2'$ and
      $C_2\ctr{e_2} C_2'$ in $\nsem{t_2}$.
    \end{enumerate}
  \end{enumerate}
  \proof A straightforward examination of the events of $\nsem{t_1\pll
    t_2}$.\qed
\end{lem}
Using the preceding lemma, the paths of the net $\nsem{t_1\pll t_2}$
on control conditions can be characterized as:
\begin{lem}
  If $\pi\ty \ic{t_1\pll t_2} \ctr{}^* C$ in $\nsem{t_1\pll t_2}$
  then any event $e$ in $\pi$ is either equal to $\pref 1 e_1$ for
  some event $e_1\in\ev{t_1}$ or equal to $\pref 2 e_2$ for some event
  $e_2\in\ev{t_2}$.  Furthermore, there exist $C_1$ and $C_2$ such
  that $C=\pref 1 C_1 \cup \pref 2 C_2$ and 
  \[\pi_1\ty \ic{t_1}\ctr{}^* C_1 ~\tand~ \pi_2 \ty \ic{t_2}\ctr{}^* C_2, \]
  where $\pi_1$ is obtained by removing events equal to $\pref 2 e_2$
  for some $e_2$ from $\pi$, and $\pi_2$ is obtained by removing
  events equal to $\pref 1 e_1$ for some $e_1$ from $\pi$.
\proof Induction on the length of path $\pi$.\qed
\end{lem}
Notably there is no analogue to Lemma \ref{lemma:seqstate} involving
the markings of state conditions for the parallel composition.

\subsubsection{Iteration}
The net $\nsem{\while b\ldo t_0\lod}$ allows runs that start with an
event that either shows that the boolean $b$ holds or an event that
shows that $b$ fails.  If $b$ fails, the net enters its terminal
marking an no further action occurs.  If the boolean $b$ passes, a run
of the net $\nsem{t_0}$ occurs, followed by the net re-entering its
initial control state.  The following lemma captures this; it is
proved by establishing an invariant in the same way as was done for
the sequential composition, though for brevity we shall omit it.

\begin{lem}
  Let $t\equiv \while b\ldo t_0 \lod$ and suppose that $\nsem{t_0}$ is
  clear.  Let $P=\{\del\}\times \pref 1 {\tc{t_1}}$, and recall that
  $P=\ic{t}$.  Assume that $\pi$ is a path such that $\pi\ty \ic{t}
  \ctr{}^* C$ in $\nsem{t}$ for some $C$.  There exists a natural
  number $n\geq 0$ and a (possibly empty if $n=0$) collection of paths
  $\pi_1,\ldots,\pi_n$ and heaps $D_1,\ldots,D_n$ such that, for each
  path $\pi_i$:
   \[
   \begin{array}{rcl@{\qquad}l}
     \pi_i &\ty& \ic{t_0}\ctr{}^* \tc{t_0} &\text{in }\nsem{t_0} \\
     \eact{\ic{b},\tc{b}}(D_i,D_i) &\ty& \ic{b}\ctr{} \tc{b} 
     &\text{in }\nsem{b}.
   \end{array}
   \]
   Write $e_i$ for the event $\eact{P,\pref 1{\ic{t_0}}}(D_i,D_i)$.
   Either:
  \begin{enumerate}[$\bullet$]
  \item $C=\ic{t}$ and $\pi=e_1\cdot (P\gluer \pref 1 \pi_1) \cdot
    \ldots \cdot e_n\cdot (P\gluer \pref 1 \pi_n)$;
  \item $C=P\gluel \pref 1 C'$ for some marking of control conditions $C'$
    and there exists a path $\pi'$ and heap $D'$ such that
    \[ \pi = e_1\cdot (P\gluer \pref 1 \pi_1) \cdot
    \ldots e_n\cdot (P\gluer \pref 1 \pi_n) \cdot
    \eact{P,\pref 1 {\ic{t_0}}}(D',D') \cdot  (P\gluer\pref 1 \pi') \]
    and  
    \[
   \begin{array}{rcl@{\qquad}l}
     \pi' &\ty& \ic{t_0}\ctr{}^* C' &\text{in }\nsem{t_0} \\
     \eact{\ic{b},\tc{b}}(D',D') &\ty& \ic{b}\ctr{} \tc{b} 
     &\text{in }\nsem{b}; \text{ or}
   \end{array}
   \]
  \item $C=\tc{t}$
    and there exists a heap $D'$ such that
    \[ \pi = e_1\cdot (P\gluer \pref 1 \pi_1) \cdot
    \ldots e_n\cdot (P\gluer \pref 1 \pi_n) \cdot
    \eact{P,\tc{t}}(D',D')\]
    and  
    $
     \eact{{\ic{\lnot b},\tc{\lnot b}}}(D',D') \ty \ic{b}\ctr{} \tc{b} 
     \qquad\text{in }\nsem{\lnot b}
   $.\qed
  \end{enumerate}
\end{lem}
The three possible cases for the control marking $C$ above correspond
to net being in its initial control state (following some number of
iterations), the net being in the body of the loop, and the net being
in its terminal control state following exit of the loop.

\subsubsection{Sums}
The behaviour of the net $\nsem{\alpha_1.t_1 + \alpha_2.t_2}$ can be
characterized as either the occurrence of an event of the action
$\alpha_1$ followed by a run of $t_1$ or the occurrence of an event of
the action $\alpha_2$ followed by a run of $t_2$.  Note that if
$C=P\gluel \pref 1 C_1$ then $C=\tc{\alpha_1.t_1 + \alpha_2.t_2}$ if,
and only if, $C_1=\tc{t_1}$, and the similar property for $t_2$.

\begin{lem}
  Let $t\equiv \alpha_1.t_1 + \alpha_2.t_2$ and $P=\pref 1 \tc{t_1}
  \times \pref 2 \tc{t_2}$ and suppose that the nets $\nsem{t_1}$ and
  $\nsem{t_2}$ are clear.  If $\pi$ is a path $\pi\ty \ic{t}\ltr{}^*
  C$ in $\nsem{t}$ for some $C$ then:
  \begin{enumerate}[$\bullet$]
  \item $C=\ic{t}$ and $\pi=()$, or
  \item $C=P\gluel \pref 1 C_1$ for some $C_1$ and 
    $\pi= \eact{\ic{t},\pref 1 {\ic{t_1}}}(D_1,D_1')
    \cdot (P\gluel \pref 1 \pi_1)$ for some $\pi_1,D_1,D_1'$ such that
    \[
    \begin{array}{rcl@{\qquad}l}
      \eact{\ic{\alpha_1},\tc{\alpha_1}}(D_1,D_1') &\ty& \ic{\alpha_1}\ctr{} \tc{\alpha_1} 
      &\text{in }\nsem{\alpha_1}\\
      \pi_1 &\ty& \ic{t_1}\ctr{}^* C_1 &\text{in }\nsem{t_1}; \text{ or}
    \end{array}
    \]
  \item $C=P\gluer \pref 2 C_2$ for some $C_2$ and 
    $\pi= \eact{\ic{t},\pref 2 {\ic{t_2}}}(D_2,D_2')
    \cdot (P\gluer \pref 2 \pi_2)$ for some $\pi_2,D_2,D_2'$ such that
    \[
    \begin{array}{rcl@{\qquad}l}
      \eact{\ic{\alpha_2},\tc{\alpha_2}}(D_2,D_2') &\ty& \ic{\alpha_2}\ctr{} \tc{\alpha_2} 
      &\text{in }\nsem{\alpha_2}\\
      \pi_2 &\ty& \ic{t_2}\ctr{}^* C_2 &\text{in }\nsem{t_2}.
    \end{array}
    \]
  \end{enumerate}
\begin{proof}
An induction following establishing an invariant in the style of Lemma
\ref{lemma:seq}.
\end{proof}
\end{lem}

\subsubsection{Resource declaration}
A consequence of the following result is that any complete run of the net $\resource
\rvar\ldo t_0 \lod$ consists first of an event that chooses a resource
$\res$ to be used for $\rvar$, then a run of $[\res/\rvar]t_0$, and
finally an event that records that $\res$ is no longer in use.
\begin{lem}
\label{lemma:resact}
Suppose that the net $\nsem{[\res/\rvar]t_0}$ is clear for any
resource $\res$ and let $t\equiv \resource\rvar\ldo t_0\lod$.  If in
the net $\nsem{t}$ we have $\pi\ty \ic{t}\ctr{}^* C$ then either:
  \begin{enumerate}[$\bullet$]
  \item $C=\ic{t}$ and $\pi=()$, or
  \item there exist $\res\in \sn{Res}$ and $C'$ and $\pi_0$ such that
    $C=\pref{\res} C'$ and \[\pi = \enew{\{\del\}, 
      \pref{\res}\ic{[\res/\rvar]t_0}}(\res)\cdot (\pref{\res}\pi_0)\] 
    and 
    \begin{center}
      \begin{tabular}{rll}
        $\enew{\{\del\},\pref{\res}\ic{[\res/\rvar]t_0}} \ty $ &
        $\ic{t} \ctr{}
        \pref{\res}\ic{[\res/\rvar]t_0}$ & in $\nsem{t}$ and\\
        $\pi_0 \ty$ & $ \ic{[\res/\rvar]t_0}\ctr{}^* C'$ & in
        $\nsem{[\res/\rvar]t_0}$, or
      \end{tabular}
      \end{center}
    \item $C=\tc{t}$ and there exist $\res\in\sn{Res}$ and $\pi_0$ such that
      \[\pi = \enew{\{\del\}, 
        \pref{\res}\ic{[\res/\rvar]t_0}}(\res)\cdot (\pref{\res}\pi_0)
      \cdot \eend{\pref{\res}\tc{[\res/\rvar]t_0},\{\cel\}}(\res)
      \] 
    and 
    \begin{center}
      \begin{tabular}{rll}
        $\enew{\{\del\},\pref{\res}\ic{[\res/\rvar]t_0}} \ty $ &
        $\ic{t} \ctr{}
        \pref{\res}\ic{[\res/\rvar]t_0}$ & in $\nsem{t}$, \\
        $\pi_0 \ty$ & $ \ic{[\res/\rvar]t_0}\ctr{}^* \tc{[\res/\rvar]t_0}$ 
        & in $\nsem{[\res/\rvar]t_0}$, and\\
        $\eend{\pref{\res}\tc{[\res/\rvar]t_0},\{\cel\}} \ty $ &
        $\pref{\res}\tc{[\res/\rvar]t_0} \ctr{} \tc{t}$ & in $\nsem{t}$.
      \end{tabular}
      \end{center}
  \end{enumerate}
\begin{proof}
By establishing an invariant on markings between the occurrences of single events,
as in Lemma \ref{lemma:seq}.
\end{proof}
\end{lem}

\subsubsection{Critical regions}
The net $\nsem{\with\res\ldo t_0\lod}$ starts by acquiring the
resource $\res$. If this action cannot proceed because the resource is
unavailable, no event will occur.  If the resource is available, the
process behaves as $t_0$, and then releases the resource $\res$ if
$t_0$ terminates.
\begin{lem}
  Let $t\equiv \with\res\ldo t_0\lod$ and suppose that the net
  $\nsem{t_0}$ is clear.  If in the net $\nsem{t}$ we have
  $\pi\ty \ic{t}\ctr{}^* C$ then either:
  \begin{enumerate}[$\bullet$]
  \item $C=\ic{t}$ and $\pi=()$, 
  \item $C=\pref \res C_0$ for some marking of control conditions $C_0$ and 
    $\pi=\eget{\ic{t},\pref \res{\ic{t_0}}}(\res)\cdot (\pref \res \pi_0)$ 
    for some path $\pi_0$ such that
    $\pi_0 \ty \ic{t_0} \ctr{}^* C_0$ in $\nsem{t_0}$, or
  \item $C=\tc t$ and 
   $\pi=\eget{\ic{t},\pref \res{\ic{t_0}}}(\res)\cdot (\pref \res \pi_0)\cdot
   \eput{\pref \res {\tc{t_0}}, \tc{t}}(\res)$ for some path $\pi_0$ such that
    $\pi_0 \ty \ic{t_0} \ctr{}^* \tc{t_0}$ in $\nsem{t_0}$.\qed
  \end{enumerate}
\end{lem}

\subsubsection{Clearness}

Now that we have established these control properties of the runs of
processes, we can show that the clearness property of Definition
\ref{def:dagger} does indeed hold in the net $\nsem{t}$ for any term
$t$.
\begin{lem}
  \label{lemma:dagger}
  For any closed term $t$, the net $\nsem{t}$ is clear.

  \proof 
  Following the observation that 
  \begin{eqnarray*}
    \pref 1 C \subseteq \pref 1 C' &\tiff& C\subseteq C'\\
    P\gluel C \subseteq P\gluel C' &\tiff& C\subseteq C'\\
    P\gluer C \subseteq P \gluer C' &\tiff& C\subseteq C' ,
  \end{eqnarray*}
  the property can be proved by induction on the size of terms
  using the above control properties.\qed
\end{lem}

\subsubsection{Preservation of consistency}

The final attribute that we aim towards is that any marking of state
conditions $\sigma$ reachable in $\nsem{t}$ from a consistent initial
marking of state conditions $\sigma_0$ is itself consistent.  The only
challenge here will be showing that if $\res\in\sigma$ then
$\curr(\res)\in\sigma$, which shall require some understanding of the
nature of the critical regions present in our semantics; the other
requirements for consistency are straightforwardly shown to be
preserved through the occurrence of the events present in $\nsem{t}$.

We shall first show that any release of a resource is dependent on the
prior acquisition of that resource: for any sequence $\pi$ and any
resource there exists an injection $f$ that associates any occurrence
of a release event to a prior occurrence of an acquisition event of
that resource, and between the two occurrences there are no other
actions on that resource.
\begin{lem}
\label{lemma:crit}
  Let $\pi$ be a sequence of events, $\pi=(e_1,\ldots,e_n)$.  For any
  closed term $t$, resource $\res$ and marking of control conditions $C$ such
  that $\pi\ty \ic{t}\ctr{}^* C$ in $\nsem{t}$, there exists a partial
  function $f\ty \mathbb{N} \rightharpoonup \mathbb{N}$ satisfying,
  for all $i,j\in\mathbb{N}$:
  \begin{enumerate}[$\bullet$]
  \item $f$ is injective,
  \item if there exist sets of control conditions $C_1,C_2$ such that
    $e_i=\eput{C_1,C_2}(\res)$ then $f(i)$ defined, and
  \item if $f(i)$ defined then $f(i)< i$ and there exist sets of
    control conditions $C_1,C_2$ such that $e_{f(i)}=\eget{C_1,C_2}(\res)$.
  \end{enumerate}
  
  Moreover, if there exist markings of state conditions $\sigma_0,\ldots,\sigma_n$ and
  markings of control conditions $C_0,\cdots,C_n$ such that
  $(C_{i-1},\sigma_{i-1})\ltr{e_i} (C_i,\sigma_i)$ for all $i$ with
  $0<i\leq n$ and $C_0=\ic{t}$, then there exists an $f$ satisfying
  the above constraints and such that, for all $k$ with $i < k <
  f(i)$, there exist no $C'$ and $C''$ such that either
  $e_k=\eget{C',C''}(\res)$ or $e_k=\eput{C',C''}(\res)$.
\begin{proof}
  The first property is shown, using the control properties of
  sequences established above, by induction on the size of terms.  The
  second property arises since if $e_i=\eget{C_i,C_i'}(\res)$ and
  $e_j=\eget{C_j,C_j'}(\res)$ for $i<j$ then there must exist $k$ such
  that $i<k<j$ and $e_k=\eput{C_k,C_k'}(\res)$, and the symmetric
  property for release events.
\end{proof}
\end{lem}

We are now able to show that the nets formed preserve the consistency
of the markings of state conditions.
\begin{lem}[Preservation of consistent markings]
\label{lemma:conspres}
  For any closed term $t$, if $(\ic{t},\sigma_0)\ltr{}^*(C,\sigma)$ in the net
  $\nsem{t}$ and the marking $\sigma_0$ of state conditions is
  consistent then $\sigma$ is consistent.
  \begin{proof}
    It is straightforward to prove by induction on the size of the
    term $t$ that the events present in that net $\nsem{t}$ are
    all of one of the following forms:
    \begin{center}
    \begin{tabular}{l@{\quad}l@{\quad}l@{\quad}l}
      $\eact{C,C'}(D,D')$ &   
      $\eall{C,C'}(\lcon,v,\lcon',v')$ &
      $\enew{C,C'}(\res)$ &
      $\eget{C,C'}(\res)$ \\ &
      $\edea{C,C'}(\lcon,\lcon',v')$ &
      $\eend{C,C'}(\res)$ &
      $\eput{C,C'}(\res)$
    \end{tabular}
  \end{center}
  It is readily shown that each form of event preserves the
  consistency of the marking of state conditions, apart from showing
  that if $\res\in\sigma$ then $\curr(\res)\in\sigma$.  
  
  Suppose, for contradiction, that $\pi'$ is a path such that $\pi'\ty
  (\ic{t},\sigma_0)\ltr{}^* (C,\sigma)$ in $\nsem{t}$ and that
  $\res\in\sigma$ but $\curr(\res)\not\in\sigma$.  Assume,
  furthermore, and without loss of generality, that any other marking
  of state conditions $\sigma'$ along $\pi$ has the property that if
  $\res\in\sigma$ then $\curr(\res)\in\sigma$.  It must be the case that
  $\pi'=\pi\cdot \eput{D_1,D_1'}(\res)$ for some $D_1,D_1'$ and $\pi$.
  By Lemma \ref{lemma:crit}, there exist $D_2,D_2',\pi_1$ and $\pi_2$
  such that $\pi=\pi_1\cdot \eget{D_2,D_2'}(\res)\cdot \pi_2$ and no
  event in $\pi_2$ is an $\lget(\res)$ or $\lput(\res)$ event.  Let
  $\pi_1\ty (\ic{t},\sigma_0) \ltr{}^* (C_1,\sigma_1)$.  We must have
  $\res\in\sigma_1$, and by assumption $\curr(\res)\in\sigma_1$.  It
  can be seen that we must have $\curr(\res)\in\sigma'$ and
  $\res\not\in\sigma'$ for all states $\sigma'$ reached along
  $\eget{D_2,D_2'}(\res)\cdot \pi_2$ from $(C_1,\sigma_1)$ since no
  $\lend(\res)$ event can have concession in such markings.
  Consequently, we must have $\curr(\res)\in\sigma_2$ for $\sigma_2$
  obtained by following the path $\pi\ty(\ic{t},\sigma_0)\ltr{}^*
  (C_2,\sigma_2)$, and therefore $\curr(\res)\in\sigma$.
  \end{proof}
\end{lem}

The structure of processes ensures that any resource initially current
remains current through the execution of the net.  The same property
working backwards from the terminal marking of the net also holds.
\begin{lem}
\label{lemma:respres}
Let $\sigma,\sigma'$ be a consistent markings of state conditions. For
any markings of control conditions $C,C'$:
\begin{enumerate}[\em(1)]
\item If $(\ic{t},\sigma)\ltr{}^*(C',\sigma')$ in $\nsem{t}$ and
  $\curr(\res)\in \sigma$ then $\curr(\res)\in\sigma'$.
\item If $(C,\sigma)\ltr{}^*(\tc{t},\sigma')$ in $\nsem{t}$ and
  $\curr(\res)\in \sigma'$ then $\curr(\res)\in\sigma$.
\end{enumerate}

\proof We shall only show (1) since (2) is similar.  An induction on
the size of terms using the control properties above gives the
following:
\begin{enumerate}[$\bullet$]
\item If there exists a sequence $\pi$ such that
  $\pi\cdot \eend{C_1,C_2}(\res)\ty \ic{t}\ctr{}^* C$ for some $C_1,C_2$
  then there exists an event $\enew{C_1',C_2'}(\res)$ in $\pi$ for
  some $C_1',C_2'$.
\end{enumerate}
Let $\pi'$ be a sequence $\pi'\ty (\ic{t},\sigma) \ltr{}^*
(C',\sigma')$ and assume that $\curr(\res)\in\sigma$. Without loss of
generality, suppose that $(C',\sigma')$ is the earliest marking along
$\pi'$ from $(\ic{t},\sigma)$ such that $\curr(\res)\not\in\sigma'$;
otherwise, we can take the initial segment of $\pi'$ with this
property.  Examination of the events given by our semantics reveals
that the last event in $\pi'$ is an $\eend{C_1,C_2}(\res)$ event,
since otherwise $\curr(\res)$ is not in the state prior to $\sigma'$.
Now, applying the result above  informs
that there is an event $\enew{C_1',C_2'}(\res)$ in $\pi'$ and this must
occur before $\eend{C_1,C_2}(\res)$.  Now, the event
$\enew{C_1',C_2'}(\res)$ can only occur in a marking $\sigma_0$ of
state conditions such that $\curr(\res)\not\in\sigma_0$, but this
contradicts our assumption that $\sigma'$ was the first marking of
state conditions reachable along $\pi'$ from $(\ic{t},\sigma)$ with
$\curr(\res)\not\in\sigma'$.  \qed
\end{lem}

\subsection{Correspondence of semantics}
\label{sec:coincidence}

As we have progressed, the event notations introduced have
corresponded to labels of the transition semantics.  Write $|e|$ for
the label corresponding to event $e$.  Before progressing to consider
separation logic, we shall give a theorem\footnote{The proof of this
  theorem is rather technical and requires a presentation of open
  maps on the category of embedded Petri nets, so we shall not present
  the proof here.  It shall appear, with the other omitted results, in
  the first author's PhD thesis.} that shows how the net and
transition semantics correspond.  It assumes a definition of open map
bisimulation \cite{wnj:openmaps,wn:petribisim} based on paths as
pomsets, $(N,M)\sim(N',M')$, relating paths of net $N$ from marking
$M$ to paths of $N'$ from $M'$.  The bisimulations that we form
respect terminal markings and markings of state conditions.

\begin{thm}[Correspondence]
  \label{theorem:semcorr}
   Let $t$ be a closed term and $\sigma$ be a consistent state.
  \begin{enumerate}[$\bullet$]
  \item If $\ang{t,\sigma}\otr\lambda \sigma'$ then there exists $e$
    such that $|e|=\lambda$ and \mbox{$(\ic t,\sigma)\ltr e
      (\tc t,\sigma')$} in $\nsem{t}$.
  \item If $\ang{t,\sigma}\otr\lambda\ang{t',\sigma'}$ then there
    exists $e$ such that $|e|=\lambda$ and $(\ic
    t,\sigma)\ltr e(C',\sigma')$ in $\nsem{t}$ and $\simbr{\nsem t,C',\sigma'} \sim
    \simbr{\nsem{t'},\ic{t'},\sigma'}$.
  \item If $(\ic{t},\sigma)\ltr e (C',\sigma')$ in $\nsem{t}$ then either
    there exists $t'$ such that
    $\ang{t,\sigma}\otr{|e|}\ang{t',\sigma'}$ and $\simbr{\nsem
      t,C',\sigma'}\sim\simbr{\nsem{t'},\ic{t'},\sigma'}$, or
    $\ang{t,\sigma}\otr{|e|}\sigma'$ and $C'=\tc{t}$.\qed
  \end{enumerate}
\end{thm}

Write $(t,\sigma)\sim (t',\sigma')$ iff there exist a label-preserving
bisimulation (in the standard sense) between the transitions systems
for $t$ from initial state $\sigma$ and $t'$ from $\sigma'$.  From the
preceding result, we obtain adequacy of our semantics:
\begin{cor}[Adequacy]
  Let $t,t'$ be closed terms and $\sigma,\sigma'$ be consistent states.
  If $(\nsem{t},\ic{t},\sigma)\sim (\nsem{t'},\ic{t'},\sigma')$
  then $(t,\sigma)\sim (t',\sigma')$.  \qed
\end{cor}
The converse property with respect to $\sim$ fails.
For instance, for any $\sigma$ we have
\[(\alpha_1 \pll \alpha_2, \sigma) \sim (\alpha_1.\alpha_2 +
\alpha_2.\alpha_1, \sigma).\] However, the definition of open
bisimulation on the nets with pomsets as paths yields
\[ (\nsem{\alpha_1\pll \alpha_2}, \ic{\alpha_1\pll \alpha_2}, \sigma)
\not\sim
(\nsem{\alpha_1.\alpha_2 +
\alpha_2.\alpha_1}, \ic{\alpha_1.\alpha_2 +
\alpha_2.\alpha_1}, \sigma).
\]
The reason why the property fails is that the transition system does
not capture the independence of actions.

 \section{Separation logic}
\label{sec:seplogic}
As discussed in the introduction, concurrent separation logic 
establishes
partial correctness assertions about concurrent heap-manipulating
programs; that  whenever a given program running from a heap
satisfying a \emph{heap formula} $\phi$ terminates, the resulting heap
satisfies a heap formula $\psi$.  The semantics of the heap logic
arises as an instance of the logic of Bunched Implications
\cite{ohearn:bi}.  At its core are the associated notions of heap
composition and the separating conjunction.  Two heaps may be composed
if they are defined over disjoint sets of locations:
\[ D_1\cdot D_2 \eqd D_1\cup D_2 \text{ if } \dom(D_1)\cap 
\dom(D_2)=\emptyset.\] 
A heap satisfies the separating conjunction $\phi_1\mand\phi_2$ if it
can be split into two parts, one satisfying $\phi_1$ and the other
$\phi_2$:
\begin{eqnarray*}
D\models\phi_1\mand\phi_2 &\tiff&
\text{there exist } D_1,D_2\text{ such that }D_1\cdot D_2 \text{ defined}~\tand\\
&&D=D_1\cdot D_2 ~\tand~D_1\models\phi_1~\tand~D_2\models\phi_2.
\end{eqnarray*}
 \newcommand{\vals}{_{\mathsf{val}}}
 \newcommand{\locs}{_{\mathsf{loc}}}
The semantics of the other parts of the heap logic is of little
significance when considering the semantics of the program logic.  For
completeness, however, it is defined by induction on the size of
formul\ae{} in Figure \ref{fig:heaplog} where the full syntax also
appears.  Unlike the heap logic presented in \cite{brookes:soundness},
we do not allow arithmetic on memory locations; this is just to
simplify the presentation, and such arithmetic could easily be added.
Since we distinguish the types of locations and values, we use
$x\locs$ as the logical variable for locations and $x\vals$ for the
logical variable for values.  We adopt the usual binding precedences,
and $\mand$ binds more tightly the standard logical connectives.  We
define the shorthand notation $\hvl{}{\something}$ for $\exists
x\vals(\hvl{}{x\vals})$.  We shall write $\models\phi$ if
$D\models\phi$ for all heaps $D$, and write $\phi\implies\psi$ if
$\models\phi\limp\psi$.

\begin{figure}[p]
  \hrule
 \[
 \begin{array}{lrcll}
   \textit{Variables:} &
   x &\coloneqq& x\locs & \text{Location variable}\\ &
   &\gor& x\vals &\text{Value variable}\\\\
   \textit{Location expressions:}
   &
   e\locs &\coloneqq& x\locs & \text{Location variable}\\ &
   &\gor& \lcon &\text{Location, }\lcon\in\sn{Loc}
   \\\\
   \textit{Expressions:}
   &
   e &\coloneqq& e\locs & \text{Location expression}\\ &
   &\gor& x\vals & \text{Value variable}\\ &
   &\gor& v &\text{Value, } v\in\sn{Val}
   \\\\
   \textit{Formul\ae{}:}
   &
  \phi & \coloneqq& e\locs\mapsto e\vals & \text{heap location}\\ &
  &\gor& \phi\mand \phi & \text{separating conjunction}\\ &
  &\gor& \mathsf{empty} & \text{empty heap}\\ &
  &\gor& \phi\land\phi & \text{conjunction}\\ &
  &\gor& \phi\lor\phi & \text{disjunction}\\ &
  &\gor& \phi\to\phi &\text{implication}\\ &
  &\gor& \lnot\phi & \text{negation}\\ &
  &\gor& \exists x.\phi &\text{existential quantification}\\ &
  &\gor& \forall x.\phi &\text{universal quantification}\\ &
  &\gor& e=e & \text{equality}\\ &
  &\gor& \top & \text{true}\\ &
  &\gor& \bot & \text{false}
 \end{array}
 \]

\begin{flushleft}
 \textit{Semantics of closed formul\ae{}:}
\end{flushleft}
 \begin{tabular}{lcl}
   $D\models \hvl{}v $ & iff & $D=\{\lcon\mapsto v\}$\\
$D\models\phi_1\mand\phi_2$ & iff & there exist $D_1,D_2$ such that $D_1\cdot D_2$ 
    defined and\\
    && $D=D_1\cdot D_2$ and $D_1\models\phi_1$ and $D_2\models\phi_2$\\
$D\models \mathsf{empty}$ &iff& $D=\emptyset$\\
$D\models\phi_1\land\phi_2$ &iff& $D\models\phi_1$ and $D\models\phi_2$\\
$D\models\phi_1\lor\phi_2$ &iff& $D\models\phi_1$ or $D\models\phi_2$\\
$D\models\phi_1\limp\phi_2$ &iff& $D\models\phi_1$ implies $D\models\phi_2$\\
$D\models\lnot\phi$ &iff& not $D\models\phi$\\
$D\models\exists x\locs.\phi$ &iff& there exists $\lcon\in \sn{Loc}$ such that $D\models[\lcon/x\locs]\phi$\\
$D\models\exists x\vals.\phi$ &iff& there exists $v\in \sn{Val}$ such that $D\models[v/x\vals]\phi$\\
$D\models\forall x\locs.\phi$ &iff& for all $\lcon\in \sn{Loc}$:  $D\models[\lcon/x\locs]\phi$\\
$D\models\forall x\vals.\phi$ &iff& for all $v\in \sn{Val}$:  $D\models[v/x\vals]\phi$\\
$D\models v=v'$ &iff& $v=v'$\\
$D\models\top$ &always\\
$D\models\bot$ &never
\end{tabular}
\hrule
\caption{Syntax and semantics of the heap logic}
\label{fig:heaplog}
\end{figure}

We now present the intuition for the key judgement of concurrent
separation logic, $\Gamma\ent\hos{}$, where $\phi$ and $\psi$ are
\formulae{} of the heap logic, and $\Gamma$ is a \emph{environment} of
\emph{resource invariants} , of the form $\res_1:\chi_1,\cdots,
\res_n:\chi_n$, associating invariants $\chi_i$ with resources $\res_i$.
(We refer the reader to \cite{ohearn:rclr} for a fuller introduction.)
Informally, the judgement means:

\begin{list}{}{
    \addtolength{\leftmargin}{-0.65\leftmargin}
    \setlength{\rightmargin}{\leftmargin}
}
\item {\it In any run from a heap satisfying $\phi$ and the invariants $\Gamma$,
the process $t$ never accesses locations that it does not own, 
 and if  the process $t$ terminates then it does so in a heap satisfying $\psi$ and  the invariants $\Gamma$.}
\end{list}

Central to this understanding is the notion of \emph{ownership},
which we capture formally in Section \ref{sec:formalownership}. 
Initially the process $t$ is considered to \emph{own} that part of the heap which satisfies $\phi$, and accordingly to \emph{own} the locations in that subheap.  
As $t$ runs the locations it owns may change 
as it acquires and releases resources, and correspondingly the locations used in justifying their invariants.  

Ownership plays a key role in making the judgements of concurrent
separation logic  compositional:  a judgement
$\Gamma\ent\hos{}$  should hold even if other (unknown)
processes are to execute in  the same heap.  It is therefore necessary to
make certain assumptions about the ways in which these other processes
might interact with the process  $t$.  This is achieved through ownership,  by
assuming that each process  {owns}, throughout its execution, a separate, though possibly changing, part
of the heap; the part of the heap that each process owns must not
be accessed by any other process;
moreover a process must not access locations it does not own.  
 
The rules of concurrent separation logic are presented in Figure
\ref{fig:seplog} in the style of \cite{brookes:soundness}.  The only
significant difference between the two systems is that we omit the
rules for auxiliary variables and for existential quantification.
Both are omitted for simplicity since they are
peripheral to the focus of our work.

\begin{figure}[p]\hrule
\small  \[
  \begin{array}{lc}
    \rn{L-Act}: 
    &
    \deriv{\begin{array}{l}
        \text{for all }D\models\phi~\tand~(D_1,D_2)\in\asem{\alpha}:\\
        \begin{array}{c}
            \dom(D_1)\subseteq\dom(D) \\\text{and}\\
            D_1\subseteq D\text{ implies } (D\setminus D_1) \cup D_2\models \psi
          \end{array}\end{array}}
    {\Gamma\ent\ho{\phi}{\alpha}{\psi}}
    \\\\
    \rn{L-Alloc}: 
    &
    {\Gamma\ent\ho{\hvl{}{\something}}{\alloc\lcon}
      {\exists x\locs(\hvl{}{x\locs}\mand x\locs\mapsto\something)}}
    \\\\
    \rn{L-Dealloc}:
    &
    {\Gamma\ent\ho{\exists x\locs(\hvl{}{x\locs}\mand x\locs\mapsto\something)}
      {\dealloc\lcon}
      {\exists x\locs(\hvl{}{x\locs})}}
    \\\\
    \rn{L-Seq}:
    &
    \deriv
    {\Gt\ho{\phi}{t_1}{\phi'}\qquad \Gt\ho{\phi'}{t_2}{\psi}}
    {\Gt\ho{\phi}{t_1; t_2}{\psi}}
    \\\\
    \rn{L-Sum}:
    &    
    \deriv    { \Gt\ho{\phi}{\alpha_1}{\phi_1}\qquad
      \Gt\ho{\phi}{\alpha_2}{\phi_2} \\
      \Gt\ho{\phi_1}{t_1}{\psi}
      \qquad\Gt\ho{\phi_2}{t_2}{\psi}}
    {\Gt\ho{\phi}{\alpha_1.t_1+\alpha_2.t_2}{\psi}}
    \\\\
    \rn{L-While}:
    &
    \deriv
    {\Gt\ho{\phi}{b}{\phi'}\qquad\Gt\ho{\phi}{\lnot b}{\psi}
      \\ \Gt\ho{\phi'}{t}{\phi}}
    {\Gt\ho{\phi}{\while b\ldo t\lod}{\psi}}
    \\\\
    \rn{L-Res}:
    &
    \deriv
    {\Gamma,\res\ty\chi\ent\ho{\phi}{[\res/\rvar]t}{\psi} }
    {\Gamma\ent\ho{\phi\mand\chi}{\resource\rvar\ldo t\lod}{\psi\mand\chi}}
    \left(\begin{array}{l}\chi\text{ precise}\\ \res\not\in \dom(\Gamma)
\end{array}\right)    \\\\
    \rn{L-CR}:
    &
    \deriv
    {\Gamma,\res\ty\chi\ent\ho{\phi\mand\chi}{t}{\psi\mand\chi}}
    {\Gamma,\res\ty\chi\ent\ho{\phi}{\with\res\ldo t\lod}{\psi}}
    \\\\
    \rn{L-Par}:
    &
    \deriv
    {\Gt\hos{1}\qquad \Gt\hos{2}}
    {\Gt\ho{\phi_1\mand\phi_2}{t_1\pll t_2}{\psi_1\mand\psi_2}}
    \\\\  
    \rn{L-Frame}:  
    &  
    \deriv
    {\Gt\hos{}}
    {\Gt\ho{\phi\mand\phi'}{t}{\psi\mand\phi'}}
    \\\\
    \rn{L-Consequence}:
    &\deriv{\phi\implies\phi' \qquad \Gt\ho{\phi'}{t}{\psi'}
      \qquad \psi'\implies\psi}
    {\Gt\hos{}}
    \\\\
    \rn{L-Conjunction}:
    &
    \deriv{\Gt\ho{\phi_1}{t}{\psi_1} \qquad
      \Gt\ho{\phi_2}{t}{\psi_2}}
    {\Gt\ho{\phi_1\land\phi_2}{t}{\psi_1\land\psi_2}}
    \\\\
    \rn{L-Disjunction}:
    &
    \deriv{\Gt\ho{\phi_1}{t}{\psi_1} \qquad
      \Gt\ho{\phi_2}{t}{\psi_2}}
    {\Gt\ho{\phi_1\lor\phi_2}{t}{\psi_1\lor\psi_2}}
    \\\\
    \rn{L-Expansion}:
    &
    \deriv{\Gt\hos{}}{\Gamma,\Gamma'\ent\hos{}}

    \\\\
    \rn{L-Contraction}:
    &
    \deriv{\Gamma,\Gamma'\ent\hos{}}{\Gamma\ent\hos{}}(\fr{t}\subseteq \dom(\Gamma))
\end{array}
\]
\hrule
\caption{Rules of concurrent separation logic}
\label{fig:seplog}
\end{figure}
\afterpage{\clearpage}

As a first example, 
the rule for heap actions \rn{L-Act} would allow the judgement
\[ \Gamma\ent \ho{\hvl{} 0}{[\lcon]:= 1}{\hvl{}1} \]
since the process is initially assumed to own the location $\lcon$
because the part of the heap that the process initially owns satisfies
$\hvl{}0$.  The resulting part of the heap owned by the process
satisfies $\hvl{}1$.  The judgement
\[ \Gamma\ent\ho{\mathsf{empty}}{[\lcon]:= 2}{\true} \]
is \emph{not} derivable however: the part of the heap initially owned by the
process satisfies $\mathsf{empty}$, and therefore the process
initially does not own the location $\lcon$.  Assignment to $\lcon$
violates the principle that the process may only act on locations that
it owns --- the so-called \emph{frame property}.

An instance of the separating conjunction is seen in the rule for
parallel composition, \rn{L-Par}:
\[
    \deriv
    {\Gt\hos{1}\qquad \Gt\hos{2}}
    {\Gt\ho{\phi_1\mand\phi_2}{t_1\pll t_2}{\psi_1\mand\psi_2}}
\]
Informally,
 the rule is sound because the
part of the initial heap that is owned by the process $t_1\pll t_2$ can be
split into two parts, one part satisfying $\phi_1$ owned by
$t_1$ and the other satisfying $\phi_2$  owned by $t_2$;  as
the processes execute the subheaps that we see each as
owning remain disjoint from each other and end up separately satisfying $\psi_1$ and $\psi_2$.

It is vital that the logic enforces the requirement that processes
only act on locations that they own.  If this requirement were not
imposed, so that the judgement
\[ \Gamma\ent\ho{\mathsf{empty}}{[\lcon]:= 2}{\true} \]
\emph{were} derivable, then the rule for parallel composition could be
applied with the other judgement above to conclude that
\[ \Gamma\ent\ho{\hvl{}0\mand \mathsf{empty}}{[\lcon]:= 1 \pll [\lcon]:= 2}
{\hvl{} 1\mand\true}.\] This flawed assertion would imply that
whenever the process $[\lcon]:=1 \pll [\lcon]:=2$ runs from a state
satisfying $\hvl{}0$, the resulting state has $\hvl{} 1$,
which is obviously wrong.
 
The notion of ownership is 
subtle since the collection of locations that a process owns may
change as the process evolves.  As seen in the rule $\rn{L-Alloc}$,
the intuitive reading is that after an allocation event has taken
place the process owns the newly current location.  Similarly,
deallocation of a location leads to loss of ownership.  
For example, it is possible to make the judgement 
\[ \Gamma\ent \ho{\hvl{}{\something}}{\alloc\lcon}
{\exists x\locs . \hvl{}{x\locs} \mand x\locs\mapsto\something}.
\]
If the new location were $\lcon'$ which initially held value $v$, this
would mean that in the the (fragment of the) resulting heap
$\{\hvl{}{\lcon'} \mand \hvl{'}{v}\}$, the locations $\lcon$
and $\lcon'$ would be owned by the process.  Consequently, an action
$[[\lcon]]:= 0$ which assigns $0$ to the location pointed to by
$\lcon$ resulting in the heap $\{\hvl{}{\lcon'},\hvl{'}0\}$
allows the judgement
\[ \Gamma\ent\ho {\exists x\locs . \hvl{}{x\locs} \mand x\locs\mapsto\something}
{[[\lcon]]:= 0}
{\exists x\locs . \hvl{}{x\locs}\mand x\locs\mapsto 0}
\]
by \rn{L-Act} since both locations would be owned by the process.  The
rule \rn{L-Seq} can now be applied to obtain
\[ \Gamma\ent\ho{\hvl{}{\something}}
{\alloc\lcon; [[\lcon]]:= 0}
{\exists x\locs.\hvl{}{x\locs} \mand x\locs \mapsto 0},
\]
indicating that the process has ownership of the location $\lcon'$,
seen in the ability to write to $\lcon'$, once it has been allocated.

To allow the logic to make judgements beyond those applicable to the
almost `disjointly concurrent' programs outlined so far, further
interaction is allowed through a system of \emph{invariants}. The
judgement environment $\Gamma$ records a formula called an invariant
for each resource in its domain, which contains all the resources
occurring in the term.  The intuition is that, whenever a resource
$\res$ with an invariant $\chi$ is available, there is part of the
heap unowned by any other process and protected by the resource that
satisfies $\chi$.  In such a situation, we shall say that the
locations used to satisfy $\chi$ are `owned' by the invariant for
$\res$.
Processes may gain ownership of these locations, and thereby the right
to access them, by entering a critical region protected by the
resource.  When the process leaves the critical region, the invariant
must be restored and the ownership of the locations used to satisfy
the invariant is relinquished.  This is reflected in the rule
$\rn{L-CR}$.  As an example, we have the following derivation:

\begin{prooftree}
\AxiomC{\rule{0pt}{6pt}}
\LeftLabel{\srn{L-Act}}
\UnaryInfC{$ \res\ty \hvl{}0 \ent
\ho{\hvl{'}{\something} \mand \hvl{}0}
{[\lcon']:= [\lcon]}{\hvl{'}0\mand \hvl{}0}$}
\LeftLabel{\srn{L-CR}}
\UnaryInfC{$\res\ty \hvl{}0 \ent
\ho{\hvl{'}{\something}}
{\with \res\ldo [\lcon']:=[\lcon]\lod}
{\hvl{'}0}$}
\end{prooftree}

The process initially owns the location $\lcon'$, and the location
$\lcon$ is protected by the resource $\res$.  We reason about the
process inside the critical region running from a state with ownership
of the locations governed by the invariant in addition to those that
it owned before entering the critical region since no other process
can be operating on them; that is, we reason about $[\lcon']:=[\lcon]$
with locations $\lcon$ and $\lcon'$ owned by the process.  However,
when the process leaves the critical region, ownership of the
locations used to satisfy the invariant is lost, indicated by the
conclusion $\hvl{'}0$ in the judgement rather than
$\hvl{'}{0}\mand \hvl{}0$.

An invariant is required to be a \emph{precise} heap logic formula.
\begin{defi}[Precision]
  A heap logic formula $\chi$ is \emph{precise} if for any heap $D$
  there is at most one subheap $D_0\subseteq D$ such that
  $D_0\models\chi$.
\end{defi}
We leave discussion of the r\^ole of precision to the conclusion,
though it might be seen to be of use since it identifies uniquely the
part of the heap 
that is owned by the invariant if the resource is available.
 Formally, $\Gamma$ ranges over finite
partial functions from resources to precise heap \formulae{}.  We
write $\dom(\Gamma)$ for the set of resources on which $\Gamma$ is
defined, and write $\Gamma,\Gamma'$ for the union of the two partial
functions, defined only if $\dom(\Gamma)\cap\dom(\Gamma')=\emptyset$.
We write $\res\ty \chi$ for the singleton environment taking resource
$\res$ to $\chi$, and we allow ourselves to write $\res\ty\chi \in
\Gamma$ if $\Gamma(\res)=\chi$.


The rules allow ownership of locations to be transferred through
invariants.  Consider the invariant $\chi$ defined as $\hvl{'}{0} \lor
(\hvl{'}1\mand \hvl{}{0})$.  If the resource is available, the
invariant is satisfied: it either protects the location $\lcon'$,
which has value $0$, or it protects location $\lcon'$, which has
value $1$, as well as location $\lcon$.  A process can acquire ownership of
$\lcon$ across a critical region by changing the value of $\lcon'$
from $1$ to $0$ and may leave ownership of $\lcon$ inside the
invariant by changing the value of $\lcon'$ from $0$ to $1$.

Assume, for example, that the process owns location $\lcon$.  The only
way in which the invariant $\chi$ can be satisfied disjointly from
the locations that the process owns is for $\lcon'$ to hold value $0$.
That is, we have
\[ \hvl{} 0\mand(\hvl{'}0 \lor (\hvl{'} 1 \mand \hvl{}0))  \implies
\hvl{}0\mand \hvl{'}0\] which is implicitly used in the instance of the rule
\rn{L-Consequence} below.  Consequently, as the process enters a
critical region protected by $\res$, it gains ownership of location
$\lcon'$.  If the process sets the value of $\lcon'$ to $1$, when the
process leaves the critical region it must restore the invariant to
the resource, and so relinquish ownership of both $\lcon'$ and
$\lcon$.  This is seen in the derivation of the following judgement,
in which we take $\Gamma= \res\ty \chi$.

\begin{prooftree}
\AxiomC{\rule{0pt}{6pt}}
\LeftLabel{\srn{L-Act}}
\UnaryInfC{$\Gamma\ent \ho{\lcon\mapsto 0\mand \lcon'\mapsto 0}
  {[\lcon']:= 1}{\lcon\mapsto 0\mand \lcon'\mapsto 1}$}
\LeftLabel{\srn{L-Consequence}}
\UnaryInfC{$\Gamma\ent\ho{\lcon\mapsto 0\mand \chi}{[\lcon']:= 1}
{\mathsf{empty}\mand\chi}$}
\LeftLabel{\srn{L-CR}}
\UnaryInfC{$\Gamma\ent\ho{\lcon\mapsto 0}{\with\res\ldo [\lcon']:=1\lod}{\mathsf{empty}}$}
\end{prooftree}
With this derivation, we can derive
\[
\Gamma\ent\ho{\lcon\mapsto 2}{[\lcon]:= 0;\with\res\ldo
[\lcon']:=1}{\mathsf{empty}}.
\]
It is also possible to acquire ownership of locations through an
invariant.  Let the action $\co{diverge}$ have the same semantics as
that of the boolean guard $\co{false}$, which is an action that can
never occur \ie{} the process is stuck.  We have the following
derivation:
\begin{prooftree}
\AxiomC{
\begin{tabular}{lr@{\}\ }c@{\ \{}l}
$\Gamma\ent$&$\{\chi$&${[\lcon']=0}$&$\lcon'\mapsto 0\}$\\
$\Gamma\ent$&$\{\lcon'\mapsto 0$ &${\co{diverge}}$ &$\lcon'\mapsto 0\mand \lcon\mapsto 0\}$\\
$\Gamma\ent$&$\{\chi$&${[\lcon']=1}$&${\lcon'\mapsto 1\mand \lcon\mapsto0\}}$\\
$\Gamma\ent$&$\{\lcon'\mapsto 1\mand \lcon\mapsto0$&${[\lcon']:=0}$
&${\lcon'\mapsto 0\mand\lcon\mapsto 0}\}$
\end{tabular}}
\LeftLabel{\srn{L-Sum}}
\UnaryInfC{$
\Gamma\ent\ho{\chi}
{([\lcon']=0.\co{diverge}) + ([\lcon']=1.[\lcon']:=0)}
{\lcon\mapsto0\mand\lcon'\mapsto 0}
$}
\LeftLabel{\srn{L-Consequence}}
\UnaryInfC{$
\Gamma\ent\ho{\mathsf{empty}\mand\chi}
{([\lcon']=0.\co{diverge}) + ([\lcon']=1.[\lcon']:=0)}
{\lcon\mapsto0\mand\chi}
$}
\LeftLabel{\srn{L-CR}}
\UnaryInfC{$
\Gamma\ent\ho{\mathsf{empty}}
{\with\res\ldo 
([\lcon']=0.\co{diverge}) + ([\lcon']=1.[\lcon']:=0)
\lod}
{\lcon\mapsto0}
$}
\end{prooftree}\vspace{0.5em}
The undischarged hypotheses at the top of the derivation are all
proved by the rule $\rn{L-Act}$.  Let $t_0$ denote the process
$([\lcon']=0.\co{diverge}) + ([\lcon']=1.[\lcon']:= 0)$.  Observe that
the process $\with\res\ldo t_0\lod$ is considered to own no part of
the initial heap.  As the process enters the critical region, it is
considered to take ownership of the part of the heap satisfying the
invariant for $\res$, \viz{} $\chi$.  There are two ways in which
$\chi$ might be satisfied:
\begin{enumerate}
\item It may be that the process gains ownership of the location
$\lcon'$ which holds value $0$.  In this case, only the guard
$[\lcon']=0$ of $t_0$ can pass, so the process must evolve to
$\co{diverge}$ and therefore never terminates.  It is therefore
trivially true that the remainder of the derivation, that if the
process $t_0$ terminates then the part of the heap that it owns
satisfies $\lcon\mapsto 0\mand \chi$ and therefore after leaving the
critical region and losing ownership of the locations satisfying
$\chi$ that the process owns location $\lcon$, is sound.

\item The process might have taken control of the locations
$\lcon$, holding value $0$, and $\lcon'$, holding value $1$.  Inside
the critical region, the process $t_0$ can be seen to change the value
of $\lcon'$ from $1$ to $0$.  The only way that the invariant $\chi$
can then be satisfied is by the location $\lcon'$ holding $0$, so
ownership of $\lcon'$ is lost as the process leaves the critical
region.  Importantly, the process retains ownership of location
$\lcon$.
\end{enumerate}
Using the derivations given above, we can give an example of ownership
of $\lcon$, as exhibited by the right to write to $\lcon$, being
transferred (we have annotated internal assertions arising from the
proofs above inside the program): \newcommand{\lco}[1]{{\color[gray]{0.4}
    #1}} \newcommand{\assert}[1]{\lco{\{#1\}}}
\[\lco{\Gamma\ent}\qquad\left.
\begin{array}{c}
\assert{\lcon\mapsto 2}\\
\begin{array}{l||l}
\assert{\lcon\mapsto 2} & \assert{\mathsf{empty}}\\
{[\lcon]}:= 0; & \with\res\ldo\\
\assert{\lcon\mapsto 0} &  \quad \quad [\lcon']=0.\ \co{diverge}\\
\with\res\ldo &  \quad +\, [\lcon']=1.\ [\lcon']:= 0\\
\quad [\lcon']:= 1 &\!\!\lod;\\
\!\!\lod &\assert{\lcon\mapsto 0}\\
\assert{\mathsf{empty}} & [\lcon]:= 1\\
\phantom{\quad\quad[\lcon']=0.\ \co{diverge}}& \assert{\lcon\mapsto 1}
\end{array}\\
\assert{\lcon\mapsto 1}
\end{array}\right.
\]
We also see that, in any terminating run of this process, it must be
the case that the process on the left terminates strictly before the
process on the right begins.

The final remark to be made on the rules of the logic is that
$\rn{L-Res}$ allows invariants to be established for newly declared
resources.  We reason about the closed term $[\res/\rvar]t$,
for an arbitrary `fresh' resource $\res$; it is sufficient to consider
only one such resource, as shall be seen in Lemma \ref{lemma:semequivar}.
The resource $\res$ is known not to occur in the domain of $\Gamma$ and
hence does not occur in the term $t$ thanks to the following lemma,
proved straightforwardly by induction on the judgement.
\begin{lem}
\label{lemma:freeres}
If $\Gamma\ent\hos{}$ then $\fr t\subseteq \dom(\Gamma)$.
\qed
\end{lem}


\subsection{Ownership model}
\label{sec:formalownership}

We now progress to give a formal interpretation of the rules presented
in the previous section.  The key idea is that the judgement
$\Gamma\ent\hos{}$ is robust against the operation of other `external'
processes (which have themselves been subject to a judgement in the
logic) on the state, so that the rule for parallel composition is
valid.  From the account presented earlier, external processes may act
on the heap providing they do not access the locations `owned' by the
process $t$, and they may act to acquire and release resources
providing they respect the invariants in $\Gamma$.  External processes
may also make non-current resources current through the instantiation
of a resource variable and might make such resources non-current.  The
semantics of judgements must therefore keep a record of how each
current location in the heap and each current resource is owned:
whether the process might access the location, whether it forms part
of an invariant protected by a resource, or whether external processes
might act on that location, along with a similar record for resources.
The semantics will include \emph{interference events} to represent
such forms of action by external processes.

Capturing these requirements, we construct an \emph{interference net}
with respect to the environment $\Gamma$ to represent the execution of
suitable external processes proved against $\Gamma$.  This involves
creating \emph{ownership conditions} $\wpr\lcon$, $\win\lcon$ and
$\wot\lcon$ for each location $\lcon$.  The intuition is that
$\wpr\lcon$ is marked if $\lcon$ is owned by the process, $\win\lcon$
if $\lcon$ is used to satisfy the invariant for an available open
resource, and $\wot\lcon$ is marked if $\lcon$ is current but owned by another
process.

To give an example, suppose that we have the judgement
\[ \Gamma\ent \ho{k\mapsto 1}{[k]:= 0}{k\mapsto 0}. \]
The proof can be composed with the judgement
$\Gamma\ent\ho{\lcon\mapsto 0}{[\lcon]:=0}{\lcon\mapsto 1}$ to obtain
\[ \Gamma\ent\ho{k\mapsto 1\mand \lcon\mapsto 0}
{[k]:= 0\pll [\lcon]:= 1}{k\mapsto 0\mand \lcon\mapsto 1}. \] The
first proof, that the assignment $[k]:=0$ changes the value at $k$
from $1$ to $0$, must take into account the possibility that the
values held at other locations may change.  In particular, it must
take into account the possibility that the value at $\lcon$ (not to
equal $k$) changes from $0$ to $1$.  We therefore reason about the net
$\nsem{[k]:=0}$ in the presence of the following interference event,
which changes the value held at $\lcon$ from $0$ to $1$:\medskip

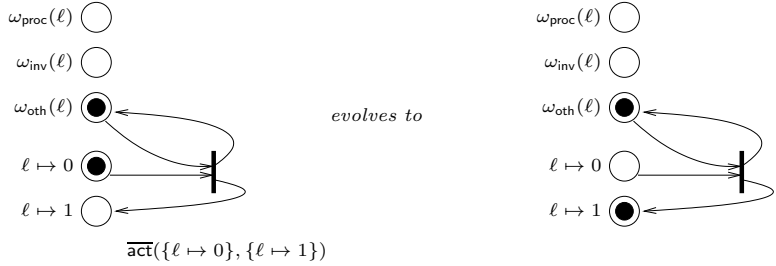
\begin{center}
\input{interfexl.pstex_t}
\end{center}
Notably, the above event requires that the location $\lcon$ is owned
by an external process, \ie{} the condition $\wot\lcon$ is marked.  

Since we do not know with which other judgements
$\Gamma\ent\ho{k\mapsto 1}{[k]:= 0}{k\mapsto 0}$ may be composed,
there are interference events present in the net for all the forms of
interference permissible according to the notion of ownership.  For
instance, the interference event which changes the value of $k$ from
$0$ to $1$\medskip

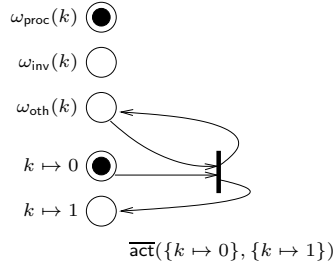
\begin{center}
\input{interfexk.pstex_t}
\end{center}
is present in the net.  However, the judgement asserts that $k$ is
owned by the process, so this interference event (and indeed any other
interference event that affects $k$) will not be able to occur because
the condition $\wpr k$ will be marked, not $\wot k$.

As mentioned above, we introduce interference events to mimic the
action of external processes on resources.  The notion of ownership is
therefore extended in this setting to resources, for example so that
an external process cannot be allowed to release a resource held by
the current process.  It is important to make a distinction between
resources in the domain of the environment $\Gamma$ (called
\emph{open} resources) and those that are not (called \emph{closed}
resources): Open resources have invariants associated with them, so
the ownership of the heap is affected by events that acquire or
release them, as presented earlier in this section; this is not the
case for closed resources.  Closed resources are those resources made
current to instantiate a local resource variable.  They may either be
used by the process being considered if it declared the resource, or
be used by some external process if some external process declared the
resource.  We shall introduce conditions $\wpr\res$, $\win\res$ and
$\wot\res$ for each resource $\res$.  The condition $\wpr\res$ will be
marked if either the resource is closed and was made current by the
process or if the resource is open and is held by the process.  The
condition $\win\res$ will be marked if $\res$ is open and available.
The condition $\wot\res$ will be marked if either the resource is
closed and was made current by an external process or if the resource
is both open and the external process holds it.

The set of ownership conditions is denoted $\cset{W}$:
\begin{eqnarray*}
  \cset{W} &\eqd& \{\wpr\lcon,\win\lcon,\wot\lcon \st \lcon\in\sn{Loc}\}\\
  &&\cup \{\wpr\res,\win\res,\wot\res \st \res\in\sn{Res}\}.
\end{eqnarray*}
We use $W$ to range over markings of ownership conditions and
introduce the notations $\pres W e$ and $\poss W e$, as before, for
the sets of pre-ownership conditions of $e$ and post-ownership
conditions of $e$, respectively.  For a set of locations $L$, we
define the notation
\[
\wpr L \eqd \{\wpr\lcon \st \lcon \in L\},
\]
and define $\win L$ and $\wot L$ similarly.  Only certain markings of
ownership conditions are consistent with a state $\sigma$:
\begin{defi}[Consistent marking]
\label{def:consown}
The marking of state and ownership conditions $(\sigma,W)$ of
$\osem{t}{\Gamma}$ is \emph{consistent} if:
\begin{enumerate}[(1)]
\item $\sigma$ is a consistent state in $\nsem{t}$,
\item for each $z\in \sn{Loc}\cup \sn{Res}$, at most one of $\{\wpr
  z,\win z, \wot z\}$ is marked,
\item for each $z\in \sn{Loc}\cup \sn{Res}$, the ownership condition
  $\curr(z)$ is in $\sigma$ iff precisely one of $\{\wpr z,\win z,\wot
  z\}$ is in $W$,
\item if $\res\in\dom(\Gamma)$ and $\res\in R$ then $\win\res\in R$,
\item if $\res\in\dom(\Gamma)$ and $\res\not\in R$ then either
  $\wpr\res\in W$ or $\wot\res\in W$, and
\item if $\curr(\res)\in\sigma$ and $\res\not\in\dom(\Gamma)$ then
  either $\wpr\res\in W$ or $\wot\res\in W$.
\end{enumerate}
\end{defi}
Requirements (2) and (3) assert that $W$ is essentially a function
from the set of current locations and resources to describe their
ownership.  Requirement (4) states that any available open resource is
owned as an invariant: it can be accessed either by the process being
considered or by an external process, and there is an invariant
associated with $\res$.  Requirement (5) states that any unavailable
open resource is either held by the process or by an external process.
Requirement (6) asserts that any closed resource is owned either by
the current process or by an external process.

Table \ref{table:ievents} defines a number of notations for events
corresponding to the permitted interference described.  To summarize,
there will be interference events to represent the following kinds of
action by external processes:
\begin{enumerate}[$\bullet$]
\item $\iact{D_1,D_2}$: 
  Arbitrary action on the heap (excluding allocation or deallocation)
  owned by external processes.
\item $\ialloc{\lcon,v,\lcon',v'}$: Allocation of a new location
  $\lcon'$ by an external process, storing the result in the location
  $\lcon$.  The location $\lcon$ must initially have been owned by an
  external process.  Ownership of the new location $\lcon'$ is taken
  by the external process.
\item $\idealloc{\lcon,v,\lcon',v'}$: Disposal of the location
  $\lcon'$ pointed to by $\lcon$.  Both locations are initially
  owned by external processes, so $\wot\lcon$ and $\wot{\lcon'}$ are
  preconditions to the event.
\item $\inew{\res}$: Declaration of a resource $\res$.  The condition
  $\curr(\res)$ is marked by the event, so the resource was not initially current.
  Ownership of $\res$ is taken by the external process, so $\wot\res$
  is in the postconditions of the event.
\item $\iend{\res}$: End of scope of a resource $\res$, only
  permissible if the resource was initially declared by an external
  process and therefore $\wot\res$ is marked.
\item $\iget{\res}$: For a closed resource $\res$, the external
  process may acquire the resource if it is not local to the process
  being considered and therefore $\wot\res$ is marked.
\item $\iput{\res}$: For a closed resource $\res$, the external
  process may release the resource if it is not local to the process
  being considered and therefore $\wot\res$ is marked.
\item $\iget{\res,D_0}$: For an open resource $\res$ with an invariant
  $\chi$ in $\Gamma$, if $D_0\models\chi$ and $D_0$ is part of the
  current heap then ownership of the locations in the domain of $D_0$ is
  changed from being protected by the resource to being owned by the external
  process, \ie{} un-marking $\win\lcon$ and marking $\wot\lcon$ for
  each location $\lcon\in\dom(D_0)$.  The ownership of $\res$ also
  changes, from $\win\res$ being marked to $\wot\res$ being marked.
\item $\iput{\res,D_0}$: The corresponding release action.
\end{enumerate}

\begin{defi}[Interference net]
  The interference net for $\Gamma$ has conditions $\cset{S}$, the
  state conditions, and $\cset W$, the ownership conditions.  It has
  the following events:
\begin{enumerate}[$\bullet$]
\item $\iact{D_1,D_2}$ for all $D_1$ and $D_2$ forming partial
  functions with the same domain
\item $\ialloc{\lcon,v,\lcon',v'}$ and $\idealloc{\lcon,\lcon',v'}$ for
  all locations $\lcon$ and $\lcon'$ and values $v$ and $v'$
\item $\inew{\res}$ and $\iend{\res}$ for all resources $\res$
\item $\iget{\res}$ and $\iput{\res}$ for all closed resources $\res$
\item $\iget{\res,D_0}$ and $\iput{\res,D_0}$ for all
  $\res\in\dom(\Gamma)$ and $D_0$ such that $D_0\models\chi$, for
  $\chi$ the unique formula such that $\res\ty\chi\in \Gamma$
\end{enumerate}
We use the symbol $u$ to range over interference events.
\end{defi}

\begin{table*}
{\small
\[\begin{array}{|l||c|c||c|c|}\hline
  \textit{Abbreviation} & \multicolumn{2}{c||}{\textit{Preconditions}} & \multicolumn{2}{c|}{\textit{Postconditions}}\\
  \multicolumn{1}{|c||}{u}  
  & \multicolumn{1}{c}{\pres S u} 
  & \multicolumn{1}{c||}{\pres W u} 
  & \multicolumn{1}{c}{\poss S u} 
  & \multicolumn{1}{c|}{\poss W u} 
\\\hline\hline&&&&\\
    \iact{D_1,D_2} 
    & D_1
    & \wot{\dom(D_1)}
    &D_2 &\wot{\dom(D_2)}
\\&&&&\\
    \ialloc{\lcon,v,\lcon',v'}
    & \{\hvl{}{v}\} & \{\wot{\lcon}\}
    & \{\curr(\lcon')\}\cup
    & \{\wot{\lcon},\wot{\lcon'}\}
\\
&&&\{\hvl{}{\lcon'}, \hvl{'}{v'}\}&
\\&&&&\\
    \idealloc{\lcon,\lcon',v'}
    & \{\curr(\lcon')\}\cup
    & \{\wot{\lcon},\wot{\lcon'}\}
    & \{\hvl{}{\lcon'}\}
    & \{\wot{\lcon}\}
\\
&\{\hvl{}{\lcon'},\hvl{'}{v'}\}&&&
\\&&&&\\
    \inew{\res}
    & \{\}
    & \{\}
    &\{\curr(\res),\res\}
    &\{\wot{\res}\}
\\&&&&\\
    \iend{\res}
    &\{\curr(\res),\res\}
    &\{\wot{\res}\}
    &\{\} & \{\}
\\&&&&\\
    \iget{\res}
    & \{\res\}
    &\{\wot{\res}\}
    &\{\}
    &\{\wot{\res}\}
\\&&&&\\
    \iput{\res}
    &\{\}
    &\{\wot{\res}\}
    &\{\res\}
    &\{\wot{\res}\}
\\&&&&\\
    \iget{\res,D_0}
    & D_0 \cup \{\res\}& \win{\dom(D_0)}\cup
    & D_0 & \wot{\dom(D_0)}\cup
\\
&&\{\win{\res}\}&& \{\wot{\res}\} 
\\&&&&\\
    \iput{\res,D_0}
    & D_0
    & \{\wot\res\}\cup
    & D_0 \cup \{\res\}
    & \{\win\res\}\cup 
\\
&& \wot{\dom(D_0)}&& \win{\dom(D_0)}
\\\hline\end{array}\]}
\vspace{-10pt}
\caption{Interference events}
\label{table:ievents}
\end{table*}

The interference events illustrate how the ownership of locations is
\emph{dynamic} and how this constrains the possible forms of
interference.  The rule for parallel composition requires that the
behaviour of the process being reasoned about itself conforms to these
constraints, allowing its action to be seen as interference when
reasoning about the other process.  This requirement may be captured
by \emph{synchronizing} the events of the process with those from the
interference net in the following way:
\begin{enumerate}[$\bullet$]
\item The process event $\eact{C,C'}(D,D')$ synchronizes with
  $\iact{D,D'}$
\item The process event $\eall{C,C'}(\lcon,v,\lcon',v')$ synchronizes
  with $\ialloc{\lcon,v,\lcon',v'}$
\item The process event $\edea{C,C'}(\lcon,\lcon',v')$ synchronizes
  with $\idealloc{\lcon,\lcon',v'}$ 
\item The process event $\enew{C,C'}(\res)$ synchronizes with $\inew{\res}$
\item The process event $\eend{C,C'}(\res)$ synchronizes with $\iend{\res}$
\item The process event $\eget{C,C'}(\res)$ synchronizes with
  $\iget{\res}$ for any closed resource $\res$, \ie{} for any
  $\res\not\in\dom(\Gamma)$
\item The process event $\eput{C,C'}(\res)$ synchronizes with
  $\iput{\res}$ for any closed resource $\res$
\item If $\res$ is an open resource with $\res\ty\chi\in\Gamma$, the
  process event $\eget{C,C'}(\res)$ synchronizes with every
  $\iget{\res,D_0}$ such that $D_0\models\chi$.  Similarly,
  $\eput{C,C'}(\res)$ synchronizes with every $\iput{\res,D_0}$ such
  that $D_0\models\chi$.
\end{enumerate}
Suppose that two events synchronize, $e$ from the process and $u$ from
the interference net.  The event $u$ is the event that would fire in
the net for the other parallel process to simulate the event $e$; it
is its dual.  Let $\sync{e}{u}$ be the event formed by taking the
union of the pre- and postconditions of $e$ and $u$, other than using
$\wpr{\lcon}$ in place of $\wot{\lcon}$, and similarly $\wpr{\res}$
in place of $\wot{\res}$.  
\begin{eqnarray*}
\pre {(\sync e u)} &\eqd&
\{b\st b\in \pre e \cup \pre u
~\tand~ \not\exists z.b=\wot z\} \cup \{\wpr z \st \wot z \in \pre u\}\\
\post {(\sync e u)} &\eqd&
\{b\st b\in \post e \cup \post u
~\tand~ \not\exists z.b=\wot z\} \cup \{\wpr z \st \wot z \in \post u\}
\end{eqnarray*}

\begin{exa}[Synchronization of heap actions]
  Define the following events:\medskip

  \begin{center}
      \input{sync1.pstex_t}
  \end{center}
  The event $e$ is an event inside the process net, with pre-control
  conditions $C$ and post-control conditions $C'$, that changes the
  value of $\lcon$ from $0$ to $1$.  It synchronizes with only one
  event, $u$, which performs the corresponding interference action.
  For the event $u$ to occur, the condition $\wot\lcon$ must be marked
  \ie{} the location $\lcon$ must be seen as owned by an `external'
  process.  The event formed by synchronizing $e$ and $u$ is $\sync e
  u$, which requires the location $\lcon$ to be owned by the current
  process for it to occur.\qed
\end{exa}

\begin{exa}[Synchronization of critical regions]
  Define the following events, where 
  the event $e$ is an event inside the process net, with pre-control
  conditions $C$ and post-control conditions $C'$, that acquires the
  open resource $\res$.  

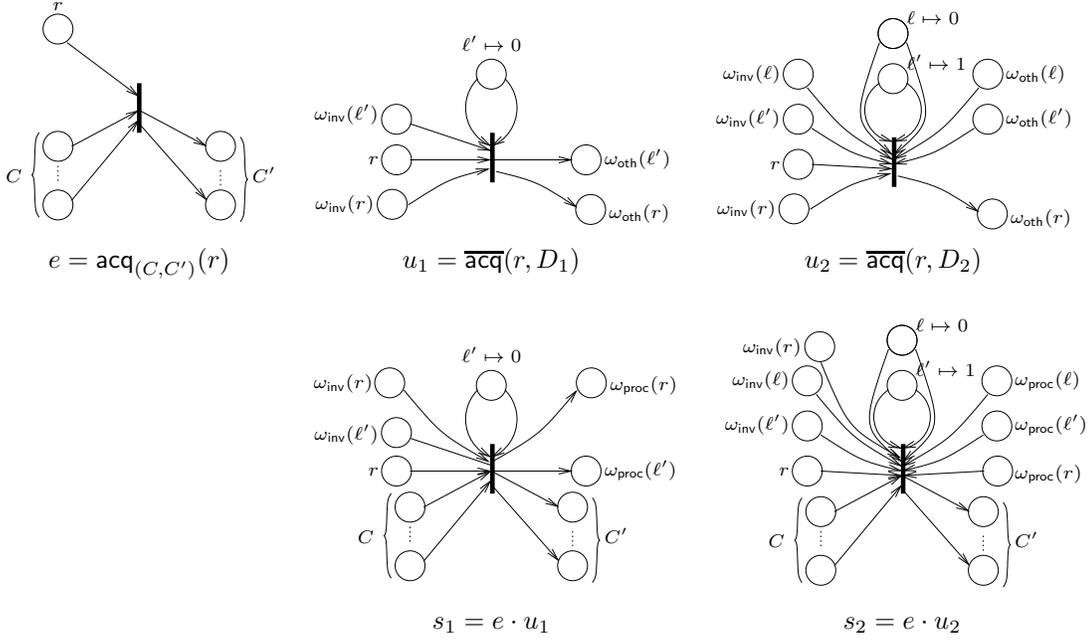
  \begin{center}
    \input{sync2.pstex_t}
  \end{center}

  Recall the invariant $\lcon'\mapsto 0 \lor (\lcon'\mapsto 1
  \mand \lcon\mapsto 0)$ used above.  There are two heaps,
  $D_1=\{\lcon'\mapsto 0\}$ and $D_2=\{\lcon'\mapsto 1,\lcon\mapsto
  0\}$ that satisfy this formula.  There are correspondingly two
  interference events $u_1$ and $u_2$ that synchronize with $e$: the
  event $u_1$ acquires the resource $\res$ and transfers the ownership
  of $\lcon'$ and $\res$ to the external process from the invariant,
  whereas the event $u_2$ acquires the resource $\res$ and transfers
  ownership of $\lcon$, $\lcon'$ and $\res$ to the external process  from the
  invariant.  The event $u_1$ requires that the heap initially has
  value $0$ at $\lcon'$; the event $u_2$ requires that the heap
  initially has value $1$ at $\lcon'$ and $0$ at $\lcon$.  The
  synchronized events $\sync{e}{u_1}$ and $\sync{e}{u_2}$ are similar,
  transferring ownership from the invariant to the process being
  considered.\qed
\end{exa}

The semantics of judgements made using the rules of concurrent
separation logic will consider a net $\osem t \Gamma$ with both
interference events to represent external processes running and
synchronized events to represent the process $t$.
\begin{defi}[Ownership net]
  The ownership net for $t$ in $\Gamma$, denoted $\osem{t}{\Gamma}$,
  is the net formed with the previous definitions of control
  conditions $\cset{C}$, state conditions $\cset{S}$ and ownership
  conditions $\cset{W}$, and events:
  \begin{titemize}
  \item Every event $u$ from the interference net for $\Gamma$, and
  \item Every event $\sync{e}{u}$ where $e$ is an event of $\nsem{t}$
    and $u$ from the interference net such that $e$ and $u$
    synchronize.
  \end{titemize}
\end{defi}
We shall continue to use the symbol $e$ to refer to any kind of event
in ownership nets, but shall reserve the symbol $s$ for those events
known in particular to be synchronized events.

A consequence of the precision of invariants is that at most one of
the synchronized events corresponding to an event in $\nsem{t}$ may be
enabled in any marking of the ownership net $\osem{t}\Gamma$.
\begin{lem}
\label{lemma:precision}
  For any marking $\sigma$ of state conditions, let $(C,\sigma,W)$ and
  $(C',\sigma,W')$ be consistent markings of the net $\osem{t}\Gamma$.
  For any event $e$ in $\nsem{t}$ and any interference events $u$ and
  $u'$ in $\osem{t}\Gamma$, if $\sync e u$ has concession in
  $(C,\sigma,W)$ and $\sync e {u'}$ has concession in $(C',\sigma,W')$
  then $u=u'$.
\begin{proof}
  Straightforwardly seen to follow from precision by an analysis of
  the possible forms of the event $e$.
\end{proof}
\end{lem}

The occurrence of a synchronized event $\sync e u$ in a marking
$(C,\sigma,W)$ of the net $\osem{t}\Gamma$ clearly gives rise to the
occurrence of the event $e$ in $\nsem{t}$.  The earlier results
describing the behaviour of $\nsem{t}$ in terms of the behaviour of
the nets representing its subterms can therefore be applied to the net
$\osem{t}\Gamma$.
\begin{lem}
\label{lemma:charev}
  If $M=(C,\sigma,W)$ and $M'=(C',\sigma',W')$ are markings of
  $\osem{t}{\Gamma}$ and $M\ltr{e}M'$ then either $e$ is an
  interference event and $C=C'$  or $e=\sync{e_1}{u}$ for an event $e_1$ of
  $\nsem{t}$ and an interference event $u$ and
  $(C,\sigma)\ltr{e_1}(C',\sigma')$ in $\nsem{t}$.
\begin{proof} 
  The events of $\osem{t}{\Gamma}$ are, by definition, only
  interference events or synchronized events.  If $e$ is an
  interference event, $C=C'$ because $\pres C e = \emptyset$ and
  $\poss C e = \emptyset$. For a synchronized event $\sync{e_1}{u}$,
  observe that $\pres C {(\sync{e_1}{u})}=\pres C {e_1}$ and that
  $\poss C {(\sync{e_1}{u})}=\poss C {e_1}$, and similarly for $\pres
  L {e_1}$, $\pres R {e_1}$, $\pres N {e_1}$, $\poss L {e_1}$, $\poss
  R {e_1}$ and $\poss N {e_1}$.  The only cases where either $\pres D
  {(\sync{e_1}{u})}\neq \pres D {e_1}$ or $\poss D {(\sync{e_1}{u})}
  \neq \poss D {e_1}$ are acquisition or release of an open resource,
  but in these cases $\pres D {e_1}=\emptyset=\poss D{e_1}$ and $\pres
  D {(\sync{e_1}{u})} = \poss D {(\sync{e_1}{u})}$.  The result
  follows as a straightforward calculation.
\end{proof}
\end{lem}

The proof that consistent markings are preserved in the net
$\osem{t}\Gamma$ is similar to that of Lemma \ref{lemma:conspres}; the
additional requirements on the marking of ownership conditions are
readily seen to be preserved by both interference and synchronized
events.
\begin{lem}[Preservation of consistent markings]
\label{lemma:oconspres}
For any closed term $t$, if in the net $\osem{t}\Gamma$ it is the case
that $(\ic{t},\sigma_0,W_0)\ltr{}^*(C,\sigma,W)$ and $(\sigma_0,W_0)$
is consistent then $(\sigma,W)$ is consistent.  \qed\end{lem}

The formulation of the ownership net permits a fundamental
understanding of when a process acts in a way that cannot be seen as
any form of interference; that is, when the process has violated its
guarantees.
\begin{defi}[Violating marking]
  Let $(C,\sigma,W)$ be a consistent marking of
  $\osem{t}{\Gamma}$.  We say that $M$ is \emph{violating} if
  there exists an event $e$ of $\nsem{t}$ that has concession in
  marking $(C,\sigma)$ but there is no event $u$ from the
  interference net that synchronizes with $e$ such that $\sync{e}{u}$
  has concession in $(C,\sigma,W)$.
\end{defi}

We shall give two examples of violating markings.  The first shall be
an example of action on an unowned location, and the second shows how
release of an open resource will cause a violation if the invariant is
not restored.
\begin{exa}
  Let $(\{\del\},\sigma,W)$ be a consistent marking of
  $\osem{[\lcon]:=1}\Gamma$ with $\hvl{} 0\in \sigma$ and
  $\wot{\lcon}\in W$. The event
  $e=\eact{\{\del\},\{\cel\}}(\{\hvl{}{0}\},\{\hvl{}1\})$ has
  concession in $(C,\sigma)$, but the only interference event that can
  synchronize with $e$ is $u=\iact{\{\hvl{}{0}\},\{\hvl{}1\}}$.  We have
  $\wot\lcon\in\pres W u$ and therefore $\wpr\lcon\in\pres W {(\sync e
    u)}$, so the event $\sync e u$ does not have concession in the
  marking $(C,\sigma,W)$ which is therefore violating: the process
  acted on the unowned location $\lcon'$.\qed
\end{exa}
\begin{exa}
  Let $\res$ be an open resource with the invariant $\chi=\hvl{'}0\lor
  (\hvl{'}1\mand \hvl{}0)$, and let $(C,\sigma,W)$ be a consistent
  marking of $\osem{t}{\Gamma,\res\ty\chi}$ with $\{\hvl{}1,\hvl{'}1\}\subseteq
  \sigma$ and $\wpr\lcon,\wpr{\lcon'}\in W$.  Suppose further that
  the event $e=\eput{C_1,C_2}(\res)$ has concession in $(C,\sigma)$ in
  the net $\nsem{t}$.  The only two interference events in
  $\osem{t}{\Gamma,\res\ty\chi}$ that synchronize with $e$ are
  \begin{eqnarray*}
    u_1 &=& \iput{\res,\{\hvl{'}0\}}\\
    u_2 &=& \iput{\res,\{\hvl{'}{1},\hvl{}0\}},
  \end{eqnarray*}
  corresponding to the two ways in which $\chi$ can be satisfied.  The
  invariant is not satisfied in the heap component of $\sigma$, so the
  preconditions of the two events
  \begin{eqnarray*}
    \pre{(\sync e {u_1})} &=& C \cup \{\hvl{'}0, \wpr{\lcon'} \}\\
    \pre{(\sync e {u_2})} &=& C \cup \{\hvl{'}1,\hvl{}0, \wpr{\lcon'},\wpr{\lcon} \}
  \end{eqnarray*}
  are not contained in the marking $(C,\sigma,W)$, which is therefore
  therefore a violating marking because there was no part of the owned
  heap that satisfied the invariant yet the resource was released.\qed
\end{exa}

If no violating marking is ever encountered, the behaviour of
$\osem{t}{\Gamma}$ encapsulates all that of $\nsem{t}$.
\begin{lem}
\label{lemma:connecsim}
  For any consistent marking $(C,\sigma,W)$ of the net
  $\osem{t}\Gamma$ and any event $e\in\ev{t}$, if $(C,\sigma)\ltr e
  (C',\sigma')$ in $\nsem{t}$ then either $(C,\sigma,W)$ is violating
  or there exists a marking of ownership conditions $W'$ and an
  interference event $u$ that synchronizes with $e$ such that
  $(C,\sigma,W)\ltr{\sync e u} (C',\sigma',W')$ in $\osem{t}\Gamma$.
  \begin{proof}
    Immediate from the definition of violating marking and the fact
    that, for any $e$ and $u$ that synchronize and any state $\sigma$
    \[
    \pres C {(\sync e u)} = \pres C e \qquad \poss C {(\sync e u)} =
    \poss C e \qquad \sigma\setminus \pres S {(\sync e u)}\cup \poss S
    {(\sync e u)} = \sigma\setminus \pres S {e}\cup \poss S { e }
    \]
    which is easily proved by inspection of the forms that $\sync e u$
    may take.
  \end{proof}  
\end{lem}

\subsection{Soundness and validity}

The rule for parallel composition permits the view that the ownership
of the heap is initially split between the two processes, so that what
one process owns is seen as owned by an external process by the other.
\begin{defi}[Ownership split]
\label{def:split}
Let $W$ be a marking of ownership conditions.  Markings of ownership
conditions $W_1$ and $W_2$ form an
\emph{ownership split} of $W$ if for all $z\in\sn{Loc}\cup\sn{Res}$:
\begin{enumerate}[$\bullet$]
\item $\wot z\in W$ iff  $\wot z\in W_1$ and $\wot z\in
    W_2$,
\item $\win z\in W$ iff  $\win z\in W_1$ and $\win z\in W_2$, and
\item $\wpr z\in W$ iff either $\wpr z\in W_1$ and $\wot z\in
    W_2$,\\\phantom{$\wpr z\in W$ iff }or  $\wpr z\in W_2$ and $\wot z\in W_1$.
\end{enumerate}
\end{defi}
If $W_1$ and $W_2$ form an ownership split of $W$, then fewer
locations and resources are owned by the process in $W_1$ than in $W$,
and similarly for $W_2$.  As one would expect, a process can act in
the same way without causing a violation if it owns more, and more
interference can occur if the process owns less.  This is the essence
of the frame property referred to earlier.
\begin{lem}
\label{lemma:framesim}
Consider markings of the net $\osem{t}\Gamma$.  Let $W_1$ and $W_2$
form an ownership split of $W$.
  \begin{enumerate}[$\bullet$]
  \item For any synchronized event $s=\sync e u$, if
    $(C,\sigma,W_1)\ltr{s} (C',\sigma',W_1')$ then there exist $W'$
    and $W_2'$ such that $(C,\sigma,W)\ltr{s} (C',\sigma',W')$ and
    $(C,\sigma,W_2)\ltr u (C,\sigma',W_2')$, and furthermore $W_1'$
    and $W_2'$ form an ownership split of $W'$.
  \item For any interference event $u$, if $(C,\sigma,W)\ltr{u}
    (C,\sigma',W')$ then there exist $W_1'$ and $W_2'$ such that
    $(C,\sigma,W_1)\ltr{u}(C,\sigma,W_1')$ and $(C,\sigma,W_2)\ltr{u}
    (C,\sigma',W_2')$, and furthermore $W_1'$ and $W_2'$ form an
    ownership split of $W'$.
  \end{enumerate}
\begin{proof}
  A straightforward (but long) analysis of the possible forms of $s$ and $u$.
\end{proof}
\end{lem}

Following Brookes' lead, we are now able to prove the key lemma upon which
the proof of soundness lies.  The effect of this lemma is that the the
terminal states of parallel processes may be determined simply by
observing the terminal markings of the net of each parallel process running in isolation if we split the
ownership of the initial state correctly.  For convenience, the lemma
is stated without intimating the particular event that takes place on
the net transition relation.
\begin{lem}[Parallel decomposition]
\label{lemma:pardecomp}
  Let $M=(\pref 1{C_1}\cup \pref 2{C_2},\sigma,W)$ be a consistent
  marking of the net $\osem{t_1\pll t_2}{\Gamma}$, and let $W_1$
  and $W_2$ form an ownership split of $W$.  The markings
  $M_1=(C_1,\sigma,W_1)$ and $M_2=(C_2,\sigma,W_2)$ are consistent, and
  furthermore:
  \begin{enumerate}[$\bullet$]
  \item If the marking $M$ is violating in $\osem{t_1\pll
      t_2}\Gamma$ then either $M_1$ is violating in
    $\osem{t_1}\Gamma$ or $M_2$ is violating in
    $\osem{t_2}\Gamma$.
  \item If neither $M_1$ nor $M_2$ is violating and $(\pref
    1{C_1}\cup\pref 2{C_2},\sigma,W)\ltr{}(\pref 1{C_1'}\cup\pref
    2{C_2'},\sigma',W')$ in $\osem{t_1\pll t_2}{\Gamma}$ then
    there exist $W_1'$ and $W_2'$ forming an ownership split of $W'$
    such that $(C_1,\sigma,W_1)\ltr{}(C_1',\sigma',W_1')$ in
    $\osem{t_1}{\Gamma}$ and
    $(C_2,\sigma,W_2)\ltr{}(C_2',\sigma',W_2')$ in
    $\osem{t_2}{\Gamma}$.
  \end{enumerate}

\proof
  \newcommand{\pro}[1]{\pref{1}{#1}}
  \newcommand{\prt}[1]{\pref{2}{#1}}
  
  It is straightforward from Definition \ref{def:consown} to see that
  $M_i$ is a consistent marking for both $i\in\{1,2\}$.

  \begin{enumerate}[(1)]
  \item Suppose that the marking $M$ is violating in $\osem{t_1\pll
      t_2}{\Gamma}$.  Without loss of generality, assume that this 
    is because there exists an event $\pro{e_1}$ of $\nsem{t_1\pll
    t_2}$ that has concession in marking $(\pro{C_1} \cup
    \prt{C_2},\sigma)$ but there is no event interference event $u$
    such that $\pro{e_1}$ synchronizes with $u$ and
    $\sync{(\pro{e_1})}{u}$ has concession in $M$.  Assume, for
    contradiction, that the marking $M_1$ is non-violating in
    $\osem{t_1}\Gamma$. The event $e_1$ has concession in marking
    $(C_1,\sigma)$ of $\nsem{t_1}$ by the first part of Lemma
    \ref{lemma:gluesim}, so there must exist $u_1$ an interference
    event of $\osem{t_1}{\Gamma}$ such that $\sync{e_1}{u_1}$ has
    concession in $M_1$.  The interference events of
    $\osem{t_1}{\Gamma}$ are precisely the interference events of
    $\osem{t_1\pll t_2}{\Gamma}$ and the tagging of control conditions
    has no effect on whether events may synchronize, so the event
    $\sync{(\pro{e_1})}{u_1}$ is in $\osem{t_1\pll t_2}{\Gamma}$.
    From Lemmas \ref{lemma:framesim} and \ref{lemma:gluesim}, the
    event $\sync{\pro{e_1}}{u_1}$ has concession in marking $M$, which
    is therefore not violating --- a contradiction.

  \item It is a straightforward consequence of Lemma
    \ref{lemma:framesim} that the second property holds if the
    transition $(\pref 1{C_1} \cup \pref 2{C_2},\sigma,W) \ltr{}
    (\pref 1{C_1'} \cup \pref 2 {C_2'},\sigma',W')$ is induced by the
    occurrence of an interference event.  Suppose instead that it is
    induced by a synchronized event. Without loss of generality, suppose that in
    $\osem{t_1\pll t_2}\Gamma$ we have
    $M\ltr{\sync{(\pro{e_1})}{u}}M'$ for $M'= (\pref 1{C_1'} \cup
    \pref 2 {C_2'},\sigma',W')$, for some event $e_1$ in $\nsem{t_1}$.
    We shall show that $M_1\ltr{\sync{e_1}u} (C_1,\sigma',W_1')$ in
    $\osem{t_1}\Gamma$ and $M_2\ltr{u}(C_2',\sigma',W_2')$ in
    $\osem{t_2}\Gamma$ for some $W_1',W_2'$ such that $W_1'$ and
    $W_2'$ form an ownership split of $W'$.  Since we have
    $M\ltr{\sync{(\pro{e_1})}u} M'$ in $\osem{t_1\pll t_2}\Gamma$, it is
    easy to see that we have $(\pref 1 C_1 \cup \pref 2
    C_2,\sigma)\ltr{\pro{e_1}} (\pref 1 C_1' \cup \pref 2
    C_2',\sigma')$ in $\nsem{t_1\pll t_2}$ and $C_2=C_2'$.  Hence in $\nsem{t_1}$ we  have
    $(C_1,\sigma)\ltr{e_1} (C_1',\sigma')$.  By assumption, the
    marking $(C_1,\sigma,W_1)$ is not a violating marking of
    $\osem{t_1}\Gamma$, so there exists an interference event $u_1$
    that synchronizes with $e_1$ such that
    $(C_1,\sigma,W_1)\ltr{\sync{e_1}{u_1}}(C_1',\sigma',W_1'')$ for
    some $W_1''$ in $\osem{t_1}\Gamma$, so in
    $\osem{t_1\pll t_2}\Gamma$ we therefore have $(\pref 1 {C_1}\cup
    \pref 2 {C_2}, \sigma, W_1)\ltr{\sync{(\pref 1 e_1)}{u_1}} (\pref
    1 {C_1'} \cup \pref 2 {C_2},\sigma',W_1'')$.  By Lemma
    \ref{lemma:precision}, we have $u_1=u$ and therefore $W_1''=W_1$
    because the occurrence of an event in a marking yields a unique
    marking.  Now, by Lemma \ref{lemma:framesim} there exist $W''$ and
    $W_2'$ such that $W_1'$ and $W_2'$ form an ownership split of
    $W''$ and $(\pref 1 {C_1}\cup \pref 2
    {C_2},\sigma,W)\ltr{\sync{(\pref 1 {e_1})}u} (\pref 1 {C_1'} \cup
    \pref 2 {C_2},\sigma',W'')$ and $(\pref 1 {C_1}\cup \pref 2
    {C_2},\sigma,W_2)\ltr{u} (\pref 1 {C_1} \cup \pref 2
    {C_2},\sigma',W_2')$.  The occurrence of an event in a marking
    leads to a unique marking, so $W''=W'$.  It is easy to see that
    $(C_1,\sigma,W_1)\ltr{\sync{e_1}u}(C_1',\sigma',W_1')$ in
    $\osem{t_1}\Gamma$ and that $(C_2,\sigma,W_2)\ltr{u}
    (C_2,\sigma,W_2')$ in $\osem{t_2}\Gamma$, so the proof is
    complete.\qed
    \end{enumerate}
\end{lem}

The ownership semantics described above has been carefully defined to
explicitly take into account the intuitions behind the rule for
parallel composition, resulting in the short proof of the parallel
decomposition lemma above.  The remaining complexity in the proof of
soundness lies in the rule for establishing an invariant associated
with a resource:
\[    \rn{L-Res}:   
    \deriv
    {\Gamma,\res\ty\chi\ent\ho{\phi}{[\res/\rvar]t}{\psi} }
    {\Gamma\ent\ho{\phi\mand\chi}{\resource\rvar\ldo t\lod}{\psi\mand\chi}}
    \left(\begin{array}{l}\chi\text{ precise}\\ \res\not\in \dom(\Gamma)
\end{array}\right)
\]
It is quite easy to see why this rule follows the intuitive semantics
for judgements presented above: Any run of the net
$\osem{\resource\rvar\ldo t\lod}\Gamma$ to a terminal marking from a
state with the heap owned by the process initially satisfying
$\phi\mand\chi$ can be seen, in conjunction with Lemma
\ref{lemma:resact}, as consisting first of an event that declares a
fresh resource $\res$ current, then a run of
$\osem{[\res/\rvar]t}\Gamma$, followed by an event that makes $\res$
non-current.  The run of $\osem{[\res/\rvar]t}\Gamma$ from a state
where the part of the heap that the process owns satisfies
$\phi\mand\chi$ is simulated by a run of
$\osem{[\res/\rvar]t}{\Gamma,\res\ty\chi}$ along which the locations
satisfying $\chi$ are owned by the invariant $\chi$ in an environment
where $\res$ is an open resource.  In particular, the run obtained has
no interference on the resource $\res$ or the locations that it
protects and $\res$ is available in the terminal state of the run.
Assuming the validity of the judgement $\Gamma,\res\ty\chi
\ent\ho{\phi}{[\res/\rvar]t}{\psi}$, the resulting state owned by the
process is therefore seen to satisfy the formula $\phi\mand\chi$.
Similarly, if there were a reachable marking in
$\osem{\resource\rvar\ldo t\lod}\Gamma$ where the process accesses a
location or resource that it does not own would result in there being
a reachable marking in $\osem{[\res/\rvar]t}{\Gamma,\res\ty\chi}$
where the process accesses an unowned location or resource.  The more
formal presentation of this intuition follows.

We shall begin by explicitly characterizing the runs of the net
$\osem{\resource\rvar\ldo t_0\lod}\Gamma$.  The result is again a
little technical, as is the following lemma, Lemma \ref{lemma:resource}; they
are used in the proof of soundness of the rule \rn{L-Res}.  The reader
may wish to pass through this result and Lemma \ref{lemma:resource}
and only take note of the following definitions of $\invar(\Gamma,R)$
and $D\rest{W}\opr$, $D\rest{W}\oin$ and $D\rest{W}\oot$.
\begin{lem}
\label{lemma:oresact}
Suppose that $\sigma_0$ and $W_0$ form a consistent marking of state
and ownership conditions and let $t\equiv \resource\rvar\ldo t_0\lod$.
For a resource $\res$, define the synchronized events
\begin{eqnarray*}
s_{\res}&=&\sync{\enew{\{\del\},\pref{\res}{\ic{[\res/\rvar]t_0}}}(\res)}{\inew{\res}}\\
s_{\res}'&=&\sync{\eend{\pref{\res}{\tc{[\res/\rvar]t_0}},\{\cel\}}(\res)}{\iend{\res}}
\end{eqnarray*}
If in the net $\osem{t}\Gamma$ we have $\pi\ty
(\ic{t},\sigma_0,W_0)\ltr{}^* (C,\sigma,W)$ then either:
  \begin{enumerate}[$\bullet$]
  \item $C=\ic{t}$ and $\pi$ consists only of interference events, or
  \item there exist $\res,C',\sigma',W',\pi_0$ and $\pi_1$ such that
    $\pi_0$ comprises only interference events,
    $C=\pref{\res} C'$ and \[\pi = \pi_0\cdot s_{\res}\cdot (\pref{\res}\pi_1)\] 
    and 
    \begin{center}
      \begin{tabular}{rll}
        $\pi_0\cdot s_{\res} \ty $ &
        $(\ic{t},\sigma_0,W_0) \ltr{}^*
        (\pref{\res}\ic{[\res/\rvar]t_0},\sigma',W')$ & in $\osem{t}\Gamma$ and\\
        $\pi_1 \ty$ & $ (\ic{[\res/\rvar]t_0},\sigma',W')\ltr{}^* (C',\sigma,W)$ & in
        $\osem{[\res/\rvar]t_0}\Gamma$, or
      \end{tabular}
      \end{center}
    \item $C=\tc{t}$ and there exist
      $\res,\sigma',\sigma'',W',W'',\pi_0,\pi_1,\pi_2$ such that
      $\pi_0$ and $\pi_2$ comprise only interference events,
      \[\pi = \pi_0\cdot s_{\res}\cdot (\pref{\res}\pi_1)
      \cdot s_{\res}' \cdot \pi_2,
      \] 
    and 
    \begin{center}
      \begin{tabular}{rll}
        $\pi_0\cdot s_{\res} \ty $ &
        $(\ic{t},\sigma_0,W_0) \ltr{}^*
        (\pref{\res}\ic{[\res/\rvar]t_0},\sigma',W')$ & in $\osem{t}\Gamma$, \\
        $\pi_1 \ty$ & $ (\ic{[\res/\rvar]t_0},\sigma',W')\ltr{}^* (\tc{[\res/\rvar]t_0},\sigma'',W'')$ 
        & in $\osem{[\res/\rvar]t_0}\Gamma$, and\\
        $s_{\res}'\cdot \pi_2 \ty $ &
        $(\pref{\res}\tc{[\res/\rvar]t_0},\sigma'',W'') \ltr{}^* (\tc{t},\sigma,W)$ & in $\osem{t}\Gamma$.
      \end{tabular}
      \end{center}
  \end{enumerate}
\begin{proof}
  Readily seen to be a consequence of Lemmas \ref{lemma:dagger}, \ref{lemma:resact},
  \ref{lemma:contsim} and \ref{lemma:charev}.
\end{proof}
\end{lem}
It can be shown, as a consequence of the preceding lemma, that during
the run of the net following the declaration event, the resource
$\res$ chosen for $\rvar$ is owned by the process until it is made
non-current at the end of the variable $\rvar$'s scope.
\begin{lem}
\label{lemma:resown}
Let $t\!\equiv \resource\rvar\ldo t_0\lod$.  If $(\pref\res C_0,\sigma,W)$ is
reachable from $(\ic{t},\sigma_0,W_0)$, which is a consistent marking
of $\osem{t}\Gamma$, then $\wpr\res\in W$.
\qed
\end{lem}

We write $\invar(\Gamma,R)$ for the formula
$\chi_1\mand\ldots\mand\chi_n$ formed as the separating conjunction of
the invariants of all the \emph{available}, according to $R$, open
resources.  It is defined by induction on the size of the domain of
$\Gamma$:
\begin{eqnarray*}
  \invar(\emptyset,R) &\eqd& \mathsf{empty}\\
  \invar((\Gamma,\res\ty\chi),R) &\eqd& \left\{
    \begin{array}{l@{\quad}l}
      \invar(\Gamma,R),& \text{if }\res\not\in R\\
      \chi\mand\invar(\Gamma,R), &\text{if }\res\in R.
    \end{array}\right.
\end{eqnarray*}
Define the notations
\begin{eqnarray*}
D\rest{W}\opr &\eqd& \{\hvl{}v\in D\st \wpr\lcon\in W\}\\
D\rest{W}\oin &\eqd& \{\hvl{}v\in D \st \win\lcon\in W\}\\
D\rest{W}\oot &\eqd& \{\hvl{} v\in D \st \wot\lcon\in W\}
\end{eqnarray*}
to represent the heap at locations owned by the process, invariants
and other processes, respectively.  In any state that we consider, we
would expect $D\rest{W}\oin \models \invar(\Gamma,R)$.  A marking
of the net $\osem{t}\Gamma$ can be converted to a marking of
$\osem{t}{\Gamma,\res\ty\chi}$ by, if $\res$ is available, regarding
ownership of the locations satisfying the invariant $\chi$ as being
owned by the invariant rather than by the process.
\newcommand{\prc}[1]{\pi_{\res}^\chi(#1)}
\begin{defi}
\label{def:resproj}
Suppose that $\chi$ is a precise heap formula.  Let
$M=({C},(D,L,R,N),W)$ be a consistent marking of
$\osem{t}{\Gamma}$ such that if $\res\in R$ then there exists
(necessarily unique) $D_0\subseteq D\rest{W}\opr$ such that
$D_0\models\chi$.  Define the projection of $M$ into the net
$\osem{t}{\Gamma,\res\ty\chi}$ to be
  \[ \prc M \eqd (C,(D,L,R,N),W'), \]
  where:
  \begin{enumerate}[$\bullet$]
  \item
    if $\res\not\in R$: $W'=W$
  \item
    if $\res\in R$: Let $D_0\subseteq D$ be such that $D_0\models\chi$.
    \[
    \begin{array}{rcl@{\st}l}
      W' &=   & \{\wot{\lcon} & \wot{\lcon}\in W  \}\\
      &\cup& \{\wot{\res'} & \wot{\res'}\in W  \}\\
      &\cup& \{\win{\lcon} & \win{\lcon}\in W  ~\tor~ \lcon\in\dom(D_0) \}\\
      &\cup& \{\win{\res'} & \win{\res'}\in W  ~\tor~ \res'=\res\}\\
      &\cup& \{\wpr{\lcon} & \wpr{\lcon}\in W ~\tand~\lcon\not\in\dom(D_0) \}\\
      &\cup& \{\wpr{\res'} & \wpr{\res'}\in W~\tand~\res'\neq\res  \}
    \end{array}
    \]
  \end{enumerate}
\end{defi}
It is clear that if $M$ is a consistent marking of $\osem{t}\Gamma$
then $\prc M$ is a consistent marking of
$\osem{t}{\Gamma,\res\ty\chi}$.  They key lemma representing the
account above, that behaviour in the net where a resource is closed is
simulated by the net where the resource is open, is now stated, though
we shall not show its proof here.

\begin{lem}
\label{lemma:resource}
  Let $\res$ be a resource such that $\res\not\in\dom(\Gamma)$ and let
  $\chi$ be a precise heap logic formula.  Let
  $M=({C},(D,L,R,N),W)$ be a consistent marking of
  $\osem{t}{\Gamma}$ such that:
  \begin{enumerate}[$\bullet$]
  \item $\wpr\res\in W$, 
  \item $D\rest{W}\oin \models \invar(\Gamma,R)$, and
  \item if $\res\in
    R$ then there exists $D_0\subseteq D\rest{W}\opr$ such that $D_0\models\chi$.
  \end{enumerate}
  Then
  \begin{enumerate}[\em(1)]
  \item If $M$ is a violating marking in $\osem{t}{\Gamma}$ then
    $\prc M$ is a violating marking in
    $\osem{t}{\Gamma,\res\ty\chi}$.
  \item For any event $u$ of $\osem{t}{\Gamma}$ that is an
    interference event, if $M$ is not a violating marking and
    $M\ltr{u} M'$ where $M'=({C'},(D',L',R',N'),W')$ and
    $\wpr\res\in W'$ then:
    \begin{enumerate}[$\bullet$]
    \item $\prc M \ltr u \prc{M'}$ in
      $\osem{t}{\Gamma,\res\ty\chi}$ and:
      \begin{enumerate}[$-$]
      \item $D'\rest{W'}\oin \models \invar(\Gamma,R')$
      \item if $\res\in R'$ then there exists $D_0\subseteq
        D'\rest{W'}\opr$ such that $D_0\models\chi$.
      \end{enumerate}
    \end{enumerate}
  \item For any synchronized event $s=\sync{{e_1}}{u}$ of
    $\osem{t}{\Gamma}$, if $M$ is not a violating marking and
    $M\ltr s M'$ where $M'=({C'},(D',L',R',N'),W')$ and
    $\wpr\res\in W'$ then either:
    \begin{enumerate}[$\bullet$]
    \item $\prc M$ is violating in
      $\osem{t}{\Gamma,\res\ty\chi}$, or
    \item there exists $u'$ such that $\prc M \ltr {\sync{e_1}{u'}}
      \prc{M'}$ in $\osem{t}{\Gamma,\res\ty\chi}$ and:
      \begin{enumerate}[$-$]
      \item $D'\rest{W'}\oin \models \invar(\Gamma,R')$
      \item if $\res\in R'$ then there exists $D_0\subseteq
        D'\rest{W'}\opr$ such that $D_0\models\chi$.\qed
      \end{enumerate}
    \end{enumerate}
  \end{enumerate} 
\end{lem}

We shall say that a state $\sigma$ with an ownership marking $W$
satisfies the formula $\phi$ and the invariants in $\Gamma$ if the
heap restricted to the owned locations satisfies $\phi$ and the
invariants are met for all the available resources.  
The rest of the heap, seen as owned by external processes, is unconstrained.
\begin{defi}
  A marking $(C,\sigma,W)$ of $\osem{t}\Gamma$ \emph{satisfies $\phi$ in $\Gamma$} if:
  \begin{enumerate}[$\bullet$]
  \item the marking $(C,\sigma,W)$ is consistent,
  \item $D\rest{W}\opr\models \phi$, and
  \item $D\rest{W}\oin\models \invar(\Gamma,R)$,
  \end{enumerate}
  where $\sigma=(D,L,R,N)$.
\end{defi}

We now attach a notion of validity to judgements $\Gamma\ent\hos{}$.
It shall assert that no violating marking is ever reached and that
whenever the process $t$ runs to completion from a state where the
part of the heap that it owns satisfies $\phi$ then the part of the
resulting heap that it owns satisfies $\psi$.
\begin{defi}[Validity]
  Let $t$ be a closed term.
  Define $\Gamma\models\hos{}$ if, for any $\sigma$ and $W$ such that
  the marking $(\ic{t},\sigma,W)$ satisfies $\phi$ in $\Gamma$:
  \begin{enumerate}[$\bullet$]
  \item any marking reachable in $\osem{t}{\Gamma}$ from
    $(\ic{t},\sigma,W)$ is non-violating, and
  \item for any $\sigma'$ and $W'$, if the marking $(\tc{t},\sigma',W')$ is
    reachable in $\osem{t}{\Gamma}$ from $(\ic{t},\sigma,W)$
    then $(\tc{t},\sigma',W')$ satisfies $\psi$ in $\Gamma$.
  \end{enumerate}
\end{defi}
It is useful to note that the occurrence of an interference event does
not affect whether a marking satisfies $\phi$ in $\Gamma$ or whether
it is violating.  Consequently, when considering validity it is
unnecessary to account for runs of the net $\osem{t}\Gamma$ that start
or end with an interference event.
\begin{lem}
  \label{lemma:interfsat}
  Let $M$ be a consistent marking of $\osem{t}{\Gamma}$ that satisfies
  $\phi$ in $\Gamma$ and is non-violating.  If $u$ is an interference
  event and $M\ltr{u}M'$ then $M'$ satisfies $\phi$ in $\Gamma$ and is
  non-violating.

  \proof Straightforward from the definition of satisfaction of $\phi$
  in $\Gamma$ by considering the possible forms of $u$. \qed
\end{lem}

In the rule $\rn{L-Res}$ which allows invariants to be established for
resources, only one resource is considered for substitution for the
variable. The following lemma shows that this is sufficient; the
semantics of judgements is unaffected by the choice of resource.
\begin{lem}
\label{lemma:semequivar}
For any resources $\res,\res'$ such that
$\res,\res'\not\in\dom(\Gamma)$ and any term $t$ with $\fv
t\subseteq\{\rvar\}$ and $\fr t\subseteq \dom(\Gamma)$,
\[ \Gamma,\res\ty\chi \models \ho{\phi}{[\res/\rvar]t}{\psi}
\text{ iff }
\Gamma,\res'\ty\chi \models \ho{\phi}{[\res'/\rvar]t}{\psi}.
\]
\begin{proof}
\newcommand{\interchange}{\leftrightarrow}
  The net $\osem{[\res/\rvar]t}{\Gamma,\res\ty\chi}$ is clearly isomorphic to
  $\osem{[\res'/\rvar]t}{\Gamma,\res'\ty\chi}$ through interchanging the conditions
  \[   \begin{array}{l}
    \res \interchange \res' \qquad \curr(\res) \interchange \curr({\res'})\\
    \wpr\res\interchange \wpr{\res'}
    \qquad \win\res\interchange \win{\res'}
    \qquad \wot\res\interchange \wot{\res'}.
  \end{array}
  \]
  The result follows from the definition of validity being insensitive
  to such  permutations.
\end{proof}
\end{lem}

We are now in a position where we the rules of concurrent separation
logic can be proved sound.  Only two important cases of the proof
shall be presented here; full details will be available in the first
author's PhD thesis.

\begin{thm}[Soundness]~
\label{theorem:soundness}
  For any closed term $t$, if $\Gamma\ent\hos{}$ then
  $\Gamma\models\hos{}$.

  \proof
    By rule induction on the judgement $\Gamma\ent\hos{}$.  Note that,
  due to Lemma \ref{lemma:interfsat}, we shall only consider runs of
  $\osem t \Gamma$ that do not start or end with an interference
  event.\medskip

\noindent\underline{\rn{L-Par}}:\ Suppose that we have
  $\Gamma\ent\ho{\phi_1\mand\phi_2}{t_1\pll t_2}{\psi_1\mand\psi_2}$
  because $\Gamma\ent\hos{1}$ and $\Gamma\ent\hos{2}$. Assume that
  marking $M=(\ic{t_1\pll t_2},\sigma,W)$ satisfies
  $\phi_1\mand\phi_2$ in $\Gamma$. It can be seen from the definitions
  that there exist $W_1$ and $W_2$ forming an ownership split of $W$
  such that $(\ic{t_1},\sigma,W_1)$ is a marking of
  $\osem{t_1}{\Gamma}$ that satisfies $\phi_1$ in $\Gamma$ and
  $(\ic{t_2},\sigma,W_2)$ satisfies $\phi_2$ in $\Gamma$.  Let marking
  $M'=(C',\sigma',W')$ be reachable from $M$; a simple induction on
  the length of path to $M$ using Lemma \ref{lemma:pardecomp} and
  Lemma \ref{lemma:paract} shows that there exist $C_1', C_2', W_1'$
  and $W_2'$ such that $C'=\pref 1{C_1'}\cup \pref 2{C_2'}$ and $W_1'$
  and $W_2'$ form an ownership split of $W'$.  Furthermore, the
  marking $(C_1',\sigma',W_1')$ is reachable from
  $(\ic{t_1},\sigma,W_1)$ in $\osem{t_1}{\Gamma}$ and
  $(C_2',\sigma',W_2')$ is reachable from $(\ic{t_2},\sigma,W_2)$ in
  $\osem{t_2}\Gamma$.
  
  Suppose that the marking $M'$ is violating.  Using Lemma
  \ref{lemma:pardecomp}, it follows that either $(C_1',\sigma',W_1')$
  or $(C_2',\sigma',W_2')$ is a violating marking.  This contradicts
  either the induction hypothesis for $\Gamma\ent\hos{1}$ or the
  induction hypothesis for $\Gamma\ent\hos{2}$, so $M'$ cannot be
  violating.
  
  Now suppose that the marking $M'$ is terminal: we have
  $C_1'=\tc{t_1}$ and $C_2'=\tc{t_2}$.  From the induction hypotheses,
  we obtain that $(C_1',\sigma',W_1')$ satisfies $\psi_1$ in $\Gamma$
  and that $(C_2',\sigma',W_2')$ satisfies $\psi_2$ in $\Gamma$.  It
  is easy to see from the definition of ownership split that therefore
  $(C',\sigma',W')$ satisfies $\psi_1\mand\psi_2$ in $\Gamma$.\medskip
  
\noindent\underline{\rn{L-Res}}:\ Let $t\equiv \resource\rvar \ldo t_0\lod$.  Suppose
  that $\Gamma\ent\ho{\phi\mand\chi}{t}{\psi\mand\chi}$ because $\Gamma,\res_0\ty\chi
  \ent\ho{\phi}{[\res_0/\rvar]t_0}{\psi}$ for some $\res_0\not\in
  \dom(\Gamma)$.  Assume that the marking $M=(\ic{t},\sigma,W)$
  satisfies $\phi\mand\chi$ in $\Gamma$, and let $M'=(C',\sigma',W')$
  be reachable from $M$ in $\osem{t}\Gamma$.  According to Lemma
  \ref{lemma:oresact}, there are three cases to consider for the
  marking $M'$.

  \begin{enumerate}[$\bullet$]
  \item The first case has $M=M'$ (we need not consider runs starting
    with an interference event according to Lemma
    \ref{lemma:interfsat}).  Since $\ic{t}\neq \tc{t}$, all that we
    must show is that $M$ is non-violating.  Using Lemma
    \ref{lemma:resact}, we can infer that the only events with
    concession in the marking $(\ic{t},\sigma)$ of $\nsem{t}$ are
    equal to $\enew{\ic{t},\pref{\res}\ic{[\res/\rvar]t_0}}(\res)$ for
    some $\res\in \sn{Res}$ such that $\curr(\res)\not\in\sigma$.  The
    marking $M$ is assumed to be consistent, so for each such $\res$
    we have $\wpr\res\not\in W$ and hence the synchronized event
    $\sync{\enew{\ic{t},\pref{\res}\ic{[\res/\rvar]t_0}}(\res)}{\inew{\res}}$
    has concession in $M$.  The marking $M'$ cannot therefore be
    violating.

  \item Secondly, there exists a resource $\res$, markings
    $\sigma_0,W_0,C_1$ and a path $\pi_1$ such that
    $C'=\pref{\res}{C_1}$ and
    \begin{center}
    \begin{tabular}{rllll}
    $s_\res\ty $ &$(C,\sigma,W)$ & $\ltr{}$ & $(\pref\res{\ic{[\res/\rvar]t_0}},\sigma_0,W_0)$
    &in $\osem{t}\Gamma$\\
    $\pi_1\ty$ &
    $({\ic{[\res/\rvar]t_0}},\sigma_0,W_0)$ & $\ltr{}^*$ & 
    $(C_1,\sigma',W')$ & 
    in $\osem{[\res/\rvar]t_0}\Gamma$,
    \end{tabular}\end{center}
  where
  $s_\res=\sync{\enew{\ic{t},\pref{\res}\ic{[\res/\rvar]t_0}}(\res)}{\inew{\res}}$.
  The marking $(C',\sigma',W')$ cannot be a terminal marking of the
  net $\osem{t}\Gamma$, so all that we must show is that it is
  non-violating.  We have $\res,\curr(\res)\in\sigma_0$ and
  $\wpr\res\in W_0$ since they are in the postconditions of $s_\res$.
 A simple induction on
  the length of $\pi$ using Lemmas \ref{lemma:resown} and
  \ref{lemma:resource} informs that $\prc{C_1,\sigma',W'}$ is
  reachable from $\prc{\ic{[\res/\rvar]t_0},\sigma_0,W_0}$ in
  $\osem{[\res/\rvar]t_0}{\Gamma,\res\ty\chi}$.  We have
  $\curr(\res)\not\in\sigma$ because the event $s_\res$ has concession
  in $M$, so $\res\not\in\dom(\Gamma)$ because the marking $M$ is
  consistent.  Since $\fr t \subseteq \dom(\Gamma)$ by Lemma
  \ref{lemma:freeres}, we may use Lemma \ref{lemma:semequivar} in
  conjunction with the induction hypothesis to obtain
  $\Gamma,\res\ty\chi \models \ho{\phi}{[\res/\rvar]t_0}{\psi}$.  It
  is an easy calculation to show that 
  $\prc{{\ic{[\res/\rvar]t_0}},\sigma_0,W_0}$ satisfies $\phi$ in
  $\Gamma,\res\ty\chi$, so the marking $\prc{C_1,\sigma',W'}$ is
  non-violating.  By Lemma \ref{lemma:resource}, the marking
  $(C_1,\sigma',W')$ of $\osem{[\res/\rvar]t_0}\Gamma$ is therefore
  non-violating.  According to Lemma \ref{lemma:resact}, there are two
  possible ways in which the marking $(C',\sigma',W')$ of
  $\osem{t}\Gamma$ might be violating.  Firstly, there might exist an
  event $e$ of $\nsem{[\res/\rvar]t_0}$ that has concession in the
  marking $(C_1,\sigma')$ but there is no interference event $u$ that
  synchronizes with $e$ such that $\sync e u$ has concession in the
  marking $(C_1,\sigma',W')$.  We have shown, however, that this is
  not the case since the marking $(C_1,\sigma',W')$ is non-violating.
  Alternatively, the event $e_\res' = \eend{\pref \res
    \tc{[\res/\rvar]t_0},\tc{t}}(\res)$ might have concession in the
  marking $(C',\sigma')$ of $\nsem{t}$ but the event $s_\res' =
  \sync{e_\res'}{\iend{\res}}$ might not have concession in
  $(C',\sigma',W')$; that is, $\wpr\res\not\in W'$.  However, we have
  $\wpr\res\in \sigma_0$ so by applying Lemma \ref{lemma:resown} along path
  $\pi_1$ we obtain $\wpr\res\in W'$.  So the event $s_\res'$ has
  concession in the marking, which is therefore not violating.    
  \item The final case is where $C'=\tc{t}$ and there exist
    $\sigma_0,\sigma_1,W_0,W_1$ and a path $\pi_1$ such that
    \begin{center}
    \begin{tabular}{rllll}
    $s_\res\ty $ &$(\ic{t},\sigma,W)$ & $\ltr{}$ & $(\pref\res{\ic{[\res/\rvar]t}},\sigma_0,W_0)$
    &in $\osem{t}\Gamma$\\
    $\pi_1\ty$ &
    $({\ic{[\res/\rvar]t_0}},\sigma_0,W_0)$ & $\ltr{}^*$ & 
    $(\tc{[\res/\rvar]t_0},\sigma_1,W,_1)$ & 
    in $\osem{[\res/\rvar]t_0}\Gamma$\\
    $s_\res'\ty $ &$(\pref{\res}{\tc{[\res/\rvar]t_0}},\sigma_1,W_1)$ 
    & $\ltr{}$ & $(\tc{t},\sigma',W')=M'$
    &in $\osem{t}\Gamma$,
    \end{tabular}\end{center}
  where
  $s_\res=\sync{\enew{\ic{t},\pref{\res}\ic{[\res/\rvar]t_0}}(\res)}{\inew{\res}}$.
  The marking $M'$ is readily seen to be non-violating since no event
  of $\nsem{t}$ has concession if the marking of control conditions is
  $\tc{t}$.  All that remains is to show that $M'$ satisfies $\psi$ in
  $\Gamma$.  As in the previous case, we have
  $\res,\curr(\res)\in\sigma_0$ and $\wpr\res\in W_0$ and $\res\not\in
  \dom(\Gamma)$.  It is easily seen that the marking
  $\prc{\ic{[\res/\rvar]t_0},\sigma_0,W_0}$ of
  $\osem{[\res/\rvar]t_0}{\Gamma,\res\ty\chi}$ satisfies $\phi$ in
  $\Gamma,\res\ty\chi$.  A simple induction on the length of the path
  $\pi$ using Lemmas \ref{lemma:resown} and \ref{lemma:resource} shows
  that the marking $\prc{\tc{[\res/\rvar]t_0},\sigma_1,W_1}$ is
  reachable in $\osem{[\res/\rvar]t_0}{\Gamma,\res\ty\chi}$ from
  $\prc{\ic{[\res/\rvar]t_0},\sigma_0,W_0}$.  Using Lemmas
  \ref{lemma:semequivar} and \ref{lemma:freeres}, from the induction
  hypothesis $\Gamma,\res_0\ty\chi \models
  \ho{\phi}{[\res_0/\rvar]t_0}{\psi}$, the marking
  $\prc{\tc{[\res/\rvar]t_0},\sigma_1,W_1}$ satisfies $\psi$ in
  $\Gamma,\res\ty\chi$.  We have $\res\in \sigma_1$ since the event
  $s_\res'$ has concession in the marking
  $(\pref{\res}{\tc{[\res/\rvar]t_0}},\sigma_1,W_1)$, so the marking
  $(\tc{[\res/\rvar]t_0},\sigma_1,W_1)$ satisfies $\psi\mand\chi$ in
  $\Gamma$, from which it is easily seen that $(\tc{t},\sigma',W')$
  also satisfies $\psi\mand\chi$ in $\Gamma$.\qed
\end{enumerate}
\end{thm}
The following result connects the definition of validity to the
execution of processes without interference or ownership.
\begin{cor}[Connection]
  \label{corr:connection}
  Let $t$ be a closed term with $\fr t=\emptyset$ and let
  $\sigma=(D,L,\emptyset,\emptyset)$ be a consistent marking of state
  conditions for which $D\models\phi$. 
  If $\emptyset\models\hos{}$ then whenever a terminal marking $(\tc{t},\sigma')$ is reachable
  from $(\ic t,\sigma)$ in $\nsem t$, the resulting heap $D'$ satisfies
  $\psi$, where $\sigma'=(D',L',R',N')$.

\proof 
A consequence of soundness and Lemma \ref{lemma:connecsim}.\qed
\end{cor}

\subsection{Fault}
\label{sec:error}
It can be seen that the rules of concurrent separation logic ensure
that processes, running from suitable initial states, only access
current locations.  The syntax of the language ensures that processes
only access current resources and that they are never blocked when
releasing a resource through it already being available.  We shall now
demonstrate that processes avoid such `faults', in which we shall say
that an event $e$ is \emph{control-enabled} in a marking $C$ of
control conditions if there exists a marking $C'$ such that $C\ctr e
C'$.

\begin{defi}[Fault]
  There is a \emph{fault} in a marking $M=(C,\sigma)$ of the net
  $\nsem{t}$ if there exists a control-enabled 
  event $e$ in $\nsem{t}$ with $\pres C e = C_1$ and
  $\poss C e=C_2$ for some $C_1,C_2$ such that either:
  \begin{enumerate}[(1)]
  \item there exist $D,D'$ such that
    $e=\eact{C_1,C_2}(D,D')$
    and there exists $\lcon\in\dom(D)$ with $\curr(\lcon)\not\in\sigma$,
  \item there exist $\lcon,v,\lcon',v'$ such that
    $e=\eall{C_1,C_2}(\lcon,v,\lcon',v')$ and
    $\curr(\lcon)\not\in\sigma$,
  \item there exist $\lcon,v,\lcon',v'$ such that
    $e=\edea{C_1,C_2}(\lcon,v,\lcon')$ and either
    $\curr(\lcon)\not\in\sigma$ or $\curr(\lcon')\not\in\sigma$,
  \item there exists $\res$ such that either $e=\eget{C_1,C_2}(\res)$
    or $e=\eput{C_1,C_2}(\res)$ and $\curr(\res)\not\in\sigma$, or
  \item there exists $\res$ such that $e=\eput{C_1,C_2}(\res)$ and
    $\res\in\sigma$.
  \end{enumerate}
\end{defi}
This definition also applies to markings $(C,\sigma,W)$ of
$\osem{t}\Gamma$ in the by ignoring the marking of ownership
conditions $W$ and considering synchronized events $\sync{e}u$.

\begin{thm}[Fault avoidance]
\label{thm:fault}
  Suppose that $\Gamma\ent\hos{}$ and that the marking
  $(\ic{t},\sigma_0,W_0)$ satisfies $\phi$ in $\Gamma$.  If
  $(C,\sigma,W)$ is reachable from $(\ic{t},\sigma_0,W_0)$ then there
  is not a fault in $(C,\sigma,W)$.
\proof By rule induction on the judgement $\Gamma\ent\hos{}$.\qed
\end{thm}
A corollary of this result and Lemma \ref{lemma:connecsim} is that if
$\emptyset\ent\hos{}$ then no fault is reachable from an initial
marking of $\nsem{t}$ if the heap initially satisfies $\phi$.

\section{Separation}
As mentioned in the introduction, the logic discriminates between the
parallel composition of processes and their interleaved expansion.  In
Brookes' trace semantics \cite{brookes:soundness}, this was accounted
for by making the notion of a \emph{race} primitive within the 
semantics: when forming the parallel composition of processes, if two
processes concurrently write to the same location, a special `race'
action occurs and the trace proceeds no further.  Our approach when
defining the semantics has been different; we do not regard a race as
`catastrophic' and have not embellished our semantics with special
race states.  Instead, we shall prove, using the semantics directly, that races
do not occur for proved processes running from suitable initial states.

Generally, a race can be said to occur when two interacting heap
actions occur concurrently.  Recall that a heap action is represented
in the net semantics by a set of events, with common pre- and
post-control conditions, representing each way in which the action can
affect the heap.  According to the net model, two actions may be
allowed to run concurrently if their events do not overlap on their
pre- or post-control conditions.  In such a situation, where $\pres C
{\poss C {e_1}}\cap \pres C{\poss C {e_2}}=\emptyset,$ we shall say
that $e_1$ and $e_2$ are \emph{control-independent}.

One way of capturing the race freedom of a process running from an
initial state is to show that there is no reachable marking in the net
where two control-independent events are control-enabled but access a
common heap location, except interaction through allocation.  We,
however, shall prove a result based on the \emph{behaviour} of
processes: that whenever two events are control-independent and can
occur, then either they are independent or they lie within a form of
prescribed class of action.

\begin{defi}[Separation of synchronized events]
  Let $M$ be a marking of $\osem{t}{\Gamma}$ and let
  $s_1=\sync{e_1}{u_1}$ and $s_2=\sync{e_2}{u_2}$ be
  control-independent synchronized events of $\osem{t}{\Gamma}$.  The
  \emph{separation property} of $s_1$ and $s_2$ at $M$ is defined as:
  \begin{enumerate}[(1)]
  \item If $M\ltr{s_1}M_1$ and $M\ltr{s_2}M_2$ and $s_1$ and $s_2$ are
    not independent then either:
    \begin{enumerate}[$\bullet$]
    \item $s_1$ and $s_2$ compete to allocate the same location:
      $e_1=\eall{C_1,C_1'}(\lcon,v,\lcon',v')$ and
      $e_2=\eall{C_2,C_2'}(k,w,\lcon',w')$ for some
      $\lcon,\lcon',k,v,v',w'$; 
    \item $s_1$ and $s_2$ compete to make the same resource current:
       $e_1=\enew{C_1,C_1'}(\res)$ and
      $e_2=\enew{C_2,C_2'}(\res)$ for some $\res$; or
    \item $s_1$ and $s_2$ compete to acquire the same resource:
            $e_1=\eget{C_1,C_1'}(\res)$ and
      $e_2=\eget{C_2,C_2'}(\res)$ for some $\res$.
    \end{enumerate}
  \item If $M\ltr{s_1}M_1\ltr{s_2}M_2$ and $s_1$ and $s_2$ are not independent then either:
    \begin{enumerate}[$\bullet$]
    \item $s_1$ deallocates a location that $s_2$ allocates:
      $e_1=\edea{C_1,C_2}(\lcon,v,\lcon',v')$ and
      $e_2=\eall{C_2,C_2'}(k,w,\lcon',w')$ for some
      $\lcon,\lcon',k,v,v',w'$; 
    \item $s_1$ makes a resource non-current that $s_2$ makes current:
      $e_1=\eend{C_1,C_1'}(\res)$ and $e_2=\enew{C_2,C_2'}(\res)$ for some
      $\res$; or
    \item $s_1$ releases a resource that $s_2$ takes:
      $e_1=\eput{C_1,C_1'}(\res)$ and
      $e_2=\eget{C_2,C_2'}(\res)$ for some $\res$.
    \end{enumerate}
  \item The symmetric statement for $M\ltr{s_2}M_2\ltr{s_1}M_1$.
  \end{enumerate}
\end{defi}
The first part of the property above tells us how the enabled events
of parallel processes \emph{conflict} with each other in a state: the
way in which one parallel process can prevent the other acting in a
particular way on the global state.  The second part dictates how the
event occurrences of parallel processes \emph{causally depend} on each
other: the way in which the ability of one process to affect the
global state in a particular way is dependent on events of the other
process.  

Importantly, whenever the two events $s_1$ and $s_2$ arise from heap
actions, they neither conflict nor causally depend on each other.
This is our net analogue of race freedom.  Theorem
\ref{theorem:separation} shows that processes proved by the logic are
race free when running from suitable initial states.  We shall make
use of the following rather technical lemmas in the proof.

For a synchronized event $s$ and an interference event $u$, define the
separation property for $s$ and $u$ at $M$ similarly, recalling that
any synchronized event is trivially control-independent from any
interference event because $\pres C {\poss C u} = \emptyset$ for any
interference event $u$.  It is always the case that a synchronized
event and an interference event satisfy the separation property in any
consistent marking.
\begin{lem}
  \label{lemma:sepint}
  If $M$ is a consistent marking of $\osem{t}{\Gamma}$ and $s$ is a
  synchronized event and $u$ is an interference event then $s$ and $u$
  satisfy the separation property in $M$.

  \begin{proof}
    A straightforward analysis of the many cases for $s$ and $u$.
  \end{proof}
\end{lem}

The following lemma relates independence from an interference event to
independence from any corresponding synchronized event.  Recall that
we write $eIe'$ if $e$ and $e'$ are independent.

\begin{lem}
  \label{lemma:sepsync}
  Let $s$ be any synchronized event of $\osem{t}{\Gamma}$ and $u$ be
  an interference event of $\osem{t}{\Gamma}$.  Suppose that $M$ is a
  consistent marking in which they both have concession.  If $e_1$ is
  an event of $\nsem{t}$ that synchronizes with $u$ and $sI u$ and $s$
  is control-independent from $e_1$ then $sI(\sync{e_1}{u})$.
  \begin{proof}
    It is easy to see that the preconditions of $\sync{e_1}{u}$ are
    simply the preconditions of $u$ along with the pre-control
    conditions of $e_1$ apart from replacing $\wot{\lcon}$ with
    $\wpr{\lcon}$ and replacing $\wot{\res}$ with $\wpr{\res}$.  The
    postconditions of $\sync{e_1}{u}$ are similar.
    
    Suppose, for contradiction, that $\lnot(s I(\sync{e_1}{u}))$.
    Since $s I u$ and $s$ is control-independent from $e_1$, it
    follows that there must exist $z\in\sn{Loc}\cup\sn{Res}$ such that
    $\wpr{z}\in \pre{\post s}\cap \pre{\post{(\sync{e_1}{u})}}$.  From
    the definition of synchronization, we therefore have $\wot z\in
    \pre{\post{u}}$.  The proof is completed by analysis of the cases
    for how $\wpr z\in\pre{\post{s}}$; we shall show only one illustrative
    case, that where $z$ is a location $\lcon$ such that $\wpr\lcon\in
    \pre s$ but $\wpr\lcon\not\in \post s$.
    
    In this case, the event $s$ must either deallocate the location
    $\lcon$ or must release a resource $\res$ with
    $\res\in\dom(\Gamma)$ and $\lcon$ forms part of the heap used to
    satisfy the invariant for $\res$.  As the event $s$ has concession
    in $M$, we have $\wpr\lcon\in M$.  By assumption, $u$ has
    concession in $M$ and $\wot\lcon\in\pre{\post u}$.  We cannot have
    $\wot\lcon\in\pre u$ since $\wpr\lcon\in M$, so $\wot\lcon\in\post
    u$.  Therefore, the event $u$ is an interference event that either
    allocates the location $\lcon$ or acquires an open resource $\res$
    and $\lcon$ is part of the heap that satisfies the invariant for
    $\res$.  If $u$ is such an event, that acquires $\res$, it must be
    the case that $\win\lcon\in\pre u$ so $\win\lcon\in M$,
    contradicting that $M$ is a consistent marking with $\wpr\lcon\in
    M$.  Consequently, $u$ must in fact be an event that allocates the
    location $\lcon$, so therefore $\curr(\lcon)\not\in M$.  We then
    arrive at another contradiction since it must then be the case
    that $\wpr\lcon\not\in M$ because $M$ is consistent.
  \end{proof}
\end{lem}

We may now show that the separation property does indeed hold for any
two events $s_1$ and $s_2$ in $\osem{t}\Gamma$ for any term $t$ and
environment $\Gamma$ such that $\Gamma\ent\hos{}$ in any marking
$M=(C,\sigma,W)$ reachable from an initial marking of $t$ that
satisfies $\phi$ in $\Gamma$.  The proof is most interesting in the
case where $t\equiv t_1\pll t_2$ and $s_1$ is an event of $t_1$ and
$s_2$ is an event of $t_2$.  The case proceeds by establishing, as in
Theorem \ref{theorem:soundness}, that there exists an ownership split
$W_1$ and $W_2$ of $W$ for which $s_1$ has concession in
$(C_1,\sigma,W_1)$, where $C_1$ is the marking of control conditions
in $C$ for $t_1$, and there exist $e_2$ and $u_2$ such that
$s_2=\sync{(\pref 2 e_2)}{u_2}$ and $u_2$ also has concession in the marking
$(C_1,\sigma,W_1)$ of $\osem{t_1}\Gamma$. By Lemma \ref{lemma:sepint},
the separation property therefore holds for $s_1$ and $u_2$ in the
marking $(C_1,\sigma,W_1)$.  It follows that the separation property
holds for $s_1$ and $s_2$ in $M$ since, by Lemma \ref{lemma:sepsync},
if the events $s_1$ and $u_2$ are independent then so are $s_1$ and
$s_2$.

\begin{thm}[Separation]
\label{theorem:separation}
  Suppose that $\Gamma\ent\hos{}$ and that $(\ic{t},\sigma_0,W_0)$
  satisfies $\phi$ in $\Gamma$.  For any events $s_1$ and $s_2$ in
  $\osem{t}\Gamma$ and any marking $(C,\sigma,W)$ reachable from
  $(\ic{t},\sigma_0,W_0)$, the separation property holds for $s_1$ and
  $s_2$ at $(C,\sigma,W)$.
  
  \proof By induction on the derivation of $\Gamma\ent\hos{}$.  We
  shall show only one case:\medskip

  \noindent\underline{\rn{L-Par}}:\ Assume that the marking $(\ic{t_1\pll
      t_2},\sigma_0,W_0)$ of $\osem{t_1\pll t_2}\Gamma$ satisfies
    $\phi_1\mand\phi_2$ in $\Gamma$ and that $M=(\pref 1 {C_1}\cup
    \pref 2 {C_2},\sigma,W)$ is reachable from this marking.  There
    exist $W_{01}$ and $W_{02}$ forming an ownership split of $W_0$ such
    that the marking $(\ic{t_1},\sigma_0,W_{01})$ of $\osem{t_1}\Gamma$
    satisfies $\phi_1$ in $\Gamma$ and the marking
    $(\ic{t_2},\sigma_0,W_{02})$ of $\osem{t_2}\Gamma$ satisfies $\phi_2$ in
    $\Gamma$.  By assumption, $\Gamma\ent\hos{1}$ and
    $\Gamma\ent\hos{2}$, so according to Theorem
    \ref{theorem:soundness} no violating marking is reachable from
    either of these markings.
  
    Let $s_1$ and $s_2$ be synchronized events in $\osem{t_1\pll
      t_2}\Gamma$.  If $s_1=\sync{(\pref 1 {e_1})}{u_1}$ and
    $s_2=\sync{(\pref 1 {e_2})}{u_2}$ for some $e_1,e_2\in \ev{t_1}$
    and interference events $u_1$ and $u_2$ in $\osem{t_1}\Gamma$, the
    result follows routinely from the induction hypothesis, and
    similarly if $s_1$ and $s_2$ both arise from events of
    $\nsem{t_2}$.  Suppose instead that there exist $e_1\in\ev{t_1}$,
    $e_2\in\ev{t_2}$ and interference events $u_1$ and $u_2$ such that
    $s_1=\sync{(\pref 1 {e_1})}{u_1}$ and $s_2=\sync{(\pref 2
      {e_2})}{u_2}$.
  
    Suppose first that in the net $\osem{t_1\pll t_2}\Gamma$ we have
    \[M\ltr{s_1}
    (\pref{1}{C_1'}\cup \pref 2{C_2'},\sigma',W')\ltr{s_2}
    (\pref{1}{C_1''}\cup \pref 2{C_2''},\sigma'',W'').
    \]
    A simple induction applying the parallel decomposition lemma
    (Lemma \ref{lemma:pardecomp}) along the path to $M$ shows that
    there exist $W_1$ and $W_2$ that form an ownership split of $W$
    such that
    \[ (C_1,\sigma,W_1)\ltr{\sync{e_1}{u_1}}(C_1',\sigma',W_1')
    \ltr{u_2} (C_1'',\sigma'',W_1'') \] in $\osem{t_1}\Gamma$ for some
    $W_1',W_1''$.  By Lemma \ref{lemma:sepint}, the separation
    property holds for $\sync{e_1}{u_1}$ and $u_2$ in
    $(C_1,\sigma,W_1)$; consider how it might hold.  If
    $\sync{e_1}{u_1}$ deallocates a location that $u_2$ allocates,
    then $s_1$ deallocates a location that $s_2$ allocates, so the
    separation property holds for $s_1$ and $s_2$.  The argument is
    similar for all the other cases where $\sync{e_1}{u_1}$ and $u_2$
    are not independent.  Suppose instead that $\sync{e_1}{u_1} I
    u_2$.  The event $u_2$ has concession in the marking
    $(C_1,\sigma,W_1)$ by virtue of the fact that the occurrence of
    independent events in a run can be interchanged (Proposition
    \ref{lemma:indepswap}).  Consider the marking $(\pref 1
    {C_1}\cup \pref 2 {C_2},\sigma,W_1)$ of $\osem{t_1\pll t_2}\Gamma$; this is
    straightforwardly seen to be consistent.  The event $s_1$ is
    readily seen using Lemma \ref{lemma:gluesim} to have concession in
    this marking, as does $u_2$.  The event $\pref 2 {e_2}$ is
    control-independent from $\pref 1 {e_1}$, so by Lemma
    \ref{lemma:sepsync} we have $s_1 I s_2$, as required.

    Now suppose that in the net $\osem{t_1\pll t_2}\Gamma$ we have
    \[M\ltr{s_1}
    (\pref{1}{C_1'}\cup \pref 2{C_2'},\sigma',W')
    \qquad \tand \qquad M\ltr{s_2}
    (\pref{1}{C_1''}\cup \pref 2{C_2''},\sigma'',W'').
    \]
    A simple induction applying the parallel decomposition lemma
    (Lemma \ref{lemma:pardecomp}) along the path to $M$ shows that
    there exist $W_1$ and $W_2$ that form an ownership split of $W$
    such that
    \[ (C_1,\sigma,W_1)\ltr{\sync{e_1}{u_1}}(C_1',\sigma',W_1')
    \qquad \tand \qquad (C_1,\sigma,W_1) \ltr{u_2}
    (C_1'',\sigma'',W_1'') \] in $\osem{t_1}\Gamma$ for some
    $W_1',W_1''$.  By Lemma \ref{lemma:sepint}, the separation
    property holds for $\sync{e_1}{u_1}$ and $u_2$ in
    $(C_1,\sigma,W_1)$; consider how it might hold.  If
    $\sync{e_1}{u_1}$ allocates a location that $u_2$ also allocates,
    then $s_1$ allocates a location that $s_2$ allocates, so the
    separation property holds for $s_1$ and $s_2$.  The argument is
    similar for all the other cases where $\sync{e_1}{u_1}$ and $u_2$
    are not independent.  Suppose instead that $\sync{e_1}{u_1} I
    u_2$.  Consider the marking $(\pref 1 {C_1}\cup \pref 2
    {C_2},\sigma,W_1)$ of $\osem{t_1\pll t_2}\Gamma$; this is readily
    seen to be consistent.  The event $s_1$ has concession in this
    marking as does $u_2$.  The event $\pref 2 {e_2}$ is
    control-independent from $\pref 1 {e_1}$, so by Lemma
    \ref{lemma:sepsync} we have $s_1 I s_2$, as required.\medskip

  \noindent The remaining cases of the proof follow relatively
  straightforwardly by induction.  The case for \rn{L-Res} requires an
  observation along the lines of Lemma \ref{lemma:semequivar}; that,
  for any term $t$ with $\fv{t}\subseteq\{\rvar\}$ and resources
  $\res,\res'\not\in\dom(\Gamma)$, if the separation property holds
  for any two synchronized events of $\osem{[\res/\rvar]t}\Gamma$ in
  any marking reachable from any initial marking satisfying $\phi$ in
  $\Gamma$ then it also holds for $\osem{[\res'/\rvar]t}\Gamma$.
  
  The proof for the rule \rn{L-Seq} follows straightforwardly by
  induction using Lemma \ref{lemma:seqpath} except in the second (and
  symmetric third) cases of the definition of the separation property,
  where there are reachable markings $M,M',M''$ such that
  $M\ltr{s_1}M'\ltr{s_2}M''$ and there exist events $e_1\in\ev{t_1}$
  and $e_2\in\ev{t_2}$ and interference events $u_1,u_2$ such that
  $s_1=\sync{(P\gluel \pref 1 {e_1})}{u_1}$ and $s_2=\sync{(P\gluer
    \pref 2 {e_2})}{u_2}$ for $P=\pref 1 {\tc{t_1}} \times \pref 2
  {\ic{t_2}}$.  In this case, it follows from Lemma
  \ref{lemma:seqpath} and Lemma \ref{lemma:ictc} that the events $s_1$
  and $s_2$ are not control-independent.
\qed
\end{thm}

The result can be applied, using Lemma \ref{lemma:connecsim} and the
observation that $\sync{e_1}{u_1} I \sync{e_2}{u_2}$ implies that $e_1 I
e_2$, to obtain a similar result for the net semantics of terms
without ownership.
\begin{cor}
  Let $t$ be a closed term.  Suppose that $\emptyset\ent\hos{}$ and
  that $\sigma_0=(D_0,L_0,\emptyset,\emptyset)$ is a state for which
  $D_0\models\phi$.  If $M$ is a marking reachable from
  $(\ic{t},\sigma_0)$ in $\nsem{t}$ and $e_1$ and $e_2$ are
  control-independent events then:
\begin{enumerate}[$\bullet$]
\item If $M\ltr{e_1} M_1 \ltr{e_2} M'$ then either $e_1$ and $e_2$ are
  independent or $e_1$ releases a resource or a location that $e_2$
  correspondingly takes or allocates, or $e_1$ makes non-current a
  resource that $e_2$ makes current.
\item If $M\ltr{e_1} M_1$ and $M\ltr{e_2} M_2$ then either $e_1$ and
  $e_2$ are independent or $e_1$ and $e_2$ compete either to make current
  the same resource, acquire the same resource or to allocate the
  same location.\qed
\end{enumerate}

\end{cor}

\subsection{Incompleteness}
\label{sec:incompleteness}
The separation result highlights an important form of possible
interaction between concurrent processes.  Observe that, although
there is neither conflict nor causal dependence arising from heap
events (and hence the processes are race-free in the sense of
Brookes), there may be interaction through the occurrence of
allocation and deallocation events.  One may therefore give judgements
for parallel processes that interact without using critical regions.
Suppose, for example, that we have a heap
\[D=\{\hvl{_0}{\lcon_1},\hvl{_1}{1},\hvl{_2}2,\hvl{_3}3,\hvl{_4}4\}.\]
For any processes $t_1$ and $t_2$ such that $t_1$ does not deallocate $\lcon_1$, if
we place the process
\[ t_1;\dealloc{\lcon_0}\] in parallel with
\[
\alloc{\lcon_2}; \quad\while{[\lcon_2]\neq \lcon_1}
  \ldo 
  \alloc{\lcon_2}\lod;\quad t_2,
  \]
  the process $t_2$ only takes place once $t_1$ has terminated, and
  possibly never, even if $t_1$ terminates.  This arises from the fact
  that the loop in the second process will only exit when location
  $\lcon_1$ is allocated by the command $\alloc{ \lcon_2}$; this can
  only occur once $\dealloc {\lcon_0}$ makes $\lcon_1$ non-current and
  therefore available for allocation by $\alloc{\lcon_2}$.  Denote
  this process $\mathrm{seq}(t_1,t_2)$.
  
  We can use this to show that concurrent separation logic is
  incomplete with respect to our definition of validity: Let $t_1$ be
  the assignment $[\lcon_3]:= 1$ and $t_2$ be $[\lcon_3]:=2$.  Define
  the formula
  \[\delta \eqd \lcon_0\mapsto\lcon_1\mand\lcon_1\mapsto\something\mand
  \lcon_2\mapsto\something\mand\lcon_3\mapsto
  \something\mand\lcon_4\mapsto\something.\] We have $\emptyset\models
  \ho{\delta}{\mathrm{seq}(t_1,t_2)}{\lcon_3\mapsto 2\mand\true}$
  since, whenever $\mathrm{seq}(t_1,t_2)$ terminates, the assignment
  $[\lcon_3]:=2$ always occurs after the assignment $[\lcon_3]:=1$.
  The separation property holds in any marking reachable from any heap
  initially satisfying $\delta$.  It can be shown that
\[\emptyset\not\ent\ho{\delta}{\mathrm{seq}(t_1,t_2)}{\lcon_3\mapsto
  2\mand\top},\] so the logic is incomplete, even for processes
satisfying the separation property.

There are also examples of incompleteness where neither process
accesses a common heap location along any run: Let
\begin{eqnarray*}
  t_1' &=& \alloc{\lcon_3};\while{[\lcon_3]\neq \lcon_5}\ldo \alloc{\lcon_3}\lod \\
  t_2' &=& \alloc{\lcon_4};([\lcon_4]=\lcon_5).\co{skip} + ([\lcon_4]\neq \lcon_5).\co{diverge},
\end{eqnarray*}
for the previous definition of $\co{diverge}$ and the obvious
definition of skip, $\asem{\co{skip}}=\{(\emptyset,\emptyset)\}$.
Since the location $\lcon_5$ is always current following termination
of $t_1'$ from $D$, process $t_2'$ always diverges.  We have
\[ \emptyset\models \ho{\delta} {\mathrm{seq}(t_1',t_2')}{\false}.\]
However, there are no $\delta_1,\delta_2$ such that
$\delta$ is logically equivalent to $\delta_1\mand\delta_2$ and
$\emptyset\models\ho{\delta_2}{t_2'}{\false}$, which would be necessary
if it were possible to prove
$\emptyset\ent \ho{\delta}{\mathrm{seq}(t_1',t_2')}{\false}$.

\section{Refinement}
\label{sec:refinement}

  \newcommand{\sv}[2]{#1\mapsto #2}

  As we remarked in the introduction, the atomicity assumed of
  primitive actions, also called their \emph{granularity}, is of
  significance when considering parallel programs.  For example,
  suppose that  the concurrent program
  \[ 
  \begin{array}{ll}
    & [\lcon]:=[\lcon']+1 \\
    \pll&
    \ ([\lcon]'\neq[\lcon]).\co{diverge} +
    ([\lcon]'=[\lcon]).\co{skip}
  \end{array}
  \]
  runs from the heap $\{\sv \lcon 0,\sv{\lcon'} 1\}$.  Given the prior
  interpretations of $\co{skip}$ and $\co{diverge}$, we might conclude
  that the program never terminates since the assignment
  $[\lcon]:=[\lcon']+1$ maintains the property through execution that
  $\lcon$ and $\lcon'$ hold different values.

  It may not, however, be reasonable to assume that the assignment is
  executed atomically.  For instance, the processor on which the
  process runs might have primitive actions for copying the values
  held in memory locations and for incrementing them, but not for
  copying \emph{and} incrementing in one clock step.  The process
  $[\lcon]:=[\lcon']+1$ might therefore be compiled to execute as
  $[\lcon]:=[\lcon'];[\lcon]:= [\lcon]+1$.  Quite clearly, the process
  \[ 
  \begin{array}{ll}
    & [\lcon]:=[\lcon'];[\lcon]:=[\lcon]+1 \\
    \pll&
    \ ([\lcon]'\neq[\lcon]).\co{diverge} +
    ([\lcon]'=[\lcon]).\co{skip}
  \end{array}
  \]
  \emph{may} terminate, so we failed to exhibit a proper degree of
  caution when asserting that it would fail to terminate.

  In \cite{REYNOLDS04C}, Reynolds proposes a form of trace semantics
  that regards the occurrence of uncontrolled interference between
  concurrent processes as `catastrophic'.  The motivation behind this
  is the race freedom property arising from concurrent separation
  logic \cite{brookes:concur}: in the semantics of a proved process
  running from a suitable initial state, no uncontrolled interference
  may occur.  Reynolds' observation is that, in this situation,
  judgements may be made that are insensitive to atomicity.

  Within our net model we can provide a form of \emph{refinement},
  similar to that of \cite{vanglabbeekgoltz} but suited to processes
  executing in a shared environment, that begins to capture these
  ideas.  Importantly, the property required to apply the refinement
  operation may be captured directly in terms of independence, with no
  changes to our semantics.  We will relate the nets representing
  processes with different levels of atomicity by regarding them as
  alternative substitutions into a \emph{context}.  We will then give
  a condition on substitutions led by Theorem \ref{theorem:separation}
  to show that any partial correctness assertion made for one of the
  nets also holds for the other.

  The treatment of substitution requires some restrictions to be
  placed on the nets we consider.  In the remainder of this section
  and in Appendix \ref{app:refinement} where we present the technical
  details of this section, we require that all embedded nets satisfy
  the structural properties described in Lemma \ref{lemma:embstruc}
  and Definition \ref{def:dagger}.

\newcommand{\khole}{\ensuremath{{[-]}}}
\newcommand{\init}{_\mathsf{i}}
\newcommand{\term}{_\mathsf{t}}
\begin{defi}[Context]
  Define a \emph{context} $K$ to be a embedded net with a
  distinguished event $\khole$.  The event $\khole$ is such that
  $\pre{\post{\khole}}\subseteq\cset{C}$ and its pre- and
  postconditions form disjoint, nonempty sets.

\end{defi}

\newcommand{\cont}[1]{\sn{Cc}(#1)} \newcommand{\state}[1]{\sn{Sc}(#1)}
We may now construct the net representing the substitution of a net
$N$ for the hole in a context $K$.  We shall assume that, as in the
semantics for terms, the two nets are formed with the same sets of
conditions.  As the nets are extensional (we regard an event simply as
its set of preconditions paired with its set of postconditions), all
that we need to specify is the events of the net and its initial and
terminal markings of control conditions.

\renewcommand{\pref}[1]{#1 \mathop{:}}
\begin{defi}[Substitution]
\label{def:contextsub}
  Let $K$ be a context and $N$ an embedded net.  Define the sets
  \[ 
  P\init\eqd \pref 1 \pre\khole \times \pref 2\ic{N} \qquad 
  P\term \eqd \pref 1 \post{\khole}\times\pref 2 \tc{N}.
  \]
  The substitution $K[N]$ is defined to be the embedded net with:
 \begin{eqnarray*}
   \ev{K[N]} &\eqd& (P\init\cup P\term)\gluel\pref 1(\ev K \setminus\{\khole\})
   \cup (P\init\cup P\term)\gluer \pref 2\ev N\\
   \ic{K[N]} &\eqd& (P\init\cup P\term)\gluel \pref 1\ic K\\
   \tc{K[N]} &\eqd& (P\init\cup P\term)\gluel \pref 1\tc K
 \end{eqnarray*}
\end{defi}
To see the definition at work, consider the following example.  We
elide details of the action of events on state conditions, which is
unaffected by the substitution operation.

\begin{exa}
  In the following example substitution, we depict the hole $\khole$
  as a hollow rectangle.
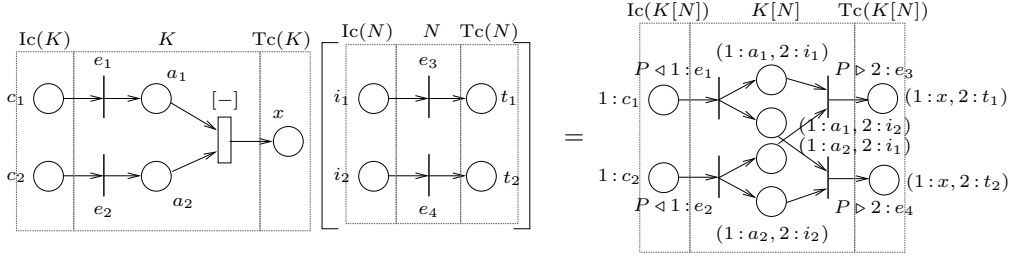
\begin{center}
  \input{subs.pstex_t}
\end{center}
\end{exa}

\newcommand{\pto}{\mathop{\Downarrow}} 
\begin{defi}
  Let $\pi$ be a sequence of events of the net $N$.  Sequence $\pi$ is
  said to be \emph{complete from $\sigma$ to $\sigma'$} if
  \[\pi \ty (\ic{N},\sigma)\ltr{}^* (\tc{N},\sigma'). \]
  Write $N\ty
  \sigma\pto \sigma'$ if there exists a complete sequence from
  $\sigma$ to $\sigma'$ in $N$.
\end{defi}

Using this definition, we can define a notion of complete trace
equivalence $\simeq$ as:
\[ N_1\simeq N_2 \quad\tiff\quad
(\forall\sigma,\sigma')N_1\ty \sigma\pto \sigma' \iff N_2\ty
\sigma\pto \sigma'.\]
We wish to constrain $K$, $N_1$ and $N_2$ appropriately so that
if $N_1 \simeq N_2$ then $K[N_1]\simeq K[N_2]$.
\begin{exa}
  Write, in the obvious way, $-$ for the action term that will be
  interpreted as forming the hole of a context.  Define
  \begin{eqnarray*}
    K &\eqd& \nsem{-\pll 
      \ ([\lcon]'\neq[\lcon]).\co{diverge} +
      ([\lcon]'=[\lcon]).\co{skip}}\\
    N_1 &\eqd& \nsem{[\lcon]:=[\lcon']+1}\\
    N_2 &\eqd& \nsem{[\lcon]:=[\lcon'];[\lcon]:=[\lcon]+1}.
  \end{eqnarray*}
  We clearly have $N_1\simeq N_2$, but $K[N_1] \not\simeq K[N_2]$
  since \[K[N_1]\ty \{\sv \lcon 0, \sv{\lcon'} 1\}\not{\!\pto} \{\sv
  \lcon 2, \sv{\lcon'} 1\}\] but
  \[K[N_2]\ty \{\sv \lcon 0, \sv{\lcon'} 1\}\pto \{\sv \lcon
  2, \sv{\lcon'} 1\}.\]
\end{exa}

Return to the general case for a substitution $K[N]$.
Intuitively, if the substituend $N$ were an atomic event, it would
start running only if the conditions $P\init$ were marked and $P\term$
were not.  There are two distinct ways in which the context $K$ can
affect the execution of $N$.  Firstly, it might affect the marking of
conditions in $P\init$ or $P\term$ whilst $N$ is running.  Secondly,
it might change the marking of state conditions in a way that affects
the execution of $N$.  An instance of the latter form of interference
is seen in the preceding example.  We now define a form of constrained
substitution, guided by Theorem \ref{theorem:separation}, so that $N$
is not subject to these forms of interference.

Say that a control condition $c$ of $K[N]$ is \emph{internal} to $N$
if $c=\pref 2{c_2}$ where $c_2$ is a pre- or a postcondition of an
event of $N$ that is not in $\ic{N}$ or $\tc{N}$.  Given a marking $M$
of $K[N]$, say that $N$ is \emph{active} if $P\init\subseteq M$ or
there exists an internal condition of $N$ in $M$.
\begin{defi}
  For a given marking of state conditions $\sigma$, we say that $K[N]$
  is a \emph{non-interfering substitution} if, for all markings $M$
  reachable from $(\ic{K[N]},\sigma)$:
  \begin{enumerate}
  \item if $P\init\subseteq M$ then $P\term\cap M=\emptyset$, and
  \item if $N$ is active in $M$ then no enabled event of $K$ has a
    pre- or postcondition in $P\init$ or $P\term$, and
  \item if $M\ltr{e_1}M_1\ltr{e_2}M'$, one of $e_1$ and $e_2$ is from
    $N$ and the other is from $K$ and $N$ is active in $M$ and $M_1$,
    then $e_1$ and $e_2$ are independent.
  \end{enumerate}
\end{defi}

\begin{thm}
  \label{theorem:refinement}
  If $N_1\simeq N_2$ and $K[N_1]$ and $K[N_2]$ are non-interfering
  substitutions from state $\sigma$, then, for any $\sigma'$:
  \[ K[N_1]\ty \sigma\pto\sigma' \quad\tiff\quad
  K[N_2]\ty\sigma\pto\sigma'.\] \proof Appendix \ref{app:refinement},
  Theorem \ref{proof:refinement}.\qed
\end{thm}

The refinement operation defined in this section allows us to change
the granularity of heap actions by substituting the occurrence of an
action in the original net with a net representing the actual
implementation of the action, but only once it has been shown that the
noninterference property holds for both the original net and for the
net formed.  The operation might be a key to proving Reynolds'
observation that an occurrence of an action $\alpha$ in the term $t$
can be replaced by a term with the same overall behaviour as $\alpha$
without affecting the validity of the judgement $\Gamma\ent\hos{}$.

\section{Related work and conclusions}
\label{sec:related}
\label{sec:conclusion}
The first component of this work provides an inductive definition of
the semantics as a net of programs operating in a (shared) state. This
is a relatively novel technique, but has in the past been applied to
give the semantics of a language for investigating security protocols,
SPL \cite{spl}, though our language involves a richer collection of
constructs.  Other independence models for terms include the Box
calculus \cite{box} and the event structure and net semantics of CCS
\cite{students,evccs,wn:mfc} (\cite{students} was, to our knowledge,
the first Petri net denotational semantics of CCS), though
these model interaction as synchronized communication rather than
occurring through shared state.  We hope that the novel Petri net
semantics presented here and in \cite{spl} can be the start of
\emph{systematic and comprehensive} methods to attribute structural
Petri net semantics to a full variety of programming languages,
resulting in a Petri net companion to Plotkin's structural operational
semantics (SOS) based on transition systems \cite{plotkin:sos}.
Paralleling the (inductive) definitions of data and transitions of SOS
would be (inductive) definitions of conditions and events of Petri
nets.

The proof of soundness of separation logic here is led by Brookes'
earlier work \cite{brookes:soundness}.  There are a few minor
differences in the syntax of processes, including that we allow the
dynamic binding of resource variables.
Another minor difference between the programming language and
logic considered here and that introduced by O'Hearn and proved sound
by Brookes is that we do not distinguish \emph{stack variables}. These
may be seen as locations to which other locations may not point and are
the only locations that terms can directly address.  In Brookes'
model, as in \cite{ohearn:rclr}, interference of parallel processes through stack variables is
constrained by the use of a side condition on the rule
rather than using the concept of ownership (the area of current
research on `permissions' \cite{bornat, bornatcalcagnoyang,
brookes:permissions} promises a uniform approach).  In particular, the
rule allows the concurrent reading of stack locations.  Though we have
chosen not to include stack variables in our model in order to
highlight the concept of ownership, our model and proofs could be
easily extended to deal with them.  Concurrent reading of memory
would be at the cost of a more sophisticated notion of independence
that allowed independent events to access the same condition providing
that neither affects the marking of that condition.

More notably, at the core of Brookes' work is a `local enabling
relation', which gives the semantics of programs over a restricted set
of `owned' locations.  Our notion of validity involves maintaining a
record of ownership and using this to constrain the occurrence of
events in the interference net augmented to the process.  This allows
the intuition of ownership in O'Hearn's introduction of concurrent
separation logic \cite{ohearn:rclr} to be seen directly as
constraining interference.  Though the relationship between our model
and Brookes' is fairly obvious, we believe that our approach leads to
a clearer parallel decomposition lemma, upon which the proof of
soundness of the logic critically stands.

The most significant difference between our work and Brookes' is that
the net model captures, as a primitive property, the independence of
parallel processes enforced by the logic.  We have used this property
to define a straightforward, yet general, form of refinement suited to
changing the atomicity of commands within the semantics of a term.
This is in contrast to \cite{brookes:footstep}, which gives a new
semantics to race-free processes that abstracts entirely away from
attaching any form of atomicity to the semantics of heap actions.  As
said at the end of the previous section, we hope to show that the
refinement operation can be applied to change the atomicity of any
action occurring within any process running from a suitable initial
state proved using to the rules of concurrent separation logic.

Our characterization of `separation' arising from the logic is much
finer than that obtained from the existing proof of race freedom, for
example showing that interaction between parallel processes may occur
through allocation and deallocation.  This is significant, as such
interaction leads to examples of the incompleteness of concurrent
separation logic.

There are a number of other areas for further research in addition to
those mentioned above.  One interesting consideration is the necessity
(or otherwise) of precision in the proof of soundness of the logic.
In forthcoming work, we hope to give a form of game semantics for the
logic and a soundness proof without precision in the absence of the
Hoare's Law of Conjunction $\rn{L-Conjunction}$. 
Another area of interest is whether symmetry present in our semantics
for allocation and resource declaration might be exploited properly
to obtain more compact nets to represent processes.

\section*{Acknowledgements}
It is a pleasure to thank Peter O'Hearn and Matthew Parkinson for a
number of helpful discussions during the development of this work.  We
would also like to thank the anonymous referees of a draft of this
paper and the anonymous referees of the conference version of this
paper \cite{lics} for their constructive suggestions.

\bibliographystyle{alpha}

\bibliography{shortbib}

  \newpage
  \appendix

\section{Refinement}
\newcommand{\IN}{\ic N}
\newcommand{\TN}{\tc N}
\newcommand{\IK}{\ic K}
\newcommand{\TK}{\tc K}
\newcommand{\IKN}{\ic{K[N]}}
\newcommand{\TKN}{\tc{K[N]}}

\label{app:refinement}

A sequence of events $\pi=(e_1,\ldots,e_n)$ considered from a marking
$M$ can be thought of equivalently as a sequence $M\ltr{e_1} M_1
\ldots \ltr{e_n} M_n$.  To describe the structure of such sequences,
we shall say that $\pi$ from marking $M$ is of form $\Pi_1\cdot \Pi_2$
if there exist $\pi_1$ and $\pi_2$ such that $\pi=\pi_1\cdot\pi_2$,
where $\cdot$ denotes the obvious concatenation of sequences, and
$\pi_1$ is of form $\Pi_1$ from marking $M$ and $\pi_2$ is of form
$\Pi_2$ from the marking obtained by following $\pi_1$ from $M$.
Sequence $\pi$ is of form $\Pi^*$ if it is the concatenation of a
finite number of sequences, each of form $\Pi$.

Throughout this section, when we consider the substitution $K[N]$ let
$P\init$ and $P\term$ be defined as in Definition
\ref{def:contextsub}:
\begin{eqnarray*}
  P\init &\eqd& \pref 1 {\pre\khole} \times \pref 2 {\ic{N}}\\
  P\term &\eqd& \pref 1 {\post\khole} \times \pref 2 {\tc{N}}.
\end{eqnarray*}
Any reachable marking of conditions of the net can be partitioned into
two sets: conditions that occur solely within $K$ and conditions that
are either $N$-internal or in $P\init$ or $P\term$.  Formally, a
condition $c$ is a \emph{$K$-condition} if $c=\pref 1 c_1$ for some
condition $c_1$ of $K$ not in $\pre{\post{\khole}}$.  A condition $c$
is an \emph{$N$-condition} if either $c\in P\init\cup P\term$ or
$c=\pref 2 c_2$ for some condition $c_2$ of $N$ not in $\ic{N}\cup
\tc{N}$.  Recall that  we call $\pref 2 c_2$ an
$N$-internal condition.  It is easy to see that, for any $\sigma_0$,
from the marking $(\ic{K[N]},\sigma_0)$ only $K$- or $N$-control
conditions may be marked: If $(C,\sigma)$ is a reachable marking of
$K[N]$, we have $C=C_N\cup C_K$ for some marking $C_N$ of
$N$-conditions and some marking $C_K$ of $K$-conditions.  We shall
frequently use the notation $(C_N, C_K)$ for a marking of control
conditions, where $C_N$ comprises only $N$-conditions and $C_K$
comprises only $K$-conditions.

Henceforth, when considering a substitution $K[N]$, we shall refer to
an event $e$ as being an $N$-event if it is equal to $(P\init\cup
P\term)\gluer \pref 2 e_2$ for some $e_2$ in $N$.  Otherwise, it is a
$K$-event.  
A little care is necessary since an event in the net
$K[N]$ might arise from both $K$ and $N$ if there are events $e$ and
$e'\neq \khole$ of $N$ and $K$, respectively, with the same effect on state
conditions and:
\[ 
\begin{array}{lcl @{\qquad} rcl}
  \pres C e &=& \ic{N}, &\poss C e &=& \tc{N}\\
  \pres C {e'} &=& \pre\khole, &\poss C {e'} &=& \post \khole.
\end{array}
\]
Throughout the remainder of this section, for simplicity we shall
require that the substitution $K[N]$ has no such events.  This
restriction may be lifted with little effect on the development
so-far by allowing the net formed to be non-extensional, or by
considering this as a special case when demonstrating properties of
the net $K[N]$

\begin{lem}
  \label{lemma:kdisj}
  In $K[N]$, no $K$ event has as a either a pre- or a postcondition an
  $N$-internal condition.
  \begin{proof}
    Immediate from the definition of substitution $K[N]$.
  \end{proof}
\end{lem}

Recall that a marking $(C_N, C_K,\sigma)$ reachable from
$(\ic{K[N]},\sigma_0)$ is $N$-active if either there is an
$N$-internal condition in $C_N$ or if $C_N=P\init$.  It is useful to
further classify the markings of conditions in $C_N$ according to
whether they support the occurrence of $N$- or $K$-events on the
conditions $P\init$ and $P\term$:
\begin{defi}
  A marking $(C_N, C_K,\sigma)$ of $K[N]$ is an $N$-marking 
if 
for all $a,a'\in
  \pre\khole$, $x,x'\in \post\khole$, $i\in\ic{N}$ and $t\in\tc{N}$:
  \begin{itemize}
  \item if $(\pref 1 a, \pref 2 i)\in C_N$ then $(\pref 1 a',\pref 2
    i)\in C_N$, and 
  \item if $(\pref 1 x, \pref 2 t)\in C_N$ then $(\pref 1 x',\pref 2
    t)\in C_N$.
  \end{itemize}
  A marking $(C_N\cup C_K,\sigma)$ of $K[N]$ is a $K$-marking if there
  is no $N$-internal condition marked, and furthermore, for all $a\in
  \pre\khole$, $x\in \post\khole$, $i,i'\in\ic{N}$ and
  $t,t'\in\tc{N}$:
  \begin{itemize}
  \item if $(\pref 1 a, \pref 2 i)\in C_N$ then $(\pref 1 a,\pref 2
    i')\in C_N$, and
  \item if $(\pref 1 x, \pref 2 t)\in C_N$ then $(\pref 1 x,\pref 2
    t')\in C_N$.
  \end{itemize}
\end{defi}

From a marking of control conditions $(C_N, C_K)$, we can extract
markings of control conditions for the nets $N$ and $K$.  We define
$\rho_N(C_N)$ to be the marking of $N$ obtained from $(C_N,C_K)$,
which is not dependent on the marking $C_K$ of $K$-conditions, and
$\rho_K(C_N,C_K)$ for the marking of $K$ obtained from $(C_N,C_K)$,
which is dependent on the marking of $N$-conditions (namely, the
marking of $N$-conditions in $P\init\cup P\term$). 

For a marking $C$ of the context $K$, we define $\theta_K(C)$ to be
the corresponding marking of $K[N]$.  For a marking $C'$ of the net
$N$, we define $\theta_N(C')$ to be the marking of $N$-conditions in
the net $K[N]$ corresponding to $C'$.
\begin{defi}
  \label{def:contproj}
  Let $K[N]$ be any substitution.  For any marking $C_N$ of
  $N$-conditions and $C_K$ of $K$-conditions, define
  \begin{eqnarray*}
    \rho_K(C_N,C_K) &\eqd&
    \begin{array}{rrll}
      \{& a \in \pre\khole &
      \st \forall i\in \ic N.(\pref 1 a,\pref 2 i)\in C_N &\}\\
      \cup 
      \{& x \in \post\khole &
      \st \forall t\in \tc N.(\pref 1 x,\pref 2 t)\in C_N &\}\\
      \cup 
      \{&c \not\in \pre\khole\cup \post\khole 
      & \st \pref 1 c \in C_K&\}
    \end{array}
    \\
    \rho_N(C_N) &\eqd &
    \begin{array}{rrll}
      \{&i \in \ic{N} &
      \st \forall a\in \pre\khole.(\pref 1 a,\pref 2 i)\in C_N &\}\\
      \cup 
      \{ &t \in \tc{N} 
      & \st \forall x\in \post\khole.(\pref 1 x,\pref 2 t)\in C_N &\}\\
      \cup 
      \{&c \not\in \ic{N}\cup \tc{N} 
      & \st \pref 2 c \in C_N&\}.
    \end{array}
  \end{eqnarray*}
  For any marking $C$ of control conditions of the net $K$ and marking
  $C'$ of control conditions of the net $N$, define
  \begin{eqnarray*}
    \theta_K(C) &\eqd& P\gluel \pref 1 C\\
    \theta_N(C) &\eqd& P\gluer \pref 2 C.
  \end{eqnarray*}
  For an event $e$ of $K[N]$, define $\rho_N(e)=e'$ for the unique
  $e'$ such that $e=(P\init\cup P\term)\gluer \pref 2 e'$.  For an
  event $e$ of $N$, define $\theta_N(e)=(P\init\cup P\term)\gluer\pref
  2 e$.  Define $\rho_K(e)$ and $\theta_K(e)$ similarly, apart from
  having $\theta_K(\khole)$ undefined.
\end{defi}
\begin{lem}
  \label{lemma:thetapres}
  For any marking $C$ of control conditions of $K$, the marking
  $\theta_K(C)$ is a $K$-marking in $K[N]$.  For any marking $C'$ of
  control conditions of $N$, the marking $\theta_N(C')$ is an
  $N$-marking in $K[N]$.
  \begin{proof}
    Immediate from the definitions.
  \end{proof}
\end{lem}

It is clear that $\rho_N$ and $\theta_N$ form a bijection between
$N$-events and $\ev N$.  It is also clear that $\rho_K$ and $\theta_K$
form a bijection between $K$-events and $\ev K \setminus\{\khole\}$.
On markings, the 
situation is a little more intricate:
\begin{lem} Let $K[N]$ be a substitution.  
  \label{lemma:contextbij}
  For any marking of control conditions $C_K\cup C_N$ of $K[N]$ that
  is a $K$-marking and any marking $C$ of control conditions of $K$:
  \[ \theta_K(\rho_K(C_K\cup C_N)) = C_K\cup C_N
  \quad\tand\quad \rho_K(\theta_K(C)) = C. \]
  For any marking of control conditions $C_K\cup C_N$ of $K[N]$ that
  is an $N$-marking and any marking $C$ of control conditions of $N$:
  \[ \theta_N(\rho_N(C_N)) = C_N
  \quad\tand\quad \rho_N(\theta_N(C)) = C. \]
\begin{proof}
  First, let $C$ be any marking of control conditions of $K$.  We
  shall show that $ \rho_K(\theta_K(C)) = C$.  Let $c$ be any control
  condition of the net $K$.  Since $K$ is an embedded net, by the
  restrictions imposed in Lemma \ref{lemma:embstruc} there are three
  distinct cases: $c\not\in\pre{\post{\khole}}$, $c\in\pre\khole$ or
  $c\in \post\khole$.  The first case is straightforward since the
  operation of $\theta_K$ on such conditions is to add a `$\pref 1$'-tag
  which is removed by $\rho_K$.  Now consider $c\in\pre\khole$; the
  case for $c\in\post\khole$ will be similar.  By the definition of
  $\theta_K$, since $\ic{N}$ is nonempty (again by Lemma
  \ref{lemma:embstruc}):
  \[
  c\in C\qquad \tiff \qquad \forall i\in \ic N.  (\pref 1 c, \pref 2
  i)\in \theta_K(C). 
  \]
  From the definition of $\rho_K$, we have $\forall i\in \ic N.
  (\pref 1 c, \pref 2 i)\in \theta_K(C)$ iff $c\in
  \rho_K(\theta_K(C))$. So $c\in C$ iff $c\in \rho_K(\theta_K(C))$.

  Now suppose that $(C_K,C_N)$ is a $K$-marking of the substitution
  $K[N]$.  Let $c$ be any condition of the net $K[N]$.  There are
  three distinct possible cases: $c\not\in P\init\cup P\term$, $c\in
  P\init$ or $c\in P\term$.  First, suppose that $c\not\in P\init \cup
  P\term$:
  \[
  \begin{array}{rclr}
    c\in C_N \cup C_K &
    \tiff&c \in C_K &\text{ (def.~of $K$-marking)}\\
    &\tiff&\exists c_1.(c_1 \in \rho_K(C_N,C_K) ~\tand~ c=\pref 1 c_1)
    &\text{ (def.~of $\rho_K$)}\\
    &\tiff&c\in\theta_K(\rho_K(C_N\cup C_K)) 
    &\text{ (def.~of $\theta_K$)}
  \end{array}
  \]
  Now suppose that $c\in P\init$, so $c=(\pref 1 a, \pref 2 i)$ for
  some $a\in \pre\khole$ and $i\in \ic N$:
  \[
  \begin{array}{rclr}
    c\in C_N \cup C_K &
    \tiff&
    \forall i' \in \ic N.\;(\pref 1 a, 2 i') \in C_N \cup C_K
    &\text{ (def.~of $K$-marking)}\\
    &\tiff& a\in \rho_K(C_N\cup C_K)
    &\text{ (def.~of $\rho_K$)}\\
    &\tiff& c\in\theta_K(\rho_K(C_N\cup C_K)) 
    &\text{ (def.~of $\theta_K$)}
  \end{array}
  \]
  We have a similar analysis if $c\in P\term$.  Hence
  $(C_K,C_N)=\theta_K(\rho_K(C_K,C_N))$.
  
  For any marking of control conditions $C$ of the net $N$ and any
  $N$-marking $(C_K,C_N)$,
  \[ \theta_N(\rho_N(C_N)) = C_N
  \quad\tand\quad \rho_N(\theta_N(C)) = C
  \]
  are shown similarly, this time with the first analysis considering
  conditions in $\ic{N}$, $\tc{N}$ and conditions not in either set.
\end{proof}
\end{lem}

\newcommand{\presc}[1]{\pres C {#1}}
\newcommand{\possc}[1]{\poss C {#1}}

\begin{lem}
\label{lemma:rhosim}
  Let $(C_K,C_N,\sigma)$ and $(C_K',C_N',\sigma')$ be markings of
  $K[N]$.  Suppose that $e$ is an event such that $(C_K,C_N,\sigma)
  \ltr e(C_K',C_N',\sigma')$.
  \begin{enumerate}
  \item If $e$ is a $K$-event and $(C_K,C_N)$ and 
        $(C_K',C_N')$ are $K$-markings then\\
    $(\rho_K(C_K,C_N),\sigma)\ltr{\rho_K(e)}
    (\rho_K(C_K',C_N'),\sigma')$ in $K$.
  \item If $e$ is an $N$-event
and $(C_K,C_N)$ and $(C_K',C_N')$ are $N$-markings
 then $C_K=C_K'$ and
    $(\rho_N(C_N),\sigma)\ltr{\rho_N(e)} (\rho_N(C_N'),\sigma')$ in
    $N$.
  \end{enumerate}

  \proof First consider (1).  The event $e$ is a $K$-event, so there
  is an event $e_1$ of $K$ such that $e_1\neq\khole$ and
  $e=(P\init\cup P\term)\gluel \pref 1 e_1$.  We have
  \[(C_N,C_K,\sigma)\ltr e (C_N',C_K',\sigma')\] in $K[N]$. By Lemma
  \ref{lemma:contextbij}, we have
  $\theta_K(\rho_K(C_N,C_K))=(C_N,C_K)$ and
  $\theta_K(\rho_K(C_N',C_K'))=(C_N',C_K')$.  From the definition of
  $\theta_K$, we therefore have 
  \[ ((P\init\cup P\term)\gluel \pref 1 \rho_K(C_N,C_K),\sigma)
  \ltr {(P\init\cup P\term)\gluel \pref 1 e_1}
  ((P\init\cup P\term)\gluel \pref 1 \rho_K(C_N',C_K'),\sigma').
  \]
  Using Lemma \ref{lemma:gluesim}, we may therefore conclude that
  \[ (\rho_K(C_N,C_K),\sigma)\ltr{e_1} (\rho_K(C_N',C_K'),\sigma')\]
  in $K$.  The proof of (2) is similar.  \qed
\end{lem}

\begin{lem}
\label{lemma:thetasim}
\begin{enumerate}
\item Let $C$ and $C'$ be markings of control conditions of $K$.  If
  $(C,\sigma)\ltr e (C',\sigma')$ in $K$ then $(\theta_K(C),\sigma)
  \ltr {\theta_K(e)} (\theta_K(C'),\sigma')$ in $K[N]$.
  
\item Now let $C$ and $C'$ be markings of control conditions of $N$.
  If $(C,\sigma)\ltr e (C',\sigma')$ in $N$ then $(\theta_N(C),
  C_K,\sigma) \ltr {\theta_N(e)} (\theta_N(C'), C_K,\sigma')$ in
  $K[N]$ for any marking $C_K$ of $K$-conditions.
\end{enumerate}
\newcommand{\tk}[1]{\theta_K(#1)}

\proof
First consider (1).  Suppose that $(C,\sigma)\ltr e (C',\sigma')$ in
$K$ for some event $e\neq \khole$.  By Lemma \ref{lemma:gluesim}, we
have
\[ ((P\init\cup P\term)\gluel \pref 1 C,\sigma)
\ltr{(P\init\cup P\term)\gluel \pref 1 e}
((P\init\cup P\term)\gluel \pref 1 C',\sigma')
\]
in $K[N]$.  Since $\theta_K(C)=(P\init\cup P\term)\gluel \pref 1 C$,
and similarly for $C'$ and $e$, we therefore have
\[ (\theta_K(C),\sigma)\ltr{\theta_K(e)} (\theta_K(C'),\sigma'),\]
as required.  The proof of (2) is similar.\qed
\end{lem}

We are now able to characterize the runs of the net $K[N]$ when a
non-interfering substitution is formed.
\begin{lem}
  \label{lemma:pathchar}
  Let $K[N]$ be a non-interfering substitution from $\sigma_0$.  Any
  complete sequence $\pi$ from $(\ic{K[N]},\sigma_0)$ is of the form
  $\Pi_0\cdot(\Pi_1\cdot\Pi_0)^*$, where:
  \begin{itemize}
  \item $\Pi_0$ ranges over sequences consisting of $K$-events between
    $K$-markings.
  \item $\Pi_1$ ranges over nonempty sequences $\pi_1$ of any events
    between $N$-markings, where no $K$-event uses any condition in
    $P\init$ or $P\term$.  If $(C_N\cup C_K,\sigma)$ and $(C_N'\cup
    C_K',\sigma')$ are the initial and final markings of $\pi_1$,
    respectively, then $C_N=P\init$ and $C_N'=P\term$.  The first
    event of $\pi_1$ is an $N$-event and the final event of $\pi_1$ is
    also an $N$-event.
  \end{itemize}

  \begin{proof}
    We first show that any sequence $\pi$ in $N$ from
    $(\ic{K[N]},\sigma_0)$ is of the form $\Pi_0\cdot(\Pi_1\cdot\Pi_0)^*$ or
    $\Pi_0\cdot(\Pi_1\cdot\Pi_0)^*\cdot\Pi_1'$ by induction on the length of
    sequence, where a sequence is of form $\Pi_1'$ if:
    \begin{itemize}
    \item it is a sequence of $K$- and $N$-events between $N$-markings
      where no $K$-event uses any condition in $P\init$ or $P\term$, and
    \item if $(C_N,C_K,\sigma)$ is the initial marking of $\pi_1$ then
      $C_N=P\init$, and the first event of $\pi_1$ is an $N$-event.
    \end{itemize}
    We shall simultaneously show that if $\pi\ty
    (\ic{K[N]},\sigma_0)\ltr{}^* (C_N,C_K,\sigma)$ and
    $(C_N,C_K,\sigma)$ is an $N$-marking then either it is $N$-active
    or $C_N=P\term$.  Furthermore, if $P\term\subseteq C_N$ then
    $P\term=C_N$.

    The base case for the induction is straightforward.  Suppose that
    $\pi\ty (\ic{K[N]},\sigma_0) \ltr{}^* M$ where
    $M=(C_N,C_K,\sigma)$ and that $e$ is an event such that $M\ltr e
    M'$.  Let $M'=(C_N',C_K',\sigma')$.  We shall show that $\pi\cdot e$
    from marking $(\ic{K[N]},\sigma_0)$ is of the correct form and
    that $M'$ satisfies the required properties.

    Suppose that $M'$ is an $N$-marking but $C_N'\neq P\term$ and $M'$
    is not $N$-active.  As $M'$ is not $N$-active, we must have
    $C_N'\neq P\init$.  From the induction hypothesis, there must
    exist a path $\pi'$ and markings $C_K''$ and $\sigma''$ such that
    \[\pi'\ty (P\init,C_K'',\sigma'') \ltr{}^* (C_N',C_K',\sigma') \]
    and $(P\init,C_K'',\sigma'')$ is reachable from
    $(\ic{K[N]},\sigma_0)$.  Furthermore, from
    $(P\init,C_K'',\sigma'')$ the path $\pi'$ is between $N$-active
    markings.  Since $K[N]$ is a non-interfering substitution from
    state $\sigma_0$, it follows from the requirement that consecutive
    $K$- and $N$-events must be independent that there must exist
    paths $\pi_1$ and $\pi_2$ made exclusively of $N$- and $K$-events,
    respectively, such that $\pi_1\cdot \pi_2 \ty
    (P\init,C_K'',\sigma'')\ltr{}^* (C_N',C_K',\sigma'')$.  Since
    $N$-events do not affect the marking of $K$-conditions and from
    the requirement that $K$-events do not affect the marking of
    $N$-conditions along the path $\pi_2$ because $K[N]$ is a
    non-interfering substitution from $\sigma_0$, there exists a state
    $\sigma_1$ such that $\pi_1\ty (P\init,C_K'',\sigma'')\ltr{}^*
    (C_N',C_K'',\sigma_1)$.  Since $\rho_N(P\init)=\ic{N}$, a simple
    induction on the length of this sequence using Lemma
    \ref{lemma:rhosim} shows that the marking
    $(\rho_N(C_N'),\sigma_1)$ is reachable from $(\ic{N},\sigma'')$ in
    $N$.  Consider the ways in which the $N$-marking
    $(C_N',C_K',\sigma')$ may fail to be $N$-active: Firstly, if
    $C_N'\subsetneq P\init$, it follows that $\rho_N(C_N')\subsetneq
    \ic{N}$.  Since $(\rho_N(C_N'),\sigma_1)$ is reachable from
    $(\ic{N},\sigma')$, this contradicts the requirement of Definition
    \ref{def:dagger}.  The proof is similar in the other
    cases, $C_N'\subsetneq P\term$ and $P\term\subsetneq C_N'$, which
    may cause the marking to fail to be $N$-active without
    $C_N'=P\term$.

    To complete the proof, it suffices to show the following
    properties:
    \begin{enumerate}
    \item $K$-events preserve $K$-markings: If  $M$ is a
      $K$-marking and $e$ is a $K$-event and $M\ltr e M'$ then $M'$ is a
      $K$-marking.
    \item $N$-events preserve $N$-markings: If $M$ is an 
      $N$-marking and $e$ is an $N$-event and $M\ltr e M'$ then $M'$
      is an $N$-marking.
    \item If $e$ is a $K$-event with no pre- or postcondition inside
      $P\init\cup P\term$ and $M$ is an $N$-marking and $M\ltr{e} M'$
      then $M'$ is an $N$-marking.
    \item The only markings that are both $N$- and $K$-markings are of
      the form $(P\init,C_K,\sigma)$ or $(P\term,C_K,\sigma)$ or
      $(P\init\cup P\term,C_K,\sigma)$ for some $C_K$ and $\sigma$.
    \item No $N$-event has concession in any reachable marking that is
      not $N$-active.
    \end{enumerate}
    Properties (1) and (2) are straightforward calculations using
    Lemmas \ref{lemma:thetapres}, \ref{lemma:contextbij} and
    \ref{lemma:rhosim}.  Property (3) follows immediately from Lemma
    \ref{lemma:kdisj}.  Property (4) is obvious from the definitions
    of $N$- and $K$-markings. Property (5) is straightforward from the
    induction hypotheses and the fact that no event has concession in
    the terminal marking of $N$ according to the requirements of Lemma
    \ref{lemma:embstruc}.

    Finally, to see that any complete run is of the form
    $\Pi_0\cdot (\Pi_1\cdot \Pi_0)^*$, observe that the terminal marking of
    control conditions $\tc{K[N]}$ is a $K$-marking.  There are no
    $C_K$ and $\sigma$ such that the marking $(P\init,C_K,\sigma)$ is
    terminal since then $\pre\khole\cap \tc{K}\neq\emptyset$,
    contradicting the requirement that $K$ should be an embedded net
    satisfying the requirements of Lemma \ref{lemma:embstruc}.  Hence
    the terminal marking is not $N$-active.
  \end{proof}
\end{lem}

\newcommand{\po}{\prec}

Having now dealt with the control structure of contexts, we return to
the idea that, given a net $K[N_1]$ which is a non-interfering
substitution from state $\sigma$, the events in any sequence may be
reordered in a way that ensures that events of $N_1$ occur
consecutively and form a ``complete run'' of the net $N_1$.  As
$N_1\simeq N_2$, the net $K[N_2]$ will therefore have a path between
the same sets of state conditions.

\newcommand{\plt}{\prec}

To formalize this, let $\pi$ be any sequential run of a
non-interfering substitution $K[N]$ from marking $M$. The set
$\mathcal P_{K[N]}(\pi,M)$ is defined to be the least set of sequences
from marking $M$ of $K[N]$ closed under the operation of swapping
consecutive independent events that contains the sequence $\pi$.  It
is easy to see that if $\pi : M \ltr{}^* M'$ and $\pi'\in\mathcal
P_{K[N]}(\pi,M)$ then $\pi':M\ltr{}^* M'$ for any paths $\pi$ and
$\pi'$.  Define the order $\plt$ on $\mathcal P_{K[N]}(\pi,M)$ as
follows:
\begin{defi}
  Let $\pi,\pi'\in \mathcal P_{K[N]}(\pi_0,M)$.
  Define $\plt$ to be the transitive closure of $\plt_1$, where
  $\pi\plt_1 \pi'$ iff there exist sequences $\pi_1$ and $\pi_2$, an
  $N$-event $e$ and a $K$-event $e'$ such that $eIe'$ and
  $\pi=\pi_1\cdot e\cdot e'\cdot\pi_2$ and $\pi'=\pi_1\cdot e'\cdot e \cdot\pi_2$.
\end{defi}
It is clear that the order $\plt$ is well-founded since any path is,
by definition, of finite length.

\begin{defi}
\label{def:ncomp}
  Say that a sequence $\pi$ of $K[N]$ from marking $M$ is
  \emph{$N$-complete} if $M=(P\init,C_K,\sigma)$ for some $C_K$ and
  $\sigma$, every event of $\pi$ is an $N$-event, and \[\pi \ty
  (P\init,C_K, \sigma) \ltr{}^* (P\term, C_K,\sigma').\]
\end{defi}

\begin{lem}
  \label{lemma:minimal}
  Let $K[N]$ be a non-interfering substitution from state $\sigma_0$
  and let $M_0=(\ic{K[N]},\sigma_0)$.  Suppose that $\pi_0$ is a
  complete sequence of $K[N]$ from $M_0$.  The $\plt$-minimal elements
  of $\mathcal P_{K[N]}(\pi_0,M_0)$ are of the form
  \[ \Pi_0\cdot (\Pi_N\cdot \Pi_0)^*, \] where $\Pi_N$ matches $N$-complete
  paths and $\Pi_0$ is as in Lemma \ref{lemma:pathchar}.
  \begin{proof}
    Suppose that $\pi$ is a $\plt$-minimal element of $\mathcal
    P_{K[N]}(\pi_0,M_0)$ but not of the form above.  The sequence
    $\pi$ is of the form of Lemma \ref{lemma:pathchar} because $\pi$
    is a complete path of $K[N]$. Consequently, there are $\pi_1$,
    $\pi_2$ and $\pi_3$ such that $\pi=\pi_1\cdot \pi_2\cdot \pi_3$ and
    $\pi_2=(e\cdot e')$ where $e$ is a $K$-event and $e'$ is an
    $N$-event.  Furthermore, the marking $M_1$ such that $\pi_1\ty
    M_0\ltr{}^* M_1$ is $N$-active.  Now, from the definition of
    non-interfering substitution, the events $e$ and $e'$ are
    independent.  Hence the sequence $\pi_1\cdot e'\cdot e\cdot \pi_3$ is in
    $\mathcal P_{K[N]}(\pi_0,M_0)$ and is beneath $\pi$, contradicting
    its minimality.
  \end{proof}
\end{lem}

This gives us the ability to prove Theorem \ref{theorem:refinement} by
induction on paths of $K[N_1]$.

\newcommand{\case}[1]{\item[(\textsc{#1})]}
\newenvironment{lcases}{
\begin{list}{}{\setlength{\itemindent}{-1em}\setlength{\leftmargin}{2em}}}
{\end{list}}

\newcommand{\rels}[1]{{\color{red} \mathcal{R}\sembr{#1}}}
\newcommand{\cp}{{\color{red}\downarrow}}

\newcommand{\pkn}[1]{\mathcal{P}_{K[N_{#1}]}}

\begin{thm}
\label{proof:refinement}
  If $K[N_1]$ and $K[N_2]$ are non-interfering substitutions from
  $\sigma_0$ and $N_1\simeq N_2$ then, for all states $\sigma$:
  \[ K[N_1]\ty \sigma_0\pto \sigma \qquad\tiff\qquad
  K[N_2]\ty\sigma_0\pto \sigma.\]

  \proof Suppose that $\pi$ is a complete sequence of $K[N_1]$ from
  $\sigma_0$ to $\sigma'$.  We shall show that, for all
  $\pi_1\in\mathcal P_{K[N_1]}(\pi,(\ic{K[N_1]},\sigma_0))$, if
  $\pi_1$ is a complete sequence from $\sigma_0$ to $\sigma'$ then
  there exists a complete sequence $\pi_2$ of $K[N_2]$ from $\sigma_0$
  to $\sigma'$.  The proof shall proceed by induction on the
  well-founded order $\plt$.  In particular $\pi\in \mathcal
  P_{K[N_1]}(\pi,(\ic{K[N]},\sigma_0))$, so, with the symmetric proof
  for the other direction, this will complete the proof of the
  required property.

    \newcommand{\sx}{^{(1)}}
    \newcommand{\sy}{^{(2)}}
    \begin{lcases}
      \case{$\pi_1$ minimal} The sequence $\pi_1$ is minimal within
      $\mathcal P_{K[N_1]}(\pi,\sigma_0)$, so, by Lemma
      \ref{lemma:minimal}, there exists an $n\in \mathbb N$ such that
      there exist sequences
      $\pi_0,\pi_{01},\pi_{11},\ldots\pi_{0n},\pi_{1n}$ with
      \[\pi_1 = \pi_0\cdot \pi_{11}\cdot \pi_{01}\ldots\pi_{1n}\cdot \pi_{0n}. \]
      Furthermore, for each $i\leq n$, the sequence $\pi_{0i}$ is of
      the form $\Pi_0$ defined in Lemma \ref{lemma:pathchar}, as is
      the sequence $\pi_0$; and, for each $i\leq n$, the sequence
      $\pi_{1i}$ is of the form $\Pi_{N_1}$, which matches
      $N_1$-complete subpaths of $K[N_1]$ as defined in Definition
      \ref{def:ncomp}.  Define:
      \[
      \begin{array}{rcl@{\qquad\qquad}rcl}
        P\init\sx & \eqd& \pref 1 \pre\khole\times \pref 2 \ic{N_1} &
        P\init\sy & \eqd& \pref 1 \pre\khole\times \pref 2 \ic{N_2} \\
        P\term\sx & \eqd& \pref 1 \post\khole\times \pref 2 \tc{N_1} &
        P\term\sy & \eqd& \pref 1 \post\khole\times \pref 2 \tc{N_2} .
      \end{array}
      \]
      Let $\rho\sx_K$ be $\rho_K$ from Definition \ref{def:contproj}
      for $K[N_1]$ and let $\rho\sy_K$ be $\rho_K$ from Definition
      \ref{def:contproj} for $K[N_2]$, and similarly for $\rho\sx_{N_1}$,
      $\rho\sy_{N_2}$, $\theta\sx_K$, \etc{} We shall show, by induction
      on $n$, that if $\pi_1$ is a sequence of this form in $K[N_1]$
      from $(\ic{K[N_1]},\sigma_0)$ to the marking $(C_1',\sigma')$
      then there exists a path $\pi_2$ from $(\ic{K[N_2]},\sigma_0)$
      to $(C_2',\sigma')$ for some $C_2'$ such that
      $\rho\sx_K(C_1')=\rho\sy_K(C_2')$.
      \begin{itemize}
      \item $n=0$: Assume that $\pi_1$ is of the form $\Pi_0$.  Let
        $\pi_1=(e_1\cdot \ldots\cdot e_m)$ and suppose that in $K[N_1]$ we
        have
        \[ (\ic{K[N_1]},\sigma_0) \ltr {e_1} (C_1,\sigma_1)
        \ltr{e_2}\ldots \ltr{e_m} (C_m,\sigma_m). \] By assumption,
        $\pi_1$ is a path from $(\ic{K[N_1]},\sigma_0)$ to $(C_1',\sigma')$,
        so $C_1'=C_m$ and $\sigma'=\sigma_m$.  Now, $\ic{K[N_1]}$ is a
        $K$-marking, and, since $\pi_1$ is of the form $\Pi_0$, for
        every $i$ such that $0< i\leq m$, the marking $C_i$ is a
        $K$-marking and $e_i$ is a $K$-event.  By Lemma
        \ref{lemma:rhosim}, in the net $K$ we have
        \[ 
        (\rho\sx_K(\ic{K[N_1]}),\sigma) \ltr {\rho\sx_K(e_1)} 
        (\rho\sx_K(C_1),\sigma_1) \ltr {\rho\sx_K(e_2)} \ldots
        \ltr {\rho\sx_K(e_m)} (\rho\sx_K(C_m),\sigma_m). 
        \]
        In the net $K[N_2]$, by Lemma \ref{lemma:thetasim}, we
        therefore have
        \begin{eqnarray*}
        (\theta\sy_K\rho\sx_K(\ic{K[N_1]}),\sigma) &\ltr
        {\theta\sy_K\rho\sx_K(e_1)} &
        (\theta\sy_K\rho\sx_K(C_1),\sigma_1)\\
& \ltr
        {\theta\sy_K\rho\sx_K(e_2)}& \ldots\\
& \ltr
        {\theta\sy_K\rho\sx_K(e_m)}&
        (\theta\sy_K\rho\sx_K(C_m),\sigma_m).
        \end{eqnarray*}
        Let $C_2'=\theta\sy_K\rho\sx_K(C_m)$.  From Lemma
        \ref{lemma:thetapres}, $\theta\sy_K$ generates $K$-markings of
        $K[N_2]$ from markings of $K$.  By Lemma
        \ref{lemma:contextbij}, we therefore have
        $\rho\sy_K(C_2')=\rho\sx_K(C_1')$ since $C_1'=C_m$, which is a
        $K$-marking.  It is an easy calculation to show that
        $\rho\sx_K(\ic{K[N_1]})=\ic{K}$ and
        $\theta\sy_K(\ic{K})=\ic{K[N_2]}$.  There therefore exists a
        path from $(\ic{K[N_2]},\sigma_0)$ to $(C_2',\sigma_m)$ in
        $K[N_2]$ and $\rho\sy_K(C_2')=\rho\sx_K(C_1')$ , which is all that
        is required since $\sigma_m=\sigma'$.
      \item $n>0$: Assume that $\pi_1 = \pi_{11}\cdot \pi_{12}\cdot \pi_{13}$ for some
        sequence $\pi_{11}$ of form $\Pi_0\cdot (\Pi_{N_1}\cdot \Pi_0)^{n-1}$, some
        sequence $\pi_{12}$ of form $\Pi_{N_1}$ and some sequence $\pi_{13}$
        of form $\Pi_0$.  Let $(C_1',\sigma')$ be the marking obtained by
        following $\pi_{1}$ from $(\ic{K[N_1]},\sigma_0)$ in $K[N_1]$.
        We wish to show that there is a path $\pi_2$ of $K[N_2]$ from
        $(\ic{K[N_2]},\sigma_0)$ to $(C_2',\sigma')$ for some $C_2'$ such
        that $\rho\sx_K(C_1')=\rho\sy_K(C_2')$.
        
        Let $(C_{11},\sigma_1)$ be the marking obtained by following
        path $\pi_{11}$ from $(\ic{K[N_1]},\sigma_0)$.  Since
        $\pi_{12}$ follows $\pi_{11}$ and  $\pi_{12}$ is of form
        $\Pi_{N_1}$, it must be the case that $C_{11}=(P\sx\init,C_K)$
        for some marking $C_K$ of $K$-conditions.  
        
        By induction, there is a path $\pi_{21}$ in $K[N_2]$ from
        $(\ic{K[N_2]},\sigma_0)$ to $(C_{21},\sigma_1)$ for some
        $C_{21}$ such that $\rho\sx_K(C_{11})=\rho\sy_K(C_{21})$.
        Now, $C_{11}=(P\sx\init,C_K)$, so
        $\rho\sx_K(C_{11})=\pre\khole\cup \{c\st \pref 1 c \in C_K\}$.
        From the definition of $\rho\sy_K$, we must therefore have
        $C_{21}=(P\sy\init,C_K)$.  Hence \[\pi_{21}\ty
        (\ic{K[N_2]},\sigma_0)\ltr{}^* (P\sy\init,C_K,\sigma_1).\]

        Suppose that in $K[N_1]$ we have $ \pi_{12}\ty
        (P\sx\init,C_K,\sigma_1) \ltr{}^* (C_{N_1}',C_K',\sigma_2)$.
        Since $\pi_{12}$ is of the form $\Pi_{N_1}$, it is an
        $N_1$-complete path, so $C_{N_1}'=P\sx\term$.  The events of
        $\pi_{12}$ are all $N$-events.  Using Lemma
        \ref{lemma:rhosim}, a simple induction shows that $C_K=C_K'$
        and that there is a path from
        $(\rho\sx_{N_1}(P\sx\init),\sigma_1)$ to
        $(\rho\sx_{N_1}(P\sx\term),\sigma_2)$ in $N_1$.  Observe that
        $\rho\sx_{N_1}(P\sx\init)=\ic{N_1}$ and
        $\rho\sx_{N_1}(P\sx\term)=\tc{N_1}$, so $N_1\ty \sigma_1 \pto
        \sigma_2$.  As $N_1\simeq N_2$, there is therefore a path of
        $N_2$ from $(\ic{N_2},\sigma_1)$ to $(\tc{N_2},\sigma_2)$.  By
        Lemma \ref{lemma:thetasim}, a simple induction on the length
        of this sequence shows that there is a sequence $\pi_{22}$ from
        $(\theta\sy_{N_2}(\ic{N_2}),C_K,\sigma_1)$ to
        $(\theta\sy_{N_2}(\tc{N_2}),C_K,\sigma_2)$ in $K[N_2]$.
        Observe that $\theta\sy_{N_2}(\ic{N_2})=P\sy\init$ and
        $\theta\sy_{N_2}(\tc{N_2})=P\sy\term$, so
        \[ \pi_{22} \ty (P\sy\init,C_K,\sigma_1) \ltr{}^*
        (P\sy\term,C_K,\sigma_2).\]
        
        As $\pi_{13}$ follows path $\pi_{12}$ in $\pi_1$, the sequence
        $\pi_{13}$ is from $(P\sx\term,C_K,\sigma_2)$ to
        $(C_1',\sigma')$ and contains only $K$-events.  Using Lemma
        \ref{lemma:rhosim}, a simple induction on the length of
        $\pi_{13}$ shows that there is a path from
        $(\rho\sx_K(P\sx\term,C_K),\sigma_2)$ to
        $(\rho\sx_K(C_1'),\sigma')$ in $K$.  A simple induction on the
        length of this path, using Lemma \ref{lemma:thetasim} shows
        that there is a path $\pi_{23}$ of $K[N_2]$ such that $
        \pi_{23} \ty (\theta\sy_K\rho\sx_K(P\sx\term,C_K),\sigma_2)
        \ltr{}^* (\theta\sy_K\rho\sx_K(C_1'),\sigma') $.  From the
        definition of $\rho\sx_K$, we have
        $\rho\sx_K(P\sx\term,C_K)=\post\khole \cup \{c\st \pref 1 c\in
        C_K\}$.  From the definition of $\theta_K\sy$, we have
        $\theta_K\sy(\post\khole \cup \{c\st \pref 1 c\in
        C_K\})=P\sy\term\cup C_K$.  Hence
        \[ 
        \pi_{23} \ty (P\sy\term,C_K,\sigma_2) \ltr{}^*
        (\theta\sy_K\rho\sx_K(C_1'),\sigma').\] Take
        $C_2'=(\theta\sy_K\rho\sx_K(C_1'),\sigma')$.  By Lemma
        \ref{lemma:contextbij}, we have
        $\rho_K\sy(C_2')=\rho_K\sx(C_1')$.  Consequently, the path
        $\pi_2=\pi_{21}\cdot\pi_{22}\cdot\pi_{23}$ satisfies
        \[ \pi_2\ty (\ic{K[N_2]},\sigma_0)\ltr{}^* (C_2',\sigma'),\]
        for some $C_2'$ such that $\rho_K(C_1')=\rho_K(C_2')$, which
        is all that is required to complete this inner induction.
      \end{itemize}
      Now, recall that $\pi_1$ is a complete sequence of $K[N_1]$, so
      \[\pi_1\ty (\ic{K[N_1]},\sigma_0)\ltr{}^* (\tc{K[N_1]},\sigma').\]
      From the immediately preceding induction, there exists a path
      $\pi_2$ of $K[N_2]$ such that
      $\pi_2\ty(\ic{K[N_2]},\sigma_0)\ltr{}^* (C_2',\sigma')$ for some
      $C_2'$ such that $\rho\sx_K(\tc{K[N_1]})=\rho\sy_K(C_2')$.  Now,
      clearly $\rho\sx_K(\tc{K[N_1]})=\tc{K}$ by the definitions of
      $\rho$ and $K[N_1]$.  Hence $\tc K = \rho\sy_K(C_2')$, so by
      Lemma \ref{lemma:contextbij} we have $\theta_K\sy(\tc K) =
      C_2'$.  The definition of $K[N_2]$ and $\theta$ gives
      $\theta_K\sy(\tc K)=\tc{K[N_2]}$.  Hence 
      \[ \pi_2 \ty (\ic{K[N_2]},\sigma_0) \ltr{}^* (\tc{K[N_2]},\sigma'),\]
      as required.
      
      \case{$\pi_1$ not minimal} Suppose that the path $\pi_1$ is not
      minimal and that $\pi_1$ is a complete path of $K[N_1]$ with
      $\pi_1\ty (\ic{K[N_1]},\sigma_0)\ltr{}^* (\tc{K[N_1]},\sigma')$.
      It is easy to see that the order $\plt$ is irreflexive, so there
      exists a path $\pi_1'$ such that $\pi_1'\plt_1 \pi_1$.  Hence
      there exist paths $\pi_2$ and $\pi_3$ and a $K$-event $e$ and an
      $N$-event $e'$ such that $\pi_1=\pi_2\cdot e\cdot e'\cdot\pi_3$ and
      $\pi_1'=\pi_2\cdot e'\cdot e\cdot\pi_3$.  Furthermore, the events $e$
      and $e'$ are independent, so $\pi_1'$ must also be a path
      $\pi_1'\ty (\ic{K[N_1]},\sigma_0)\ltr{}^*
      (\tc{K[N_1]},\sigma')$.  By induction, there exists a path
      $\pi_2'\ty (\ic{K[N_1]},\sigma_0)\ltr{}^*
      (\tc{K[N_1]},\sigma')$, as required to complete the case.
    \end{lcases}
    Hence, if $K[N_1]\ty \sigma_0 \pto \sigma'$, there exists a path
    $\pi\ty (\ic{K[N_1]},\sigma_0)\ltr{}^*(\tc{K[N_1]},\sigma')$ in
    $K[N_1]$.  Since $\pi\in\mathcal P_{K[N_1]}(\pi,\sigma_0)$, we
    have a path
    $\pi_2\ty(\ic{K[N_2]},\sigma_0)\ltr{}^*(\tc{K[N_2]},\sigma')$ in
    $K[N_2]$, so $K[N_2]\ty \sigma_0\pto \sigma'$.  The proof for the
    reverse implication is symmetric.\qed
\end{thm}


\end{document}